\documentclass{aa_5}
\usepackage{rotating}   
\usepackage{longtable}   
\usepackage{graphicx, natbib, psfrag, latexsym, amssymb}
\bibstyle{aa}  
\def \bb{ }       

\def \etal   {{~et~al.~}}
\def\ltsim{\mathrel{<\kern-1.0em\lower0.9ex\hbox{$\sim$}}}
\def\gtsim{\mathrel{>\kern-1.0em\lower0.9ex\hbox{$\sim$}}}

\def\WSRT{\scriptscriptstyle WSRT}
\def\Eff{\scriptscriptstyle Effelsberg}
\def\chiW{\chi_{\rm \WSRT}}
\def\chiE{\chi_{\rm \Eff}}
\begin{document}
\title{The B3-VLA CSS sample. \\
       VII: WSRT  Polarisation Observations  
       and the Ambient Faraday Medium Properties Revisited }

\author{ A. Rossetti  \inst{1}   \and 
         D. Dallacasa \inst{1,2} \and 
	 C. Fanti     \inst{1}   \and 
	 R. Fanti     \inst{1}   \and
	 K.-H. Mack   \inst{1} 
         }

\offprints{A. Rossetti,\\
  \email{rossetti@ira.inaf.it}}

\institute {Istituto di Radioastronomia -- INAF, Via Gobetti 101,
  I-40129 Bologna, Italy  
\and Dipartimento di Astronomia, Universit\`a di Bologna,
Via Ranzani 1, I-40127 Bologna, Italy}

\date{Received \today; Accepted ???}

      %
      %
      \titlerunning{B3-VLA CSSs: polarisation}
      \authorrunning{Rossetti \etal}

\abstract {
{\bb We present new polarisation  observations at 13~cm, acquired
  using the Westerbork Synthesis Radio Telescope (WSRT), of 65 sources,}
  from the B3-VLA {\bb sample} of Compact Steep-Spectrum sources. These new
  data are combined with {\bb our} VLA polarisation data, at 3.6, 6 and, 21~cm,
  presented in a previous paper. 
Due to the multi-channel frequency capabilities of the WSRT, these new
  13~cm observations enable {\bb a more reliable determination of integrated
  Rotation Measures, and of depolarisation behaviour with
  wavelength. The new data are inconsistent with the  depolarisation
  models that we used earlier, and we propose an alternative model
  which seems to work properly. 
  We also revise our previous model for the
  external Faraday screen, and its dependence on the source redshift.} 

}
\maketitle

\section {Introduction}

 Compact Steep-Spectrum (CSS) sources and GHz-Peaked Spectrum (GPS)
sources, because of their small size, are fully embedded in the
Interstellar Medium (ISM) of the host galaxy.  
Their radio properties are affected  by the
properties of the ISM, unlike large-size radio sources which extend
well beyond the optical dimensions of their host galaxies, 
and can therefore be used to probe the conditions of the ISM.
Polarisation characteristics are a useful tool for this purpose.

The measurable quantities from polarisation observations are:
\begin{itemize}
\item The Faraday Rotation Measure ({\it RM}), defined as the slope of
  a linear fit of the polarisation angles, as a function of
  $\lambda^2$: \[\chi(\lambda)  = RM~\lambda^2 + \chi(0).\] If the
  medium is homogeneous (both its density and magnetic field), or if
  {\bb the} inhomogeneities are resolved by the observing beam,
  $\chi(\lambda)$ is strictly proportional to $\lambda^2$ at all
  wavelengths, and $RM = k \int  B_{||} n d\ell$  (this quantity is
  also called Faraday Depth), where  $B_{||}$ is the component of the
  magnetic field along the line of sight, $n$ is the electron density
  of the medium, $\ell$ is the geometrical depth of the medium along
  the line of sight (los), and $k$ is a constant. 
  If the medium is unresolved, or partially resolved by the observing
  beam, {\it RM} changes from point to point across the source, different
  contributions of polarised radiation are rotated differently,
  and  $\chi(\lambda)$  can deviate, more or less strongly, from the
  $\lambda^2$-linear law. 

\item The change of the fractional polarisation $m{_\lambda}$ with
   $\lambda$, described by the quantity $DP(\lambda) =
   m_{\lambda}/m_0$. If the medium is uniform, or if its
   inhomogeneities are well resolved by the observing beam,
   $DP(\lambda) = 1$. 
   In an inhomogeneous medium, the variations of {\it RM} from point to
   point  will produce  changes in $m_{\lambda}$ versus
   $\lambda$. $DP(\lambda)$ will exhibit different behaviours with
   $\lambda$, depending on the properties of the inhomogeneities inside the medium.
\end{itemize}

Several models have been developed to interpret  the
Faraday effects (e.g. ~\citealt{Burn66, Tribble91}; see also
~\citealt{Laing84} for an excellent {\bb and concise} review).   
The behaviour of the polarisation angle, $\chi(\lambda)$, and
fractional polarisation, $m_{\lambda}$, as a function of $\lambda^2$,
enables the ``average'' Faraday Rotation Measure, and, from 
the ``screen models'', its 
dispersion, $\sigma_{\rm RM}$, to be determined. 
Using these data, we can obtain information on the density
distribution of the ISM that surrounds the radio source, on its
clumpiness, and  on both the ordered and tangled components of the
magnetic field. 
  
Polarisation studies of CSS and GPS source samples have been conducted
by several authors ~\citep[e.g.][]{vanbru84,AG95,Stang98,Peck00}. 
They have found that GPS sources are almost
unpolarised, while the larger-size CSS sources can show 
large {\it RM}s and/or large depolarisations, as a function of $\lambda^2$.

In a previous paper ~\citep[][hereafter Paper IV]{Fanti04}, we
used ``low resolution'' polarisation measures at 8.5 and 4.9~GHz
($\approx 3.6$ and 6~cm) from ~\citet[][hereafter Paper I]{Fanti01}, and
at 1.4~GHz ($\approx 21$~cm) from the NVSS ~\citep{nvss}, to
{\bb derive } the {\it RM} and the depolarisation properties of a complete
sample of CSSs, the B3-VLA CSS sample (\citeauthor*{Fanti01}). Our
main results were:
\begin{enumerate}
\item In general, the total source depolarisation was found to follow
  either the Burn or the Tribble model. 

\item In $\ge 50$\% of the cases, the integrated $\chi(\lambda)$
  follows the $\lambda^2$-linear law from 3.6 to 20~cm. The derived
  {\it RM}s have {\bb  values} of up to a few hundred rad\,m$^{-2}$. After
  subtraction of the Galactic Rotation, and correction for the source
  redshift, {\it z},  
  we found that $\approx 20$\% of the sources have intrinsic
  $RM$s of up to 1000~rad\,m$^{-2}$. 

\item There is a wavelength-dependent characteristic scale (from
  $\approx 3$~kpc at 3.6~cm to $\ge 6$~kpc at 21~cm), below which radio
  sources are almost totally depolarised ~\citep[see also][]{Cotton03}.


\item $\sigma_{\rm RM}$ increases with redshift; a similar, but less
  significant, dependence was suggested  for {\it RM} (Figs.~13 and 14
  in ~\citeauthor*{Fanti04}). 
\end{enumerate}

To explain our results, we proposed a simple model, based on 
Faraday effects, with {\it an appropriate spatial} 
distribution of the ambient gas density and magnetic field.

The results of ~\citeauthor*{Fanti04}, however, were based on 
polarisation data with a non-optimal wavelength coverage, because
the gap in $\lambda^2$, between 6 and 20~cm, is too wide.
In a  number of cases there were remaining {\bb ambiguities } in both {\it RM} 
and $\sigma_{\rm RM}$. In fact, for the few sources for which a
polarisation measurement was available at the intermediate wavelength
of 11~ cm ~\citep{Klein03} the initial model was not 
supported in a number of cases (see Figs.~4 and 10 in ~\citeauthor*{Fanti04}).

{\bb To improve our polarisation information we performed} new
polarisation observations at 13~cm, using
the Westerbork Synthesis Radio Telescope (WSRT) for 65 radio
sources, 58 of which were detected in polarisation at
one or more of the three available VLA frequencies. The remaining 7
unpolarised objects were observed as control sources.

The new observations fill the large gap in $\lambda^2$
between the 6 and 21~cm VLA data.  
They  allow us to improve the reliability of the
{\it RM}s significantly  by reducing the ambiguities, and   
constrain the {\bb depolarisation behaviour as a function of $\lambda^2$}.

\smallskip
Section~\ref{the-sample} provides a short description  of the B3-VLA CSS 
sample and of the
previous polarisation data (3.6, 6, and 21~cm, VLA), and presents the
selection criteria for the WSRT sub-sample.

Sections~\ref{pol_data} describes the new WSRT polarisation
observations at 13~cm, the data reduction strategy, and the derived results.

Section~\ref{poldataset} summarizes the polarisation status of the
WSRT sub-sample.

Section~\ref{results} presents the results on Rotation Measure ({\it RM}) and
its dispersion  ($\sigma_{\rm RM}$).

Section~\ref{discussion} revisits the model of the ambient
magneto-ionic medium.

Section~\ref{concl} provides our conclusions.

Appendix~\ref{app_1} contains the data table and comments on
individual sources.

Appendix~\ref{pol_mod} describes a simple two-polarised-component
model, which has been applied to a minority of radio sources.

\section{The WSRT sample }
\label{the-sample}

The sources discussed in this paper were selected from the B3-VLA CSS
sample ~\citep{vig1}, described in ~\citeauthor*{Fanti01}, 
which consists of 87 CSSs/GPSs (three of which do not have polarisation
data) with {\bb flux density  $\ge$ 0.8~Jy} at 408~MHz, with projected Linear 
Sizes ({\it LS})\footnote{In this paper we have kept H$_0=
  100\:h$~km\,s$^{-1}$\,Mpc$^{-1}$, and $q_0=0.5$ for consistency with
  previous papers. We have also used the Concordance Cosmology, with
  H$_0 =71\:h$~km\,s$^{-1}$\,Mpc$^{-1}$, $\Omega_{\rm M}=0.27$,
  $\Omega_{\rm{vac}}= 0.73$, and found that the results discussed in
  Sect.~\ref{discussion} remain largely unchanged.} in the range
 $0.2\,h^{-1} \leq LS$~(kpc) $\leq 20\,h^{-1}$. 
Their radio luminosity, at the selection frequency, is 
P$_{0.4\,\rm{GHz}} \geq 10^{26}\,h^{-2}$~W\,Hz$^{-1}$.
The sources were observed using the VLA in A configuration, at 6
and 3.6~cm in total intensity and  polarisation (see \citeauthor*{Fanti01}). 
A detailed description of the polarisation data reduction was provided in
\citeauthor*{Fanti04}.
In addition,  polarisation data at 21~cm are available from the NVSS 
\citep{nvss}.

A sub-sample of 65 sources, hereafter referred to as the ``WSRT
sub-sample'', was observed using the WSRT at 13~cm. At this wavelength,
most sources, according to {\bb the} previous VLA 
total polarisation measurements, {\bb were} expected to have
a polarised flux density $S_{\rm P}\gtsim 1$~mJy.
Seven sources, undetected in polarisation at 3.6, 6 and 21~cm,
were observed as a control sample.

The 19 B3-VLA CSS sources that were not observed at 13~cm, were either
unpolarised, or strongly depolarised at 6~cm and 20 cm.
 
The WSRT sub-sample includes 45 out of the 54 B3-VLA CSS sources of
Linear Sizes larger than 2.5~kpc. Of the  missing sources, 2 are
unpolarised at all frequencies, 6 are polarised at 3.6~cm only, and
one, polarised at 3.6 and 6~cm, does not meet the 1-mJy selection
criterion.

\section{The Polarisation Data}
\label{pol_data}

\subsection{The WSRT Observations} 
\label{data-red}

The observations were {\bb carried out} in  November 2004, using the WSRT, in
dual circular polarisation mode, at the mean frequency of 2263~MHz
($\approx 13$~cm). Eight Intermediate Frequencies (IFs) were used,
{\bb  each one} divided into 64 identical frequency channels. 

Using the full available bandwidth of 128~MHz,
the expected root mean square (rms) noise in the I, U, Q Stokes parameters
($\mu_{{\rm{\scriptscriptstyle I,U,Q}}}$), for a 20-minute integration,
is $\sim 0.14$~mJy/beam. 
The $8\times 64=512$ independent frequency channels enable radio
interferences to be removed reliably, which helps to bring the noise
level close to its theoretical value. 

Thanks to its sensitivity and high resolution ($\sim8\arcsec$
at 13~cm), the WSRT is at present the only instrument, in the northern
hemisphere, that can detect such low-levels of polarised flux-density,
at 13~cm. 
The Effelsberg telescope, for example, is
confusion-limited in polarisation at a level of $\approx 0.5$~mJy 
~\citep{Klein03} at the closeby wavelength of 11~cm (beam size
$\approx 4\farcm3$). 
 
We observed each target source in snapshot mode at three
well-spaced hour-angles of approximately $(0^{\rm{h}}$, $\pm4^{\rm{h}}$), 
for a total integration time of $\approx 20$~minutes. 
Flux-density and phase-calibrator sources were observed, on average,
every 4~hours,  for 15~minutes.

\subsection{Data reduction}
\label{data_red}

All data reduction (editing, calibration, imaging and analysis)
was performed with the NRAO package AIPS (Astronomical Image
Processing System).   
Interference spikes were removed using the task UVLIN. 

\subsubsection{Calibration}
3C\,286 was used as a primary calibrator, for flux-density, phase, 
bandpass and polarisation angle. Secondary calibrators were 3C\,147 
and CTD\,093 (known to be unpolarised at 13~cm).
The flux-density calibration uncertainty was $\sim 1$\%. The flux
density {\bb scale is  within 3\%  of that of ~\citet{Baars78}}.

To determine the residual instrumental polarisation
(D-term), the task LPCAL was applied to the observation of CTD\,093. 
The measured value was approximately 0.1\% of the source
flux density ($S_{\rm I}$), which {\bb is} consistent with {\bb the}
results for {\bb the} 7 unpolarised sources of the sample 
(Sect.~\ref{the-sample}).
After the D-term calibration, an arbitrary offset in the
polarisation angle $\chi$ remains, which was determined by using
integrated measurements of the source 3C\,286. Each IF was corrected 
separately. We were unable to obtain a good calibration for IF1 and
IF7. The data reduction was, therefore, based on 6~IFs out of 8
for polarisation data, while total intensity data were derived using
all 8~IFs.

Based on the r.m.s. of the polarisation angles
obtained for each IF and each scan of 3C\,286, we estimate that the
polarisation-angle calibration is accurate to within 1\fdg4.

\subsubsection{Imaging}
\label{flux_d}
The three  snapshots of each source were combined (task IMAGR), to
produce  two-dimensional  ``dirty images'' of  $1024\times1024$ pixels
($\sim 43\arcmin \times 43\arcmin$) for  the Stokes parameters I, Q,
U, and occasionally V. This size was usually sufficiently wide to identify 
and remove all confusing field sources occasionally present in
polarisation. We then cleaned the images  down to the theoretical
noise level.

At the WSRT resolution ($\approx 8\arcsec$), all sources were
unresolved.
We derived {\it total-band} flux densities  ($S_{\rm_I}$, $S_{\rm_Q}$,
$S_{\rm_U}$) for the Stokes parameters I, Q, U by fitting
a bidimensional Gaussian to the brightness distribution (task IMFIT),
and setting the search boxes for Q and U about the source, {\bb as}
visible on the total  intensity image (I). When the Q and/or U signal
were too weak for reliable values to be produced using IMFIT, we used
instead IMEAN, which integrates the surface brightness inside the
box. Using these measurements, we derived the polarised flux density
$S_{\rm P} = \sqrt {S_{\rm_Q}^2+S_{\rm_U}^2}$.

A noise-dependent statistical correction was applied to the polarised
flux-densities to correct for polarisation bias
~\citep{Wardle74,Simmons85,nvss}. {\bb The de-biased polarised
flux-density is} $S_{\rm P}^{\rm db} = \sqrt{S_{\rm P}^2 - \mu_{\rm P}^2}$, 
where $\mu_{\rm P}$ is the noise error of $S_{\rm P}$
(see Sect.~\ref{noise}). According to ~\citet{Wardle74}, this {\bb formula}
is appropriate for $S_{\rm P}/ \mu_{\rm P} \gtsim 1.2$. 
The fractional polarisation, $m=S_{\rm P}^{\rm db}/S_{\rm I}$, was
also obtained.  

Thirty-three sources were detected in polarisation at 13~cm, at
levels $>3\mu_{\rm P}$, using the total bandwidth.

\subsubsection{Noise estimate and errors}
\label{noise}
The pixel histograms in ``empty'' regions of the I, Q and U images,
and the pixel statistics of the V images, are approximately Gaussians, and
are {\bb in} agreement with each other. We can therefore assume that the
rms is a reliable noise
estimate. The typical  r.m.s. error is $\approx 0.15$~mJy/beam for
all Stokes parameters, in agreement with expectations.

{\bb We adopted noise errors appropriate to each individual source,
  and we used the statistics of the
residuals provided by AIPS, after the Gaussian fit, to derive
the values of $\mu_{\rm  I},\mu_{\rm  Q},\mu_{\rm U}$.} We assume that
this procedure takes account of the fit quality, and the possible
confusion {\bb by} residual sidelobes from field sources. The distribution
of the individual noise errors is approximately Gaussian, {\bb 
 ranging in value between 0.08~mJy/beam and 0.24~mJy/beam,
 with a peak at $\approx 0.15$~mJy/beam, in agreement with the above 
 estimate.}
 
For $\mu_{\rm Q}$ and $\mu_{\rm U}$, when IMEAN had to be used,
the noise errors were computed using the above r.m.s. $\approx
0.15$~mJy/beam, scaled by the square-root of the number of beam areas
included in the search area. The total noise error, $\mu_{\rm P}$, was
computed by taking into account the different values of $\mu_{\rm Q}$
and $\mu_{\rm U}$.  
The uncertainties of the flux-density calibration, and  of
the residual instrumental polarisation, were quadratically added to the
noise error, to derive the final errors of $S_{\rm I}$ and $S_{\rm P}^{\rm db}$
($\sigma_{\rm I}$ and $\sigma_{\rm P}$). 
The error of $m$ was computed using error propagation:

\[\sigma_{m}\approx \frac{1}{S_{\rm I}} \sqrt{\sigma_{\rm P}^2 + \left
  (S_{\rm P}^{\rm db} \frac{\sigma_{\rm I}}{S_{\rm I}}\right)^2.}\]

The noise error of $\chi$ was {\bb also} computed, using error propagation:
\[\mu_\chi\approx\frac{1}{2S_{\rm P}^2} \sqrt{(S_{\rm Q}\mu_{\rm
    U})^2+(S_{\rm U}\mu_{\rm Q})^2.}\]
According to ~\citet{Wardle74}, this {\bb formula} is correct only 
for  $m \gtrsim 2 \sigma_m$. 

The total error, $\sigma_\chi$, was computed by quadratically adding the
calibration error of $\approx 1\fdg4$ to the noise error.

\subsubsection{In-band Polarisation Data}
\label{in_band}

For 24 sources with signal-to-noise ratios $\gtsim 10$,
we measured the polarisation parameters for each IF 
(in-band data). For all but two sources, we detected signal 
in {\bb the} individual IFs, with a signal-to-noise ratio $\gtsim 5$.
 
For the determination of the flux densities  of the Stokes parameters 
in individual IFs, we used the same procedure as for the total band
(Sect.~\ref{flux_d}). 
The errors of $S_{\rm Q}$ and $S_{\rm U}$ were estimated from the
r.m.s. of the six $S_{\rm P}$ of the individual IFs. 
The average value was $\sigma_{\rm P}^*= 0.5$~mJy/beam, rather than the
expected 0.4~mJy/beam. 
This value is consistent with the r.m.s. of the in-band polarisation
angles, computed for sources that show no significant gradient
of the angle across the band.
We therefore empirically adopted the value $\sigma_{\rm P}^*=0.5$~mJy/beam
as the error  of the polarised flux density $S_{\rm P}$ of each IF. 

No use was made of the in-band polarised flux density, while the
$\chi$ from the individual IFs were linearly fitted, and the derived
in-band $RM_{13}$ used as a guide, to solve the $n \pi$ ambiguities of
the polarisation angles, when determining the total {\it RM}.

\subsubsection{``Recalibration'' of  the Polarisation  position angle}
\label{chi_wsrt}

The polarisation position angle  was computed  initially 
{\bb as} $\chi = \frac{1}{2} \tan^{-1} (S_{\rm U}/S_{\rm Q})$. 
{\bb When we compared these angles with those obtained by
  interpolation at 13~cm, with a $\lambda^2$ law, of the VLA data from 
~\citeauthor*{Fanti04},  we found large
  disagreements. At this point we decided to use, as internal calibrators, the
  8 sources B3\,0110+401, B3\,0213+412, B3\,0754+396, B3\,0800+472, 
B3\,0805+406, B3\,0955+390, B3\,1220+408, and B3\,1343+386, which were also
  observed by ~\citet{Klein03}, using the Effelsberg telescope, at
  the close wavelength $\lambda = 11$~cm. The polarisation angles
  measured with the  WSRT,
  $\chiW$, and those measured at Effelsberg,  $\chiE$, are plotted in
  Fig.~\ref{fig:chi_test}. 


    \begin{figure}[htbp]
      \centering
      \resizebox{\hsize}{!}
      {\includegraphics{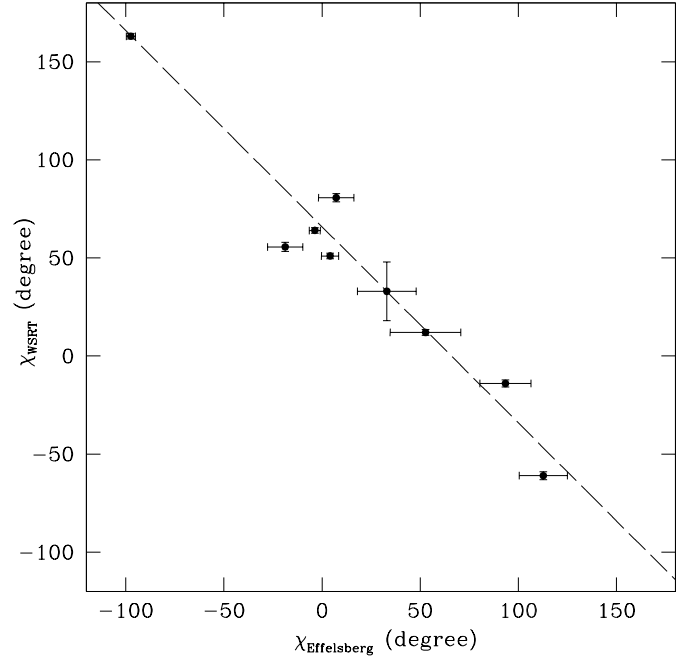}}          
      \caption{Comparison between  $\chiW$ and $\chiE$. 
	Errorbars of WSRT data do not usually exceed the plotted
	symbol size. The sources in the plot are: 3C\,286 (primary
	calibrator; large cross at [33,33]) and 8 sources observed
	using the Effelsberg telescope by ~\citet{Klein03} (see
	text). The r.m.s. about the $\lambda^2$-linear fit is
	$\approx 10\degr$, compatible with the errors.} 
      \label{fig:chi_test}
    \end{figure}

 The two sets of angles  are
 clearly related to each other by the relation $\chiE=-
 \chiW+66\degr$.} Obviously $\chiE$ and 
 $\chiW$ coincide for 3C\,286, because the polarisation angles
 were calibrated using this source. 
  
Given the AIPS definition for crossed circular  polarisation,
RL = Q $+i$U, LR = Q $-i$U,  i.e. U = $i$(LR - RL), Q = RL $+$ LR,
($i = \sqrt{-1}$), our result  implies that there is a swap between RL
and LR.

\begin{figure*}[htbp]
      \centering
      {\includegraphics[width=8.5cm]{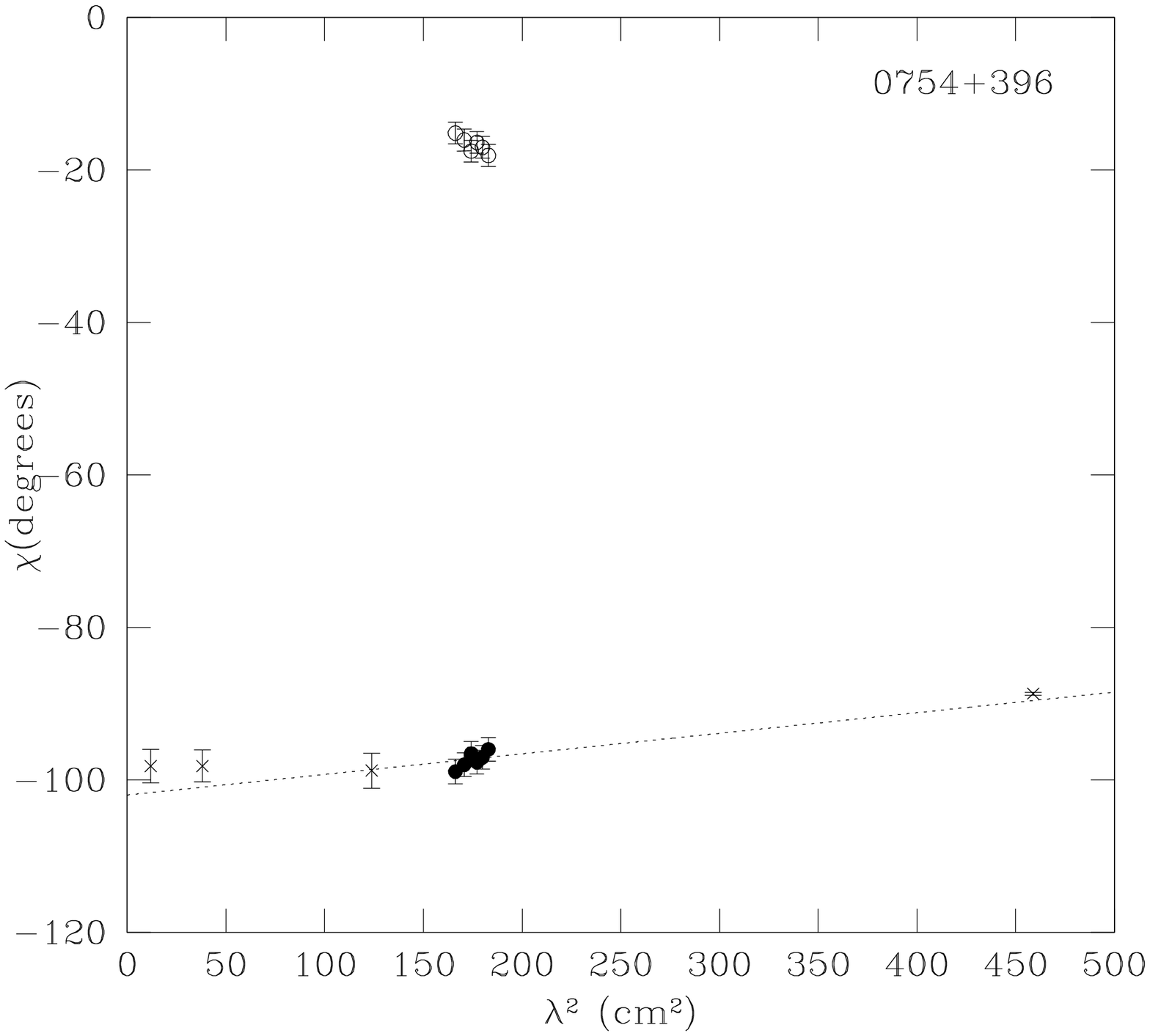}}
      {\includegraphics[width=8.5cm]{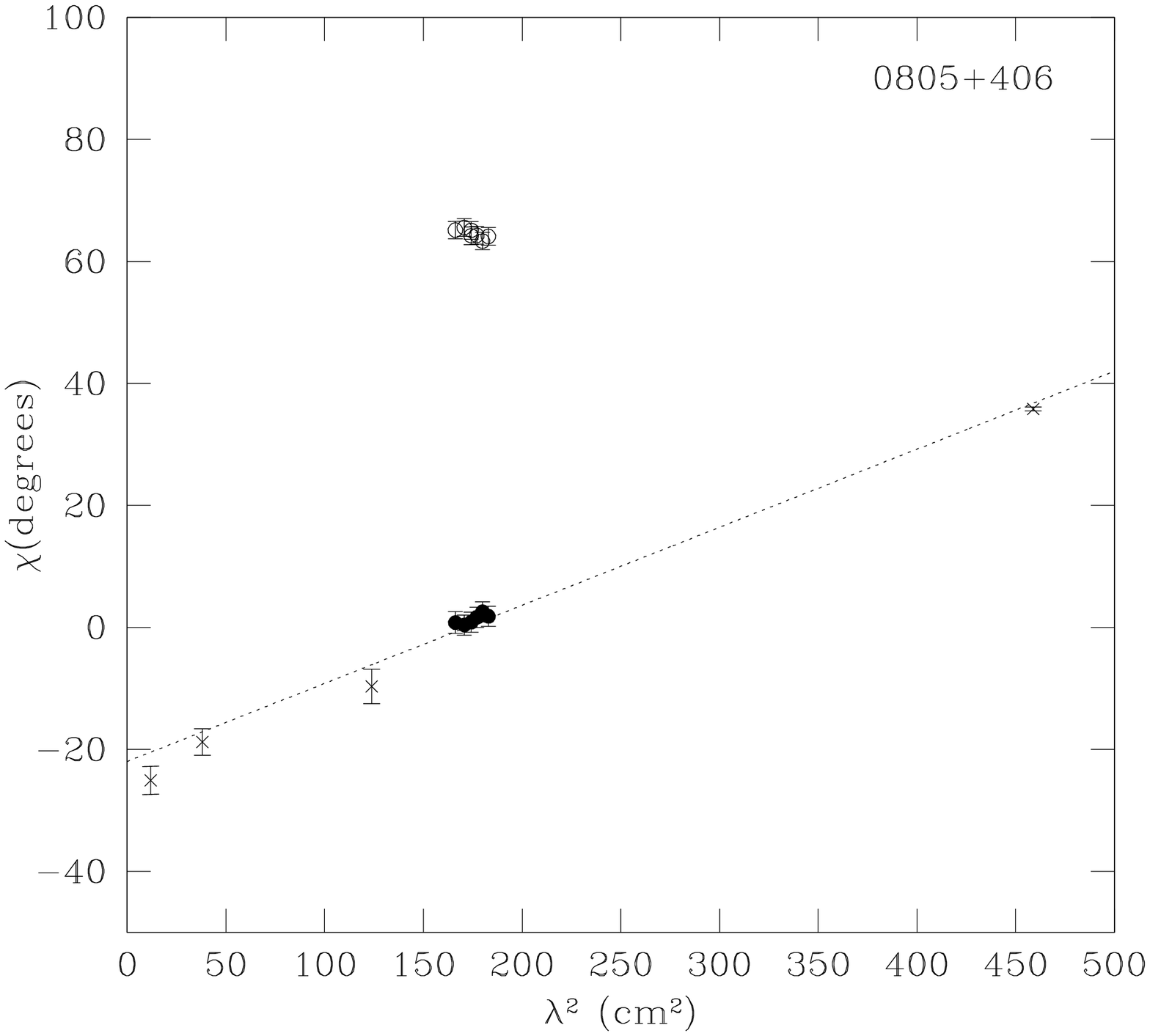}}
      {\includegraphics[width=8.5cm]{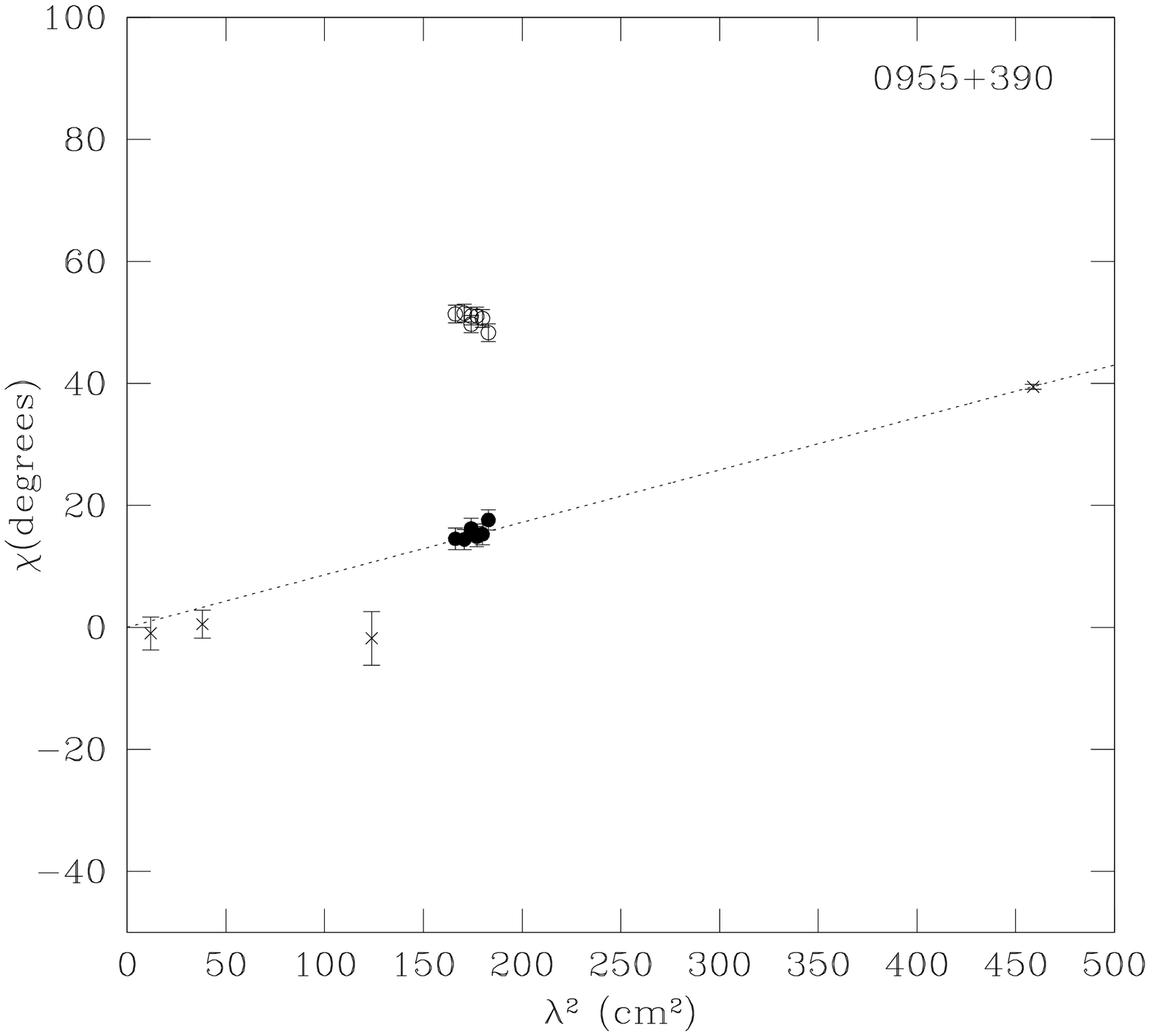}}
      {\includegraphics[width=8.5cm]{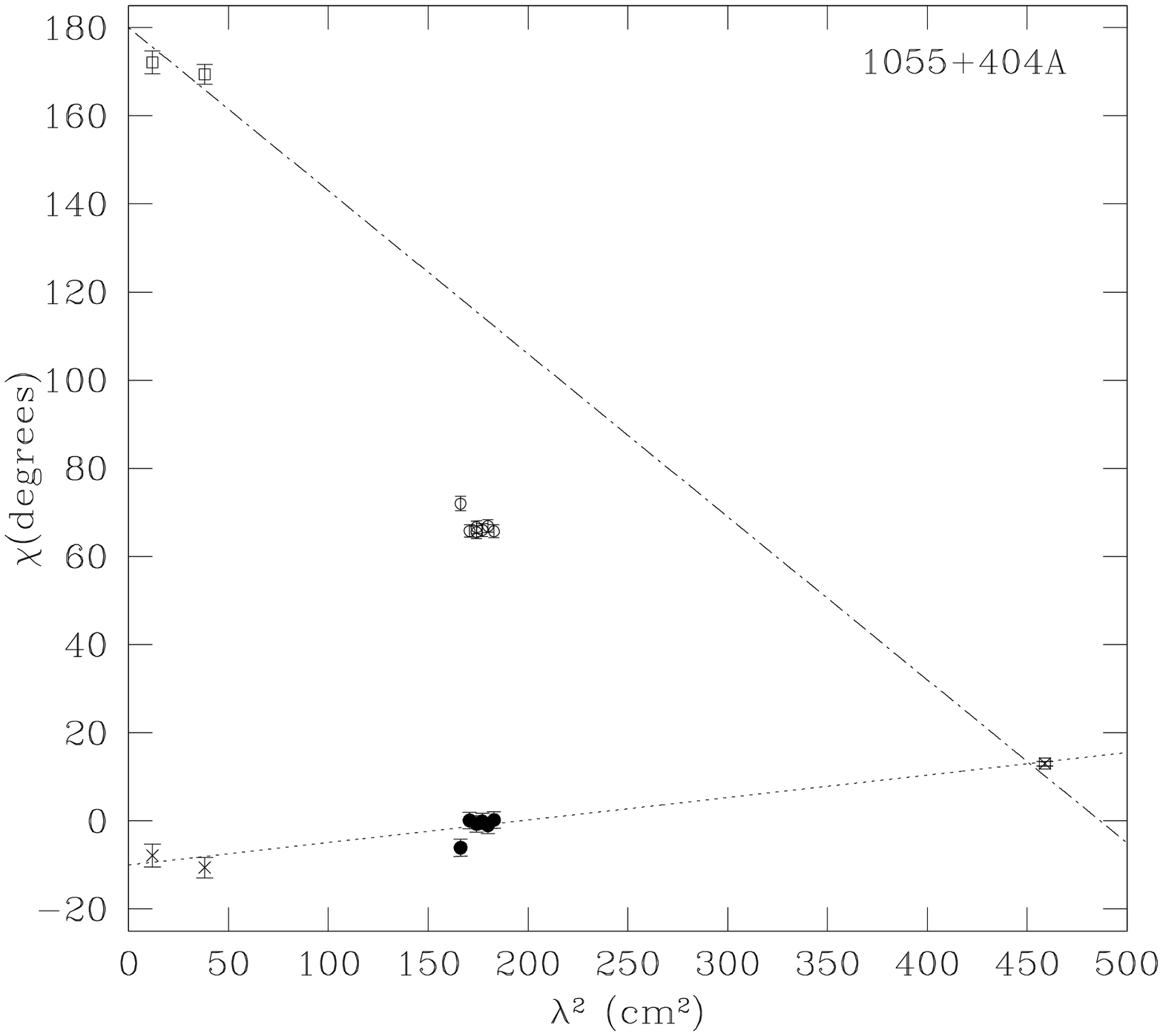}}
	\caption{Examples of sources with $\chi_{13}$ before
	  ({\large $\circ$}) and after ({\large $\bullet$}) the empirical
	  correction of Sect.~\ref{chi_wsrt}. The 
	  original in-band 13-cm angles are largely displaced from the
	  general $\chi(\lambda)$ behaviour {\bf($\times$)} and also
	  generally have a wrong slope.
	  Moreover, the trends {\it before} the empirical correction
	  have a $\lambda^2$ slope with opposite sign to that {\it
	  after} the correction. The corrected  ones have the slope in
	  agreement with the general {\it RM}. The last panel
	  (B3\,1055+404A) shows how the 13-cm 
	  data helped in solving the $n \pi$ ambiguity of the fit
	  obtained with three wavelengths only ($\Box$ represent a
	  1-$\pi$ ambiguity in $\chi$).} 
	\label{fig:chi_exa}
    \end{figure*}

We corrected empirically all of the WSRT angles, using  the above relation.
In Fig.~\ref{fig:chi_exa}, we show a few examples of sources before and after
the swap of the cross-hand visibilities on the 13~cm angles. We also plot 
the in-band angles (see Sect.~\ref{in_band}), and
confirm the correctness of our approach. 
 
\section{The WSRT Complete Polarisation Data Set}
\label{poldataset}

The polarisation status of the 65 sources, observed using the WSRT at
13~cm,  is provided below:
\begin{itemize}
\item 26 sources have been detected at $\ge 3 ~\sigma_{\rm P}$, at all four 
  wavelengths (labelled {\it P1} in Table~\ref{bt});
\item 4 sources have been detected at $\ge 3~ \sigma_{\rm P}$, at 3.6,
  6, and 13~cm  (labelled  {\it P2} in Table~\ref{bt});
\item 6 sources have been detected at $\ge 3~ \sigma_{\rm P}$, at 3.6 and 6~cm 
  (labelled  {\it P3} in Table~\ref{bt});
\item 5 sources have been detected at $\ge 3~ \sigma_{\rm P}$, at 3.6~cm only 
  (labelled  {\it P4} in Table~\ref{bt});
\item 8  sources have been detected $\ge 3~ \sigma_{\rm P}$, at two or three
  non-contiguous wavelengths (labelled  {\it P5} in Table~\ref{bt});
  only 3 of these have been detected at 13~cm;
\item 9 sources have been detected at 21~cm only (labelled  {\it P6}
  in Table~\ref{bt});  
\item the 7 remaining sources {\bb are} undetected in polarisation at all
  wavelengths and are labelled  {\it NP} (not polarised) in Table~\ref{bt}.
\end{itemize}  
We remark that bandwidth depolarisation is not a problem
at 3.6 and 6~cm, unless very high Rotation Measures ($\ge 10^4$~rad\,m$^{-2}$)
are present. At 13 cm,  bandwidth depolarisation is $\ltsim
10$\% for $RM \ltsim 400$~rad\,m$^{-2}$, while at 21~cm it
can be  large ($\ge 40\%$ for $RM \ge 200$~rad\,m$^{-2}$; see ~\citealt{nvss}).

The polarisation data are presented in Table~\ref{bt}. 
We provide: the total flux density ($S_{13}$), the fractional
polarisation  ($m_{13}$) with the corresponding error, the
polarisation position angle ($\chi_{13}$, 
 defined within $\pm 90\degr$), the polarisation parameters derived
 in Sect.~\ref{results}, i.e. Rotation Measure ({\it RM}) and
 Rotation Measure dispersion ($\sigma_{\rm RM}$), both observed (obs) 
and in the source frame (sf), and the intrinsic fractional
polarisation ($m_0$), 
and  Covering Factor (hereafter $f_c$) (see Sections~\ref{depol}, and
~\ref{rot-angl}). 
To provide all of the data used in this paper we include
the redshift, either spectroscopic {\bb or photometric from $K$ or $R$
band data (\citeauthor*{Fanti01})}, and the source {\it projected} total
Linear Size (taken from \citeauthor*{Fanti04}). When no 
redshift was available, $z=1.05$ was assumed\footnote{As discussed
  in \citeauthor*{Fanti01}, this is the average  of the
  spectroscopically-determined redshifts for the objects of the B3-VLA
  sample, not detectable at the limit of the POSS plates, which were
  later identified using deeper observations.}. 
\label{ave_z}
We provide notes on individual sources, in Appendix~\ref{app_1}.

We emphasize that these data are integrated over the entire source. All
sources are unresolved at both 13~cm (WSRT) and  21~cm (NVSS)
wavelengths. In contrast, the majority of sources were resolved, or
partially resolved, by the VLA at 3.6 and 6~cm. Hence, at these short
wavelengths, the polarisation parameters were vectorially added over
the source extension.

\section{Results}
\label{results}

\subsection{The ``Cotton Effect'' at 13~cm}
\label{Coteff}

\citet{Cotton03} found that, at 21~cm, the fractional polarisation 
of radio sources with $LS \le6$~kpc, is {\bb very small, typically
$\le0.4\%$.  
For $LS \approx$ 6~kpc}  the fractional polarisation  rises
abruptly to a median value of approximately $1\%$, reaching values of
up to 4-5\%.  
In ~\citeauthor*{Fanti04}, we extended  the analysis of fractional 
polarisation versus Linear Size to shorter wavelengths, and found the
same effect at 3.6~cm and 6~cm. The critical size, defined visually,
that separates polarised from unpolarised sources, was {\bb estimated} to
be $\approx 2.5$~kpc at 3.6~cm, and $\approx 4$~kpc at 6~cm. 

\begin{figure}[htbp]
      \centering
      \resizebox{\hsize}{!}
      {\includegraphics{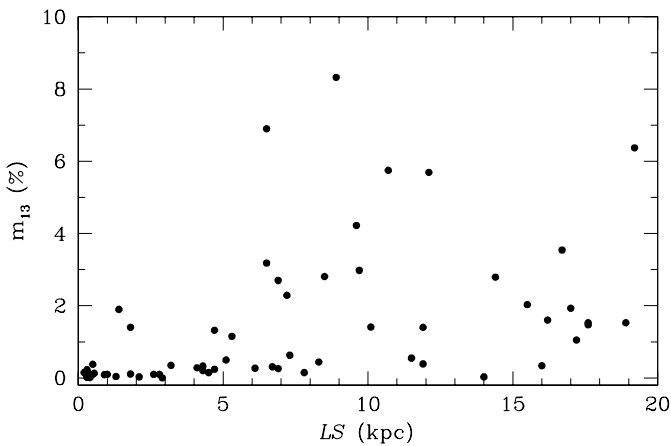}} 
      \caption{The ``Cotton Effect'' at 13~cm. We note that  most of the
      19 unpolarised sources (not observed by us; see 
      Sect.~\ref{the-sample}) would fall at small $LS$.}
     \label{fig:Cotton_1}
    \end{figure}

In Fig.~\ref{fig:Cotton_1}, we show a plot similar to {\bb those} discussed
above, for {\bb the} data at 13~cm presented in this paper. 
The critical scale for the drop in fractional polarisation is about 
5~kpc.

In Sect.~\ref{farCurt}, we interpret the abrupt change in the
distribution of $m_{\lambda}$, as a function of {\it LS}, in terms of
a dependence of $\sigma_{\rm RM,sf}$  (see Sect.~\ref{depol}) on {\it LS}.

\subsection{Modelling the Depolarisation of individual sources}
\label{depol}

The analysis of  the fractional polarisation, as a function of 
$\lambda^2$, indicates that the majority of sources have a fractional
polarisation that decreases with increasing wavelength. 
This implies that an inhomogeneous magneto-ionic medium, a Faraday
curtain, is present in front of the source. The {\bb effects} of such a
curtain  depend on the properties of  the inhomogeneities in the
magnetized medium, which are often referred to as ``cells''. 
In ~\citeauthor*{Fanti04}, we interpreted the data using the models
of both ~\citet{Burn66} and ~\citet{Tribble91}. 
In each model, the screen completely covers the source ($f_c=1$).

\citet{Burn66} {\bb assumes that the  ``cells''} are much smaller
than the beam size, and produce a random distribution of {\it RM}s,
across the source, which are zero on average and have {\bb a dispersion 
$\sigma_{\rm RM}$}.
The fractional polarisation follows a Gaussian law (in~$\lambda^2$):
\begin{equation}
m_{\lambda}=m_0e^{-2\sigma_{\rm RM}^2\lambda^4},
\label{eq:delta}
\end{equation}

\noindent
where $m_0$ is the intrinsic ($\lambda = 0$) fractional polarisation. 
We {\bb performed} a simple Monte Carlo analysis that {\bb shows} 
that the model
is applicable up to $\sigma_{\rm RM}\lambda^2\approx 1$~rad, provided that
the cell area is much smaller than a hundredth of the entire source area. 
It is well known, however, that  Eq.~\ref{eq:delta} predicts 
a fractional polarisation that is too low at long-wavelengths, to be
able to reproduce the observations. {\bb This discrepancy is reduced 
in Tribble's model ~\citep{Tribble91}. Also this model  assumes that 
the {\it RM} is randomly distributed, and has a dispersion $\sigma_{\rm
  RM}$, but in addition it introduces a broad distribution of ``cell'' sizes.} 

\begin{figure}[ht]
      \centering
      \resizebox{\hsize}{!}
      {\includegraphics{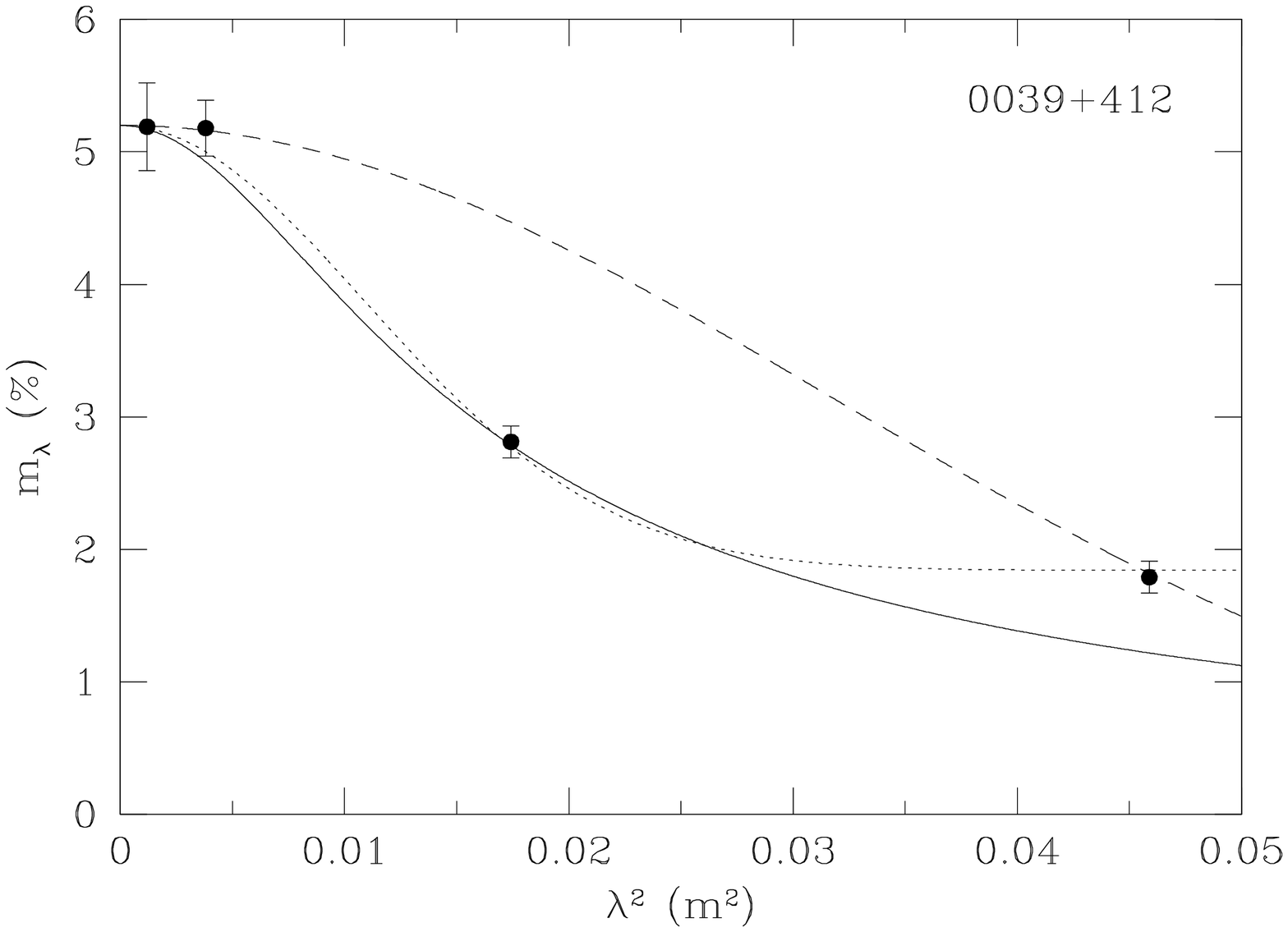}}
      \resizebox{\hsize}{!}
      {\includegraphics{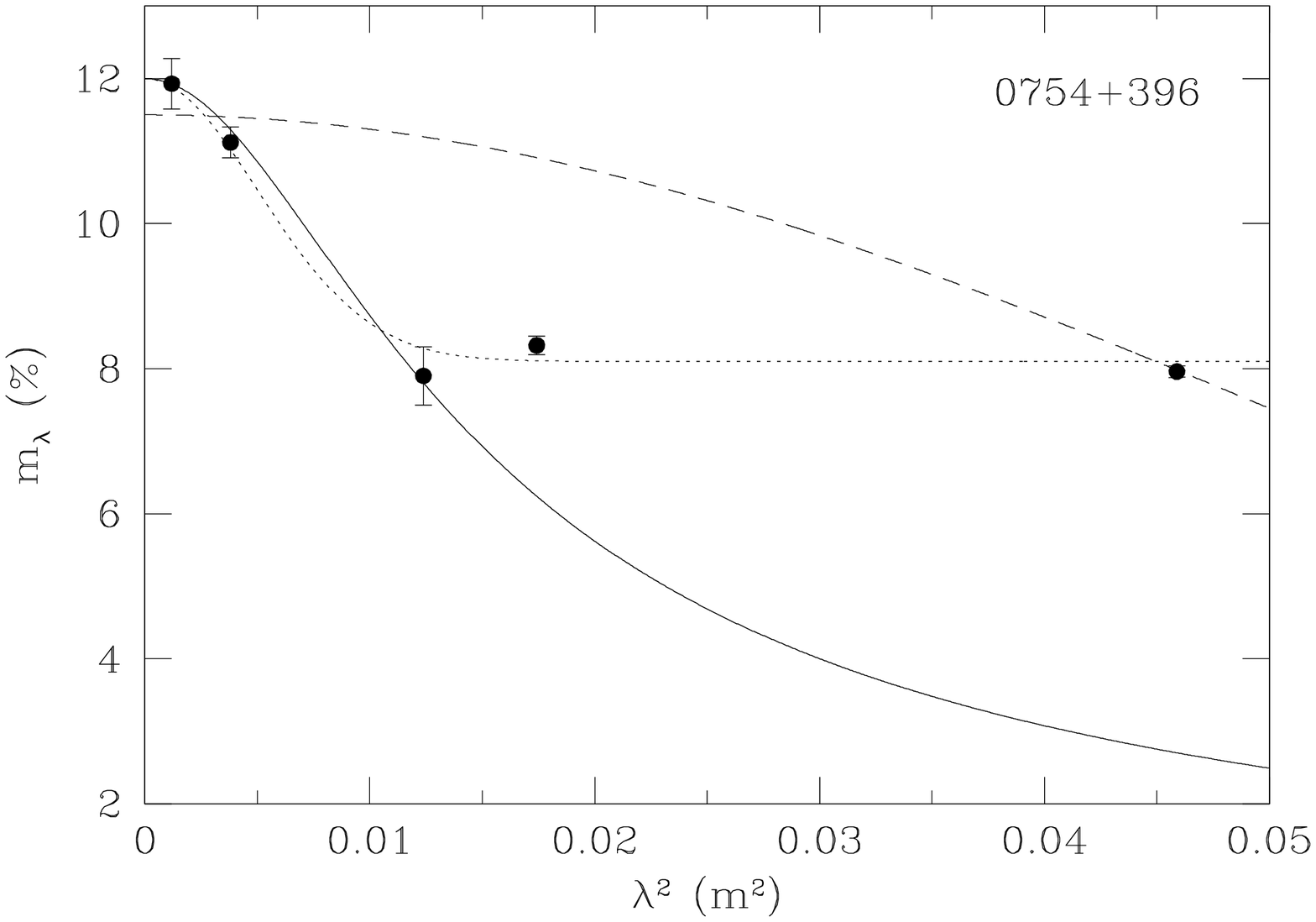}}
      \caption{Examples of fits from ~\citeauthor*{Fanti04} (3
  wavelengths only) with Burn's (Eq.~\ref{eq:delta}, dashed line),
  Tribble's (solid line) and a partial coverage (Eq.~\ref{eq:cover_f}, 
  dotted line) models which  use also the 13~cm data. The {\it
  top} panel shows  
  that with the
  old VLA data B3\,0039+412 could have been fitted with a Burn model,
  while the addition of the 13~cm data indicates that
  Eq.~\ref{eq:cover_f} gives a much better fit. Note that below 13~cm also
  Tribble's model could have been satisfactory within the measurement
  errors. In the {\it lower} panel the only acceptable fit for
  B3\,0754+396 is with
  Eq.~\ref{eq:cover_f}. The data point at $\lambda=11$~cm is from
  ~\citet{Klein03}. For $\lambda < 11$~cm Tribble's model would have
  been satisfactory as well.}
      \label{fig:es_mod}
    \end{figure}

In \citeauthor*{Fanti04}, we fitted the three-wavelength VLA
polarisation data {\bb with either model, solving for the
long-wavelength fractional polarisations, which are often too high for 
the Burn model. 
We now have data for a fourth wavelength, between 6~cm and 21~cm, that
enables {\bb far tighter} model fits to be achieved 
(see Fig.~\ref{fig:es_mod}).}

Using the new 13-cm data, the number of sources with polarisation data
that can be fitted using the Gaussian Burn model, is reduced with
respect to the findings in ~\citeauthor*{Fanti04}. {\bb We  
also find that fits in ~\citeauthor*{Fanti04}, made with the
Tribble model, are generally in bad agreement with 13-cm data.} 
{\bb As a matter of fact}, for a large fraction of sources, the
fractional polarisation drops quickly between 3.6~cm and $6-13$~cm, and  
remains approximately constant. This is inconsistent with the
predictions of either model
described above (see e.g. plot for the source B3\,0754+396 in
Fig.~\ref{fig:es_mod}).  
In such cases, the following empirical modification of Eq.~\ref{eq:delta}
\begin{equation}
m = m_0 \left [f_c e^{-2\sigma_{\rm RM}^2\lambda^4}+(1 - f_c)\right ]
\label{eq:cover_f}
\end{equation}
\noindent
is  more successful in reproducing the data.

The obvious interpretation of Eq.~\ref{eq:cover_f} is that, if a source is
only partially covered by the screen, a fraction ($1 - f_c$) of the
source radiation  emerges non-depolarised, {\bb and}  maintains  a
constant level of fractional polarisation at long wavelengths. 
We remark that  $\sigma_{\rm RM}$ only concerns the source
``covered fraction''. At this stage, we however consider the
introduction of $f_c$ to be a mathematical artifice that improves the
fit. We discuss the possible origins of this ``partial coverage'' in
Sect.~\ref{covering-factor}. 

We stress that, for $2\sigma_{\rm RM}^2 ~ \lambda^4 \ll 1$~rad$^2$, the
three models provide similar results and differences appear only
at longer wavelengths. $\sigma_{\rm RM}$ derived in the 
short-wavelength regime is therefore  {\bb largely} model-independent.

\begin{figure}[ht]
      \centering
      \resizebox{\hsize}{!}
      {\includegraphics{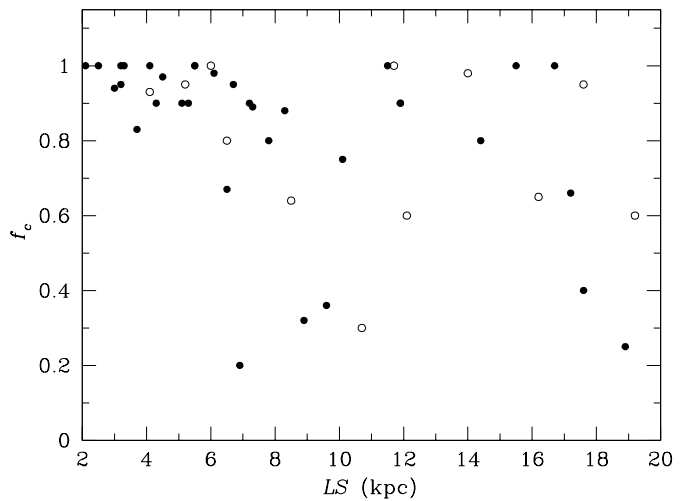}}    
      \caption{Covering Factor $f_c$ vs {\it LS}. Open symbols represent
	optically-unidentified radio sources. Their {\it LS} have been
	computed by assuming that $z = 1.05$ (see note on
	page~\pageref{ave_z}).} 
      \label{fig:cf_LS}
    \end{figure}

We  applied Eq.~\ref{eq:cover_f}  to  {\bb the} 36 sources in the WSRT
sub-sample (labelled {\it P1}, {\it P2}, {\it P3} in
Sect.~\ref{poldataset}) that were well detected at 3.6~cm and 6~cm, using
the 13~cm and 21~cm fractional polarisations or upper limits as well. A
large fraction of these sources show a flattening in fractional
polarisation (as, e.g., in Fig.~\ref{fig:es_mod}), which requires that
$f_c<1$.   
The remaining sources do not show the flattening of $m_{\lambda}$
within our wavelength range and can be  fitted using
Eq.~\ref{eq:delta}. {\bb However}, for homogeneity, we also fitted 
{\bb their} data using 
Eq.~\ref{eq:cover_f}. In these cases, we provide a lower limit
to the Covering Factor, and an upper limit to the corresponding
$\sigma_{\rm RM}$.  
In $\le 50 \%$ of the radio sources, $f_c \ge 0.9$; in a number of
sources, however, $f_c$ can be as low as 0.3. Figure~\ref{fig:cf_LS}
shows that the fraction of sources with small $f_c$ is larger for $LS
\ga 6$~kpc. 

\begin{figure*}[ht]
      \centering
      {\includegraphics[width=8.5cm]{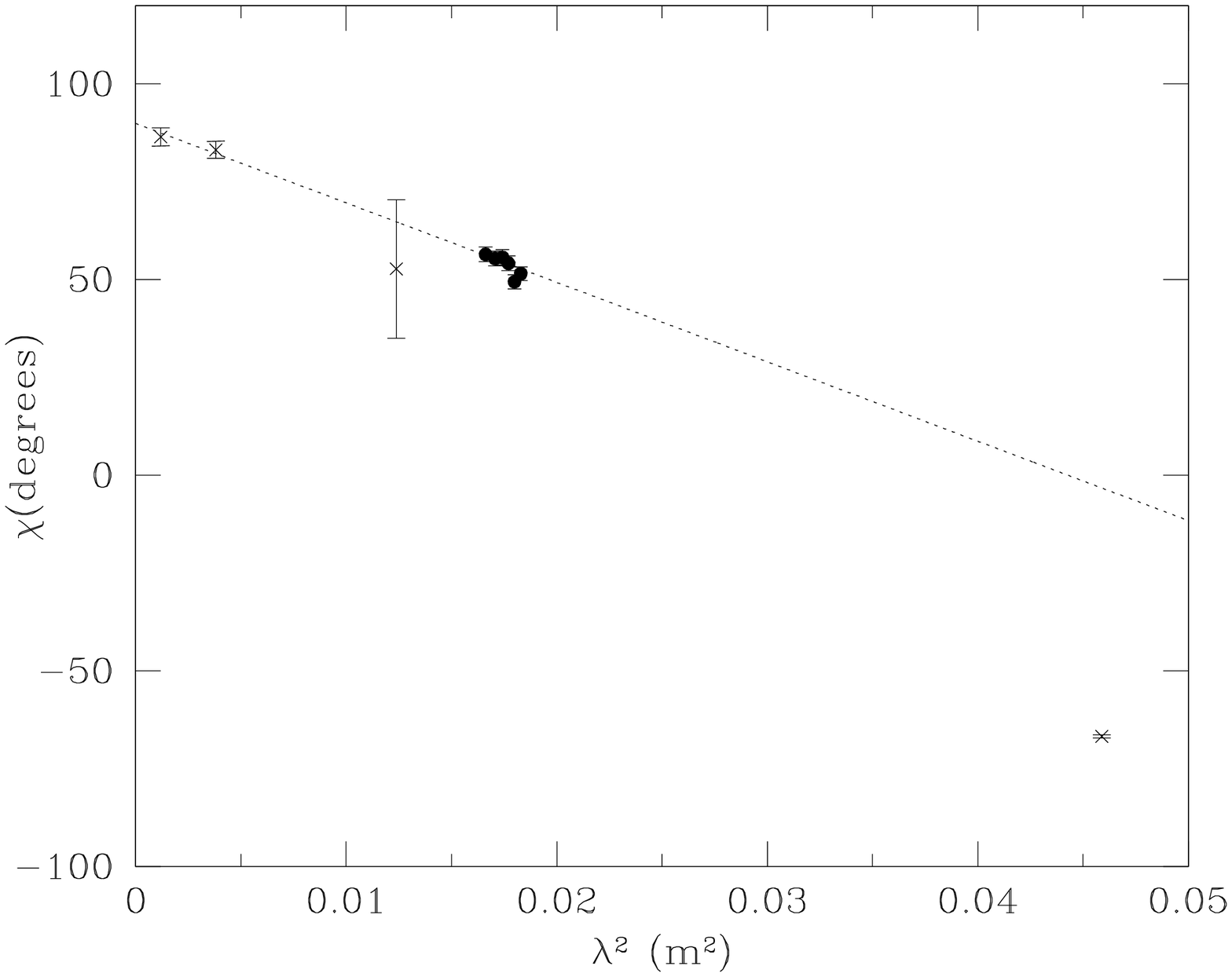}}
      {\includegraphics[width=8.5cm]{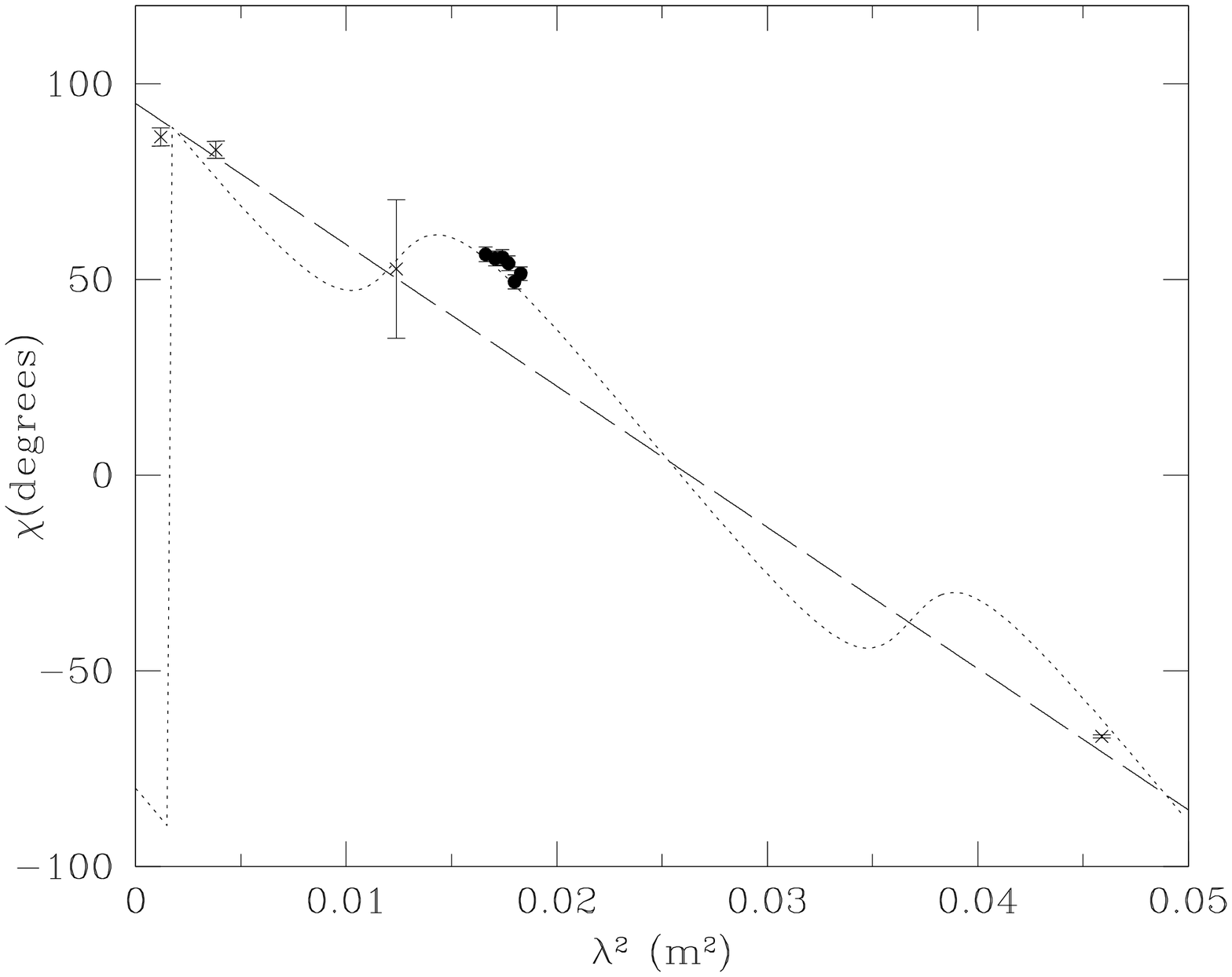}}
      {\includegraphics[width=8.5cm]{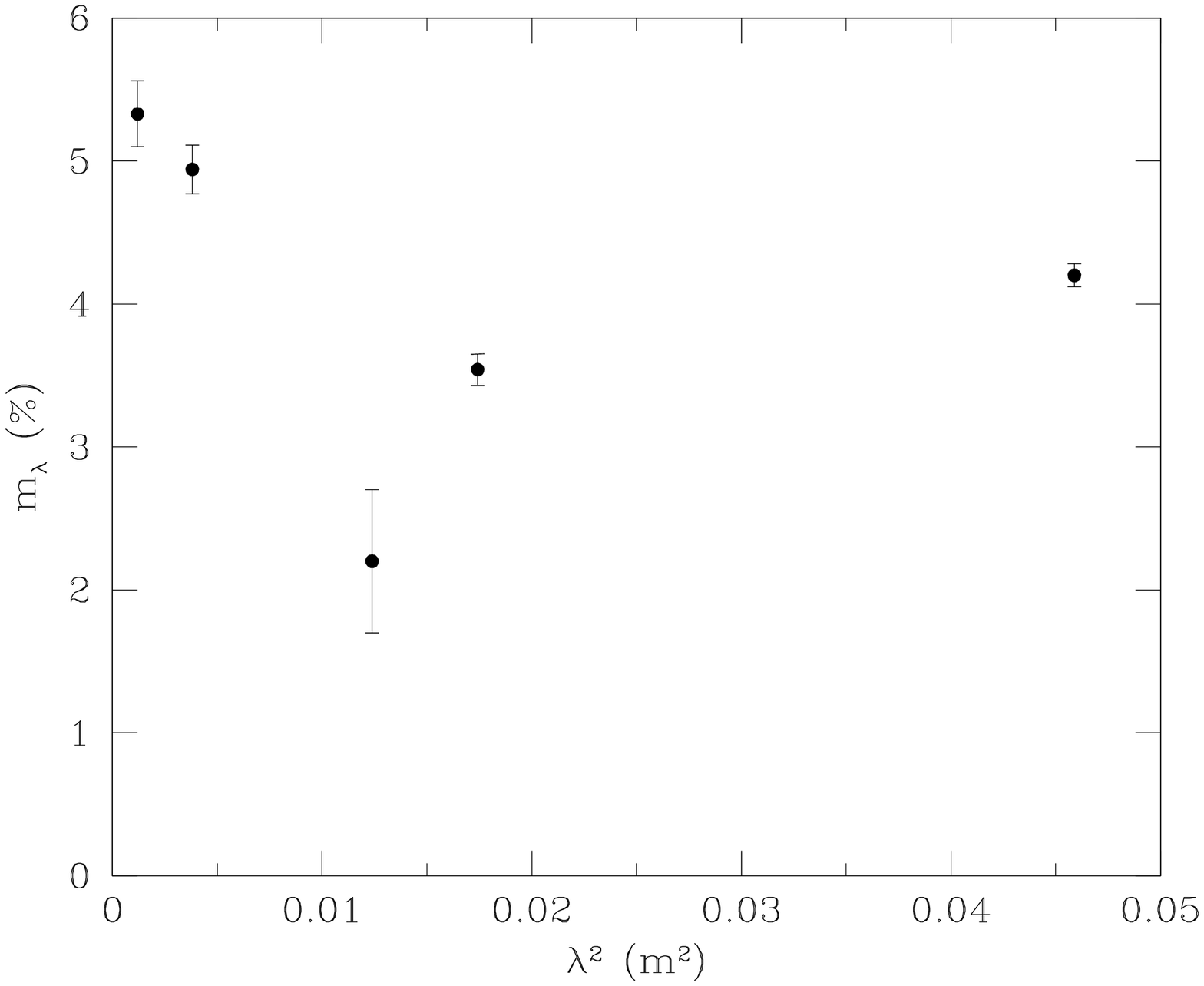}}
      {\includegraphics[width=8.5cm]{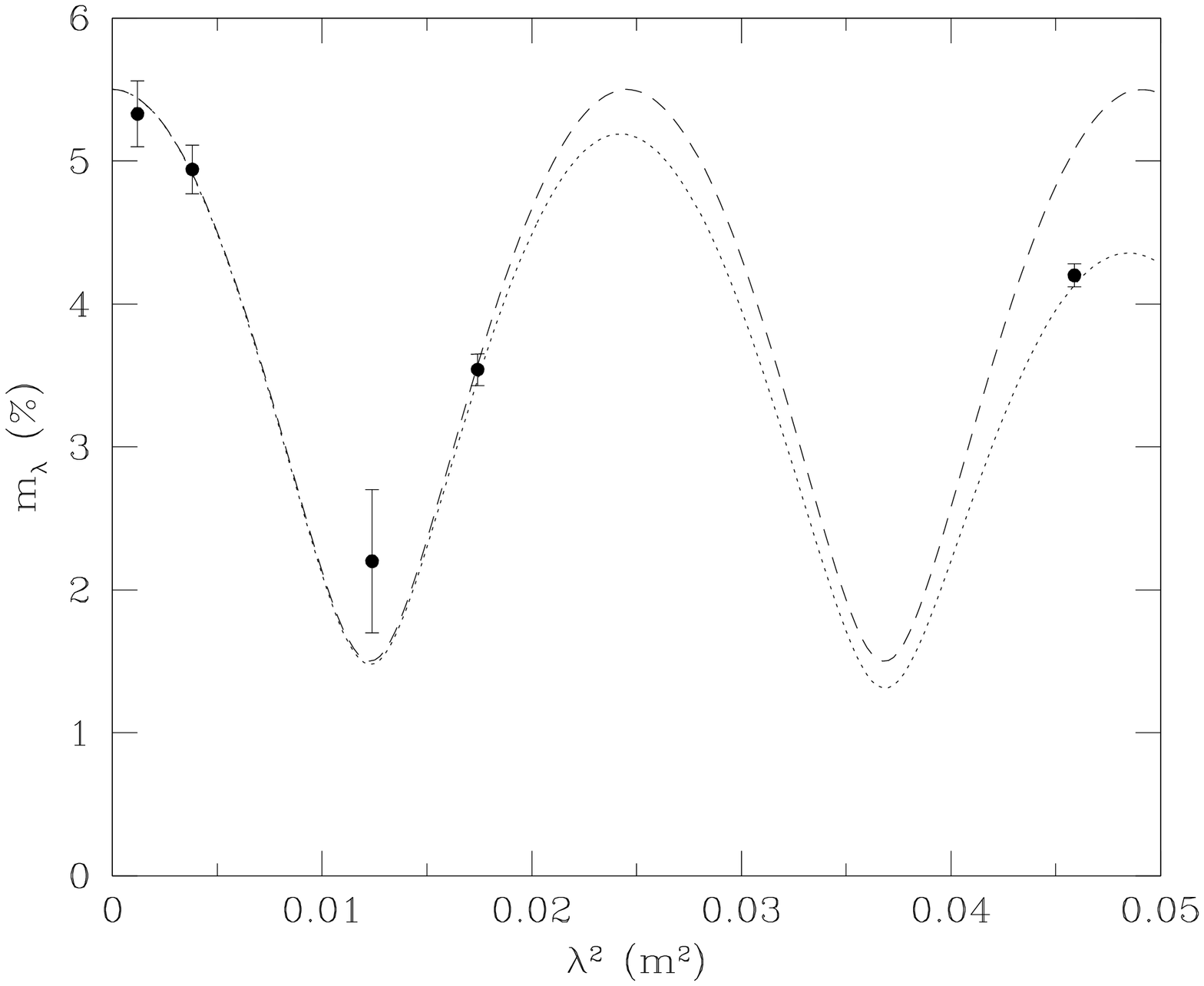}}
      \caption{B3\,0110+401. {\it Left Panels}: $m_{\lambda}$ 
        and $\chi(\lambda)$  
	data. Note that $\chi_{21}$ is largely off the
        $\lambda^2$-linear fit of the higher frequency data. 
	{\it Right Panels}: 
	Fit of $m_{\lambda}$ and $\chi(\lambda)$ with the
	two-component-model of Appendix~\ref{pol_mod}. The  dotted curve
	in the modelled $m_{\lambda}$ takes into account a little 
        depolarisation of the two components. The dashed
        straight line in the $\chi(\lambda)$ plot represents the {\it
        average RM}.}  
      \label{fig:es_0110}
    \end{figure*}

Three sources (B3\,0110+401, B3\,1025+390B, B3\,1216+402) out of {\bb these}
36, show an oscillatory behaviour in their fractional polarisation, 
even with repolarisation at long wavelengths. This behaviour could be
due to the beating of sub-components with different Rotation
Measures (see Appendix~\ref{pol_mod}).
{\bb The data for B3\,0110+401 were} modelled in this way, and a good fit
was achieved for both $m_{\lambda}$ and $\chi(\lambda)$ (see
Fig.~\ref{fig:es_0110}). For the other two sources, the fit is more
poorly constrained (see notes to Table~\ref{bt}).

In addition to the above 36 sources, another 5 were detected  {\bb only
at 3.6~cm}  ({\it P4}). We applied the Gaussian Burn model
($f_c\approx 1$) to these source data, and derived  lower limits to
$\sigma_{\rm RM,obs}$, which are consistent with the sources being strongly depolarised at $\lambda\ga 6$~cm.

Data for another 8 sources, detected at two or three non-contiguous
frequencies ({\it P5}), were not well fitted by either
Eq.~\ref{eq:delta} or~\ref{eq:cover_f}.
{\bb Several of them might have oscillations as a function of
  $\lambda^2$ in the fractional polarisation.} 
For these data, we applied the model described in 
Appendix~\ref{pol_mod} with some success, 
(see notes to Table~\ref{bt}). Because of the limited amount
of data, the parameters may however be weakly constrained.

Of the 9 sources found to be polarised only at 21~cm ({\it P6}), one source
(B3\,2302+402) has a fractional polarisation much larger than the
upper limits at the other wavelengths. 
The remaining 8 have upper limits at 3.6~cm, 6~cm, and 13~cm close to
$m_{21}$, and are therefore {\bb very little} polarised at all frequencies,
consistent with a large  $\sigma_{\rm  RM}$, and an $f_c$  
slightly less than unity. 

{\bb Also the 7 sources that are unpolarised at all frequencies ({\it
  NP}), are
likely to have a large $\sigma_{\rm RM,obs}$ ($\gtsim
500$~rad\,m$^{-2}$)}. 

Table~\ref{bt} provides the best-fit Rotation Measure dispersions  
in the observer's frame  ($\sigma_{\rm RM,obs}$) and in the source frame
($\sigma_{\rm RM,sf}$).
 The latter are computed using the redshift, either
spectroscopic or photometric, from Table~\ref{bt}. Formal errors are
typically $\le 20\%$. 

\begin{figure}[htbp]
      \centering
      \resizebox{\hsize}{!}
      {
       \psfrag{N}{\huge{N}}%
       \psfrag{m_0}{\huge{$m_0$ (\%)}}%
      \includegraphics[angle=-90]{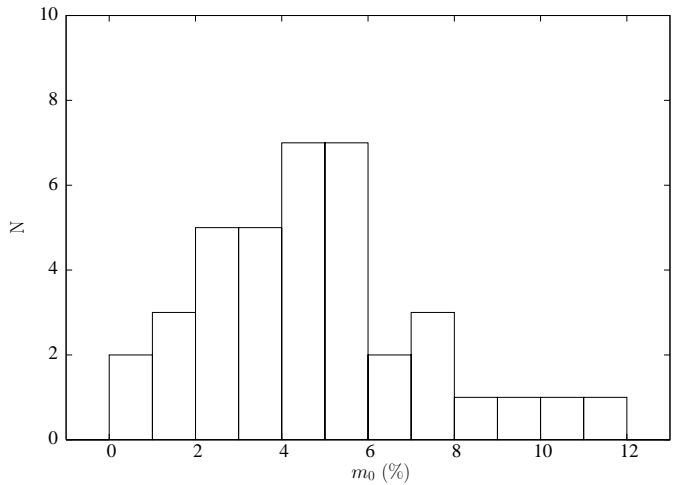}}
      \caption{Distribution of the intrinsic polarisation, $m_0$, for
      the WSRT sub-sample.} 
      \label{fig:histo_m_0}
    \end{figure}

From the 45 sources of the WSRT sub-sample for which depolarisation
parameters (or limits) could be compute,  38 have 
$2.5 < LS({\rm kpc}) < 20$. Their intrinsic degree of polarisation
has a median value of approximately $ 4.3\%$, an r.m.s. of 2.5\%, and
a distribution tail
extending up to 12\% (see Fig.~\ref{fig:histo_m_0}). As already noted
in ~\citeauthor*{Fanti04}, these intrinsic fractional polarisations do
not differ significantly from those of radio sources of tens to
hundreds kpc sizes. 

\subsection{Rotation Measures}
\label{rot-angl}

We used the E-vector position angles at 13~cm ($\chi_{13}$)
and those at 3.6~cm, 6~cm, and 21~cm, reported in \citeauthor*{Fanti04}, to
derive the Faraday Rotation Measure, 
{\it RM}, by weighted linear interpolation of the data, with the 
$\lambda^2$-linear law
$\chi(\lambda)=\chi_0+\lambda^2 RM$. We are aware that the 
assumption of a $\lambda^2$-linear law could be incorrect. As
discussed  in  Sect.~\ref{depol}, the presence of a few polarised
sub-components  with different {\it RM} 
would produce modulations of the $\chi(\lambda)$ versus 
$\lambda^2$-linear relation (see Appendix~\ref{pol_mod}). 
A $\lambda^2$-linear fit would provide an {\it average RM} that
corresponds  approximately to that of the component with the highest
polarised flux-density. Individual data points may then  deviate
significantly from an average $\lambda^2$-linear law (e.g.,
Fig.~\ref{fig:es_0110}, top-right panel).

\begin{figure}[htbp]
      \centering
      \resizebox{\hsize}{!}
      {\includegraphics{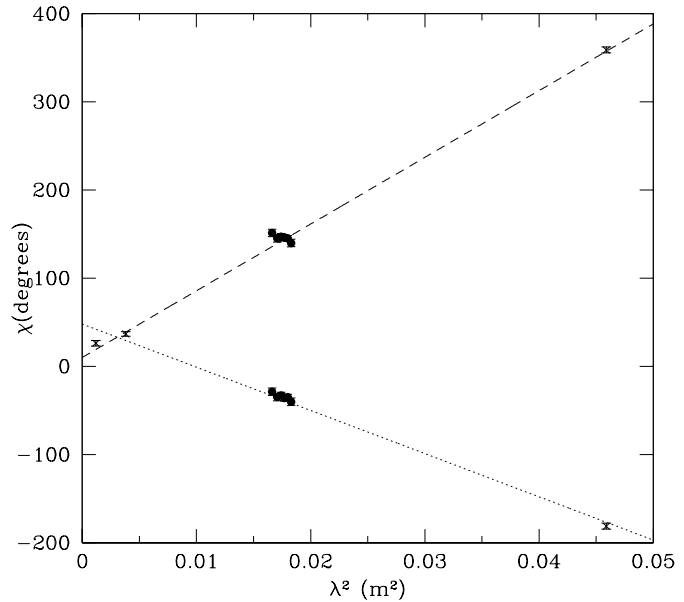}} 
      \caption{B3\,0034+444. Example of how the in-band polarisation angles
	help in choosing the {\it RM}. The positive {\it RM} (dashed
	line) is clearly ruled out. With the total band $\chi_{13}$
	only, instead, both fits would have been acceptable.}
      \label{fig:es_0034}
    \end{figure}

The  fitting procedure was applied  to the  32 sources that were 
detected in polarisation at 13~cm, with  $S_{\rm P} \ge 10 \sigma_{\rm
  P}$, and have at least two additional detections at the
$3\,\sigma_{\rm P}$ level.    
As stated in Sect.~\ref{in_band}, the 24 sources {\bb with $S_p \ge 10
\sigma_p$} were
analysed by means of the individual IFs. Twenty-two of these sources
were detected  at a level of more than 
$5\,\sigma_{\rm  P}$ in the individual IFs, and we were able to derive the 
in-band  Rotation Measure, $RM_{13}$, with typical uncertainties in
the range of 25 -- 70~rad\,m$^{-2}$. Then, we used $RM_{13}$ as a
guide to resolve possible $n \pi$-ambiguities at the other
wavelengths. For four sources (B3\,0034+444 in Fig.~\ref{fig:es_0034},
B3\,0137+401, B3\,0814+441, and B3\,0930+389) the first fitted {\it RM}
had to be changed drastically, to reach agreement with $RM_{13}$. For
two additional sources (B3\,1216+402 and B3\,1350+432) the
disagreement between  {\it  RM} and $RM_{13}$ remains; however this is
probably due to substructures in polarisation that produce a modulation 
of $\chi(\lambda)$, over the $\lambda^2$-linear law (see the
individual notes for these sources). 
Another source (B3\,0120+405) shows strong disagreement
between {\it RM} and $RM_{13}$. Although we have been unable to develop a 
model to explain this discrepancy, we suppose that the explanation
proposed for the previous two sources may also apply to {\bb this  one}.
The value of {\it RM} provided in Table~\ref{bt}, in this case, was
derived by fitting the individual IFs, together with data at
other wavelengths.   

The observed {\it RM}s (uncorrected for the Galactic Faraday
Rotation, $RM_{\rm{obs}}$), are provided in Table~\ref{bt}. {\bb A code
``a'', marks the sources}
for which a $\lambda^2$-linear fit provides a good chi-square
(probability $\ge 0.05$ of being exceeded 
because of random fluctuations). 
Only 12 sources out of 32 are found to be in this class.
The remaining sources have $\lambda^2$-linear fits of {\bb lower 
chi-square quality}.  
In a number of cases, this can be attributed to a single discrepant
point; the exclusion of this point would significantly improve the
chi-squared fit, {\bb with moderate changes in RM}.  
We suspect that, in a number of cases, we observe 
modulations, {\bb about the $\lambda^2$-linear  law, caused by
polarised sub-components of a source that have different {\it RM}s}. 
For this reason, we excluded no data point from the fit.

The formal errors of {\it RM} are  small, typically $\la
10$~rad\,m$^{-2}$. The
actual uncertainties are related to a few residual ambiguities of $n \pi$
at 21~cm, and to the assumption that the $\lambda^2$-linear law is
valid over the entire wavelength range.

We compared the present {\it RM}s with those provided in
\citeauthor*{Fanti04}. 
For 9 out of the 30 sources in that paper, the old Rotation Measure 
is rejected, while  for 4 sources  not detected at
13~cm,  we can neither confirm nor disprove the old values.

{\bb The source-frame Rotation Measure, ($RM_{\rm sf}$), is
calculated  multiplying  the Galaxy-corrected $RM$} \footnote{We estimated the
  Milky Way contribution from the data set of Rotation Measures of
  ~\citet{Klein03}, 
which contains about 200 sources carefully measured at 4 frequencies
in the same sky area of our sources. From these data, the Galactic
Rotation Measure is $(-33\pm 7)$~rad\,m$^{-2}$ in the range $23^{\rm
  h} \le {\rm RA} \le 0^{\rm h}$,  $(-80\pm 8)$~rad\,m$^{-2}$ in the
range $0^{\rm h} \le {\rm RA} \le 3^{\rm h}$, and $(+9\pm
1)$~rad\,m$^{-2}$ in the range $7^{\rm h} \le {\rm RA} \le 15^{\rm h}$.}
{\bb  by the factor $(1+z)^2$.}

\section{Discussion}
\label{discussion}

\subsection{The depolarising Faraday curtain: an empirical model}
\label{farCurt}

To describe the Cotton Effect at the different 
observing wavelengths, in ~\citeauthor*{Fanti04}, we outlined a model
that relates the source frame {\it RM}  dispersion, $\sigma_{\rm
  RM,sf}$, to the source projected  Linear Size {\it LS}, with a
suggestion also of a dependence on $z$.  

More accurate values of $\sigma_{\rm RM,sf}$, determined in the
present paper, and further data from~\citeauthor*{Fanti04}, are {\bb now}
available for 33 out of 38 of the B3-VLA CSS sources, which have
Linear Sizes $LS \ge 2.5$~kpc, and known redshifts (either spectroscopic or
photometric). 
Four  of the missing sources (B3\,0255+460, B3\,1128+455, B3\,1241+411,
and B3\,2349+410) are strongly depolarised {\bb at $\lambda \ge$ 6 cm}
or {\bb not polarised already at 3.6 cm}, and their $\sigma_{\rm
  RM,sf}$ are likely to have a
high value ($\gtsim 1000$~rad\,m$^{-2}$). For the last source
(B3\,1025+390B), $\sigma_{\rm RM,sf}$ is not well constrained (see
notes in Sect.~\ref{notes}). 
A remaining sixteen sources with $LS \ge 2.5$~kpc do not have
a redshift and cannot be used in this analysis. 
 We are however confident that their absence does not introduce
 bias in the results. 
Thus the sub-sample for which $LS \ge 2.5$~kpc, and  {\bb known $z$}, is
adequate to investigate  the dependence of $\sigma_{\rm RM,sf}$ on
both Linear Size  (projected) and redshift.

Sources smaller than 2.5~kpc are mostly depolarised at all
frequencies. Therefore in the following analysis, we explore only the
range $LS \ge 2.5$~kpc.

    \begin{figure}[htbp]
      \centering
      \resizebox{\hsize}{!}
     {\includegraphics{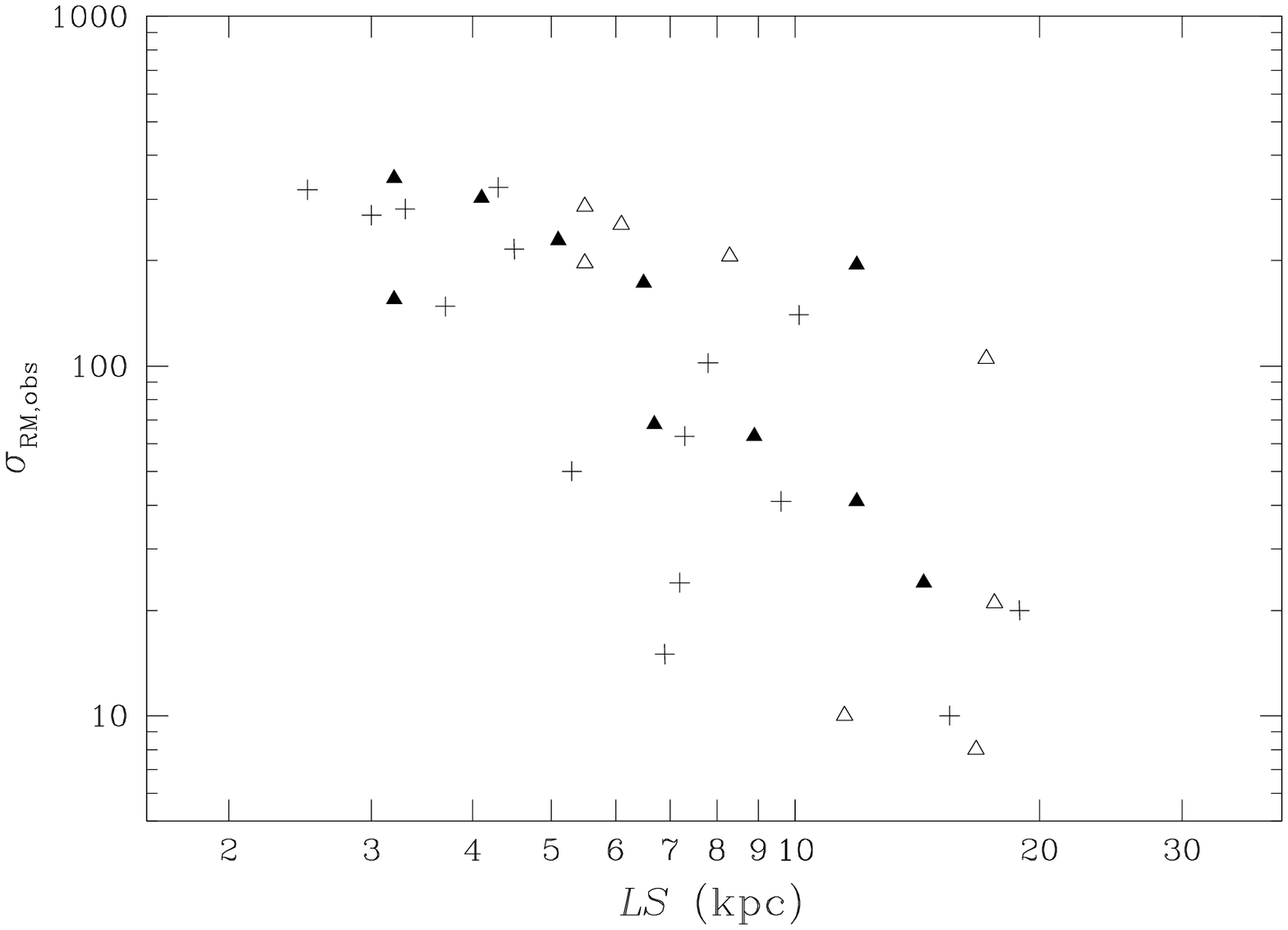}} 
      \resizebox{\hsize}{!} 
     {\includegraphics{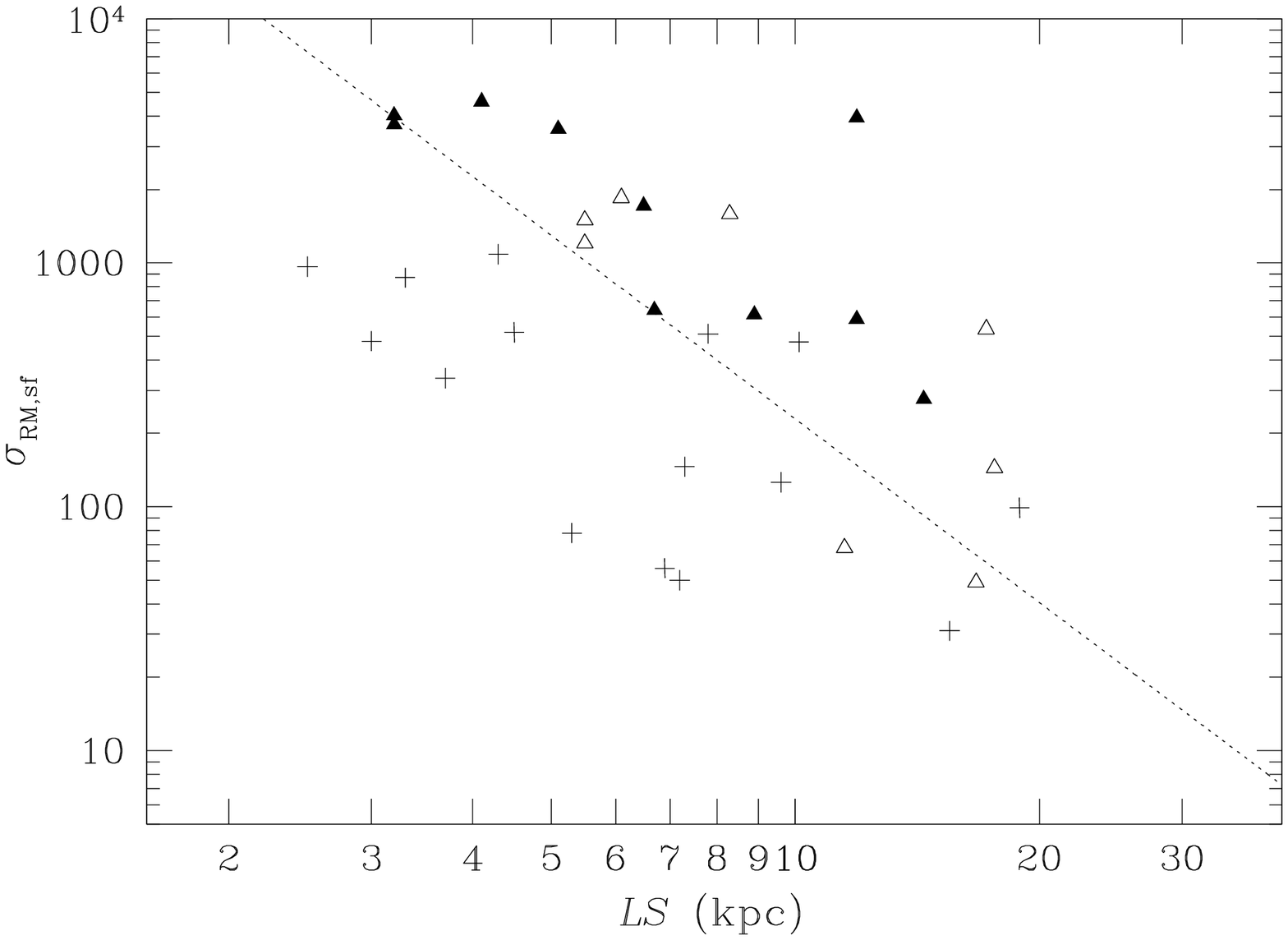}}
     \resizebox{\hsize}{!} 
    {\includegraphics{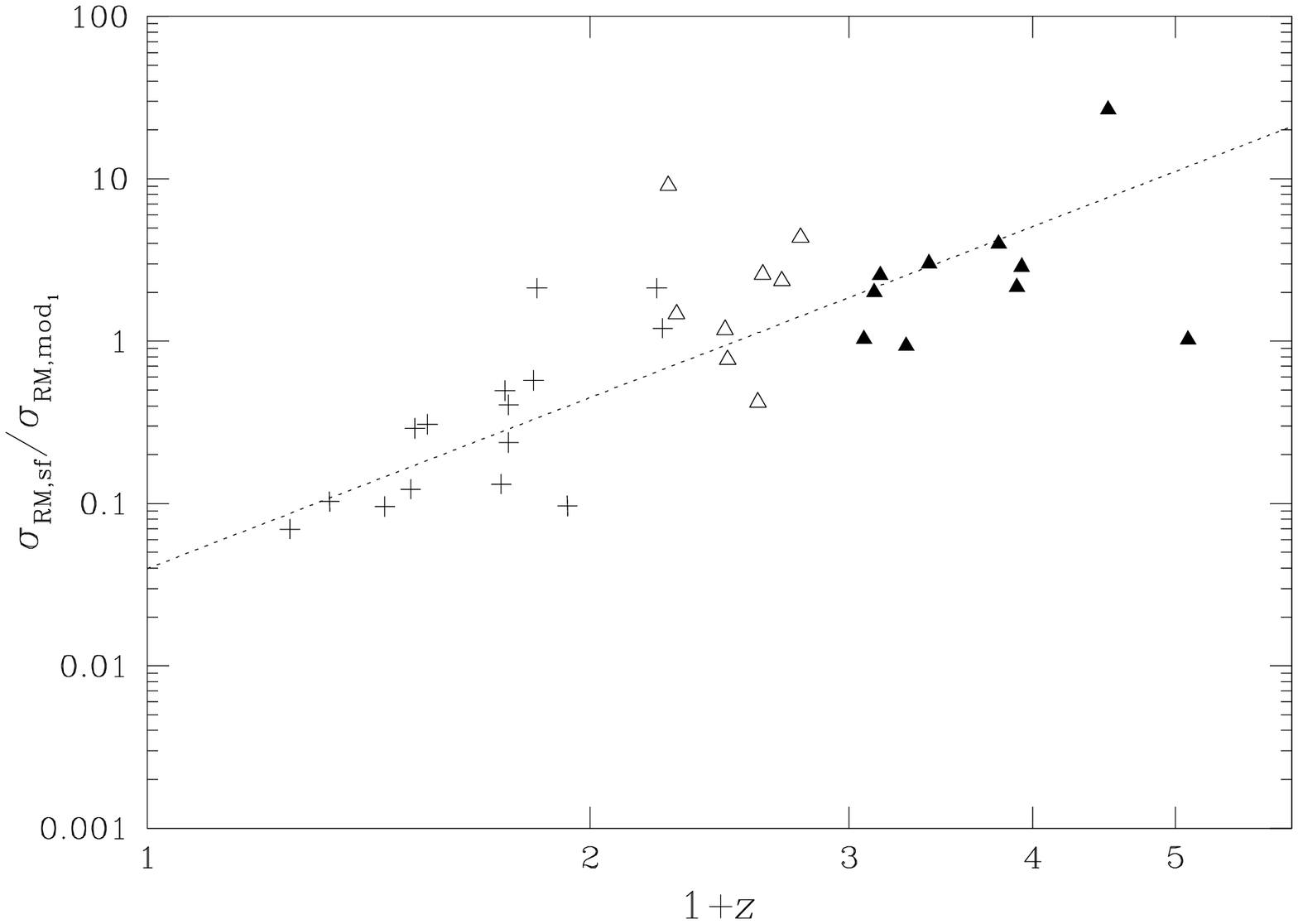}}
      \caption{{\it Top}: Plot of $\sigma_{\rm RM,obs}$ vs {\it LS}. 
        {\it Middle}: Plot of $\sigma_{\rm RM,sf}$ vs {\it LS}; the
        dotted line is the first order fit.  
	{\it Bottom}: Plot of $\sigma_{\rm RM, sf}$, corrected for the
        {\it LS} dependence, vs $(1+z)$; the dotted line is the first order
        fit. Different symbols are:
	   (+) for $z <1.25$, ($\vartriangle$) for $1.25 < z <1.8$ and 
	  ($\blacktriangle$) for $z > 1.8$.}  
	\label{fig:sigma_mod_1}
    \end{figure}

\smallskip
We proceed as follows:
\begin{description}
\item[{\mdseries a)}] The plot of $\sigma_{\rm RM,sf}$ versus {\it LS} (see
    Fig.~\ref{fig:sigma_mod_1}, middle panel) shows that:
\begin{enumerate}
\item $\sigma_{\rm RM,sf}$ decreases with {\it LS}, with a large
   dispersion (about a factor 8 at $\approx 90\%$ level);
\item there is a clear segregation in redshift, the 
   high-redshift sources exhibiting, at the same {\it LS}, larger
   values of $\sigma_{\rm RM,sf}$.  A hint of this  dependence
   on redshift is also seen in the distribution of $\sigma_{\rm RM,obs}$ (see
   Fig.~\ref{fig:sigma_mod_1}, top panel).
   
   To parametrize these findings, we assumed a model for the
   source frame {\it RM} dispersion that has a  power-law
   dependence on $LS$  and $(1+z)$,
   $\sigma_{\rm RM,mod} \propto LS^{-a} (1+z)^b$,
   and made a first estimate of $a$  from the plot, neglecting
   at this stage the dependence on ($1+z$). In this way, we {\bb obtained} a
   first approximation model: $\sigma_{\rm RM,mod_1} \propto LS^{-2.5}$.
\end{enumerate}
   
\item[{\mdseries b)}] Using this law, we computed for each source a
  $\sigma_{\rm RM,mod_1}(LS)$,  
  corresponding to its Linear Size. These $\sigma_{\rm RM,mod_1}$ take into
  account the dependence on {\it LS}, but not on redshift. The ratios
  $\sigma_{\rm RM,sf}/\sigma_{\rm RM,mod_1}$ show a large spread, part
  of which is due  to the redshift dependence.  
  A plot of these ratios versus redshift (see
  Fig.~\ref{fig:sigma_mod_1}, bottom panel) allows us to estimate the
  parameter $b$. We find $b \approx 3.5$. We note, however, that this
  power-law fit slightly underestimates $\sigma_{\rm RM,sf}$ at 
  intermediate redshifts, and overestimates $\sigma_{\rm RM,sf}$ at high redshifts. 
   
\item[{\mdseries c)}] We used a second-order model,  
$\sigma_{\rm RM,mod_2} \propto LS^{-2.5}(1+z)^{3.5}$, computed for each
  source   $\sigma_{\rm RM,mod_2}(LS, 1+z)$, and
 analysed the ratios $\sigma_{\rm RM,sf}/\sigma_{\rm RM,mod_2}$, as a
 function of both {\it LS} and $(1+z)$, and adjusted the two parameters 
$a$ and $b$ to derive {\bb another} $\sigma_{\rm RM,mod}$.
   
\item[{\mdseries d)}]  We find $a = 2.0$ and $b = 3.3$. 
 However, (see Fig.~\ref{fig:sigma_mod_pl}) the $(1+z)$ power-law fit
 still  underestimates $\sigma_{\rm RM,sf}$ at intermediate redshifts,
 and  overestimates  {\bb them}  at
 high redshifts:
at intermediate redshifts 8 out of 10 data points are 
above the $\sigma_{\rm RM,sf}/\sigma_{\rm RM,mod} = 1$ line, while at high
redshifts 8 out of 10 are below this line. We  
evaluated the statistical significance of these systematic effects
by using both a contingency table and by analysing the medians and their
variance in the two redshift bins. We find that the probability that 
the observed systematic effects are due to chance is $\le 5\%$.

 \begin{figure}[ht]
     \centering
     \resizebox{\hsize}{!}
     {\includegraphics{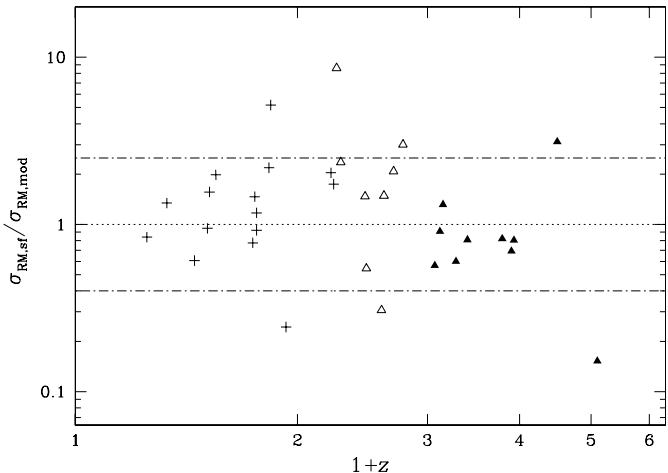}}
      \caption{Plot of
        the ratios $\sigma_{\rm RM,sf}/\sigma_{\rm mod}$ vs $(1+z)$ for
        the power law model $(1+z)^{3.3}$.
        Symbols are as in Fig.~\ref{fig:sigma_mod_1}}
    \label{fig:sigma_mod_pl}
    \end{figure}

\item[{\mdseries e)}] To remove this systematic effect, we split the sample
{\bb into two  redshift bins: (i) $z \le 1.8$, 
and (ii) $ z > 1.8$}.  We
fitted the data for the lower redshift bin using the power-law
relation $(1+z)^b$, where $b = 4.0$; data for the higher redshift bin
were fit using a constant, i.e.  $b = 0$ in the power-law relation.
\end{description}

The adopted model is:

\begin{eqnarray}\label{eq:sigmafit}  
\sigma_{\rm RM,mod} &=& {\cal A}\cdot \left(\frac{LS}{{\rm kpc}}\right)^{-2} (1+z)^4
\mbox{ for } z<1.8 \nonumber\\
                &=& {\cal A}\cdot \left(\frac{LS}{{\rm kpc}}\right)^{-2}(2.8)^4
~~~\mbox{   for } z\geq 1.8    
\end{eqnarray}

\noindent
with ${\cal A} = 1.0 \cdot 10^3$~rad\,m$^{-2}$\,kpc$^{-2}$.
The uncertainties on $a$ and $b$ are $\pm 0.25$ and $\pm 0.4$, respectively.
The parameters $a$ and ${\cal A}$ may be {\bb slightly
underestimated, because we did not account for the 4 small size 
depolarised sources, which  probably have a  high $\sigma_{\rm
  RM,\rm obs}$.}

\begin{figure}[ht]
      \centering
      \resizebox{\hsize}{!}
      {\includegraphics{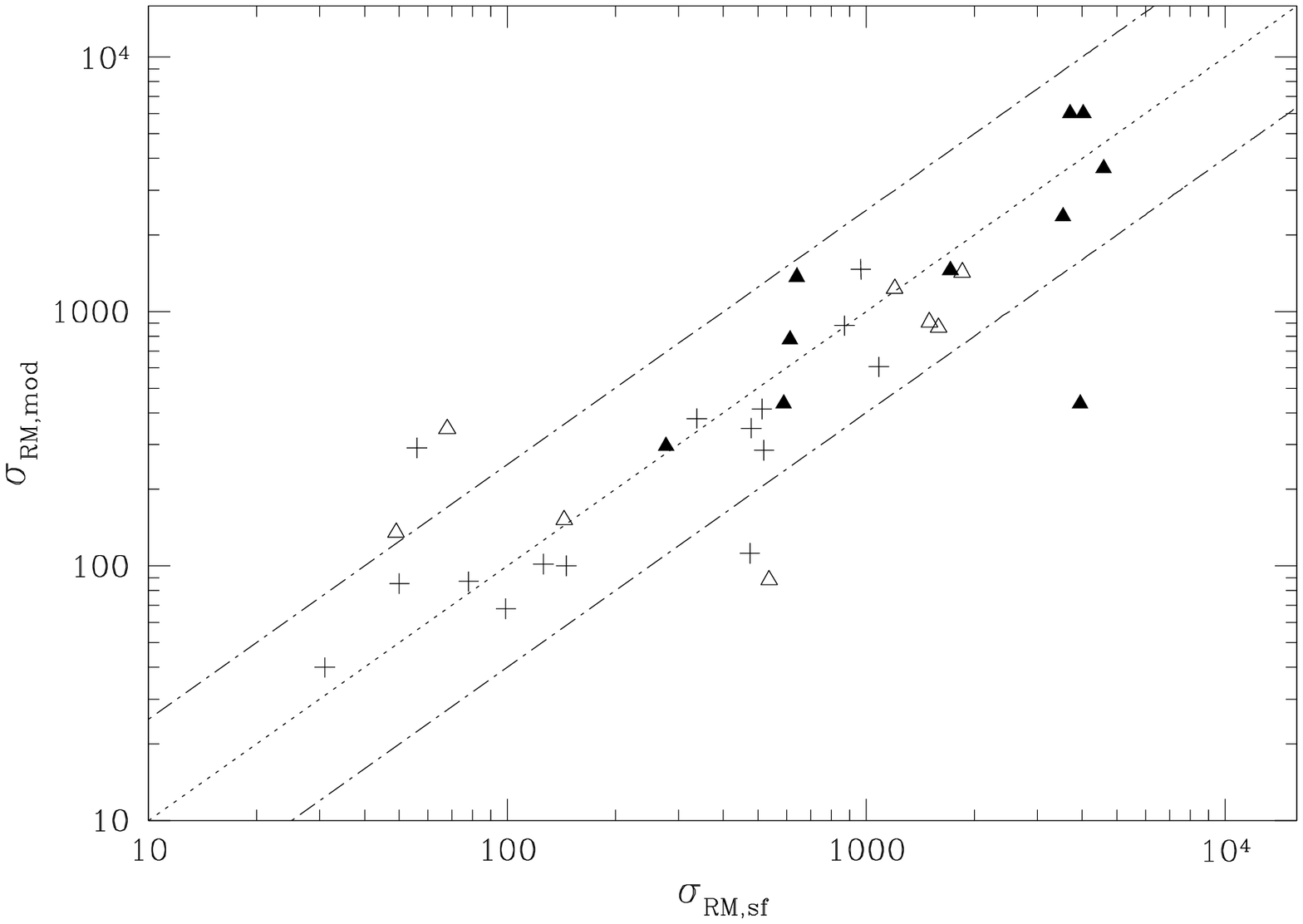}}     
       \resizebox{\hsize}{!}
      {\includegraphics{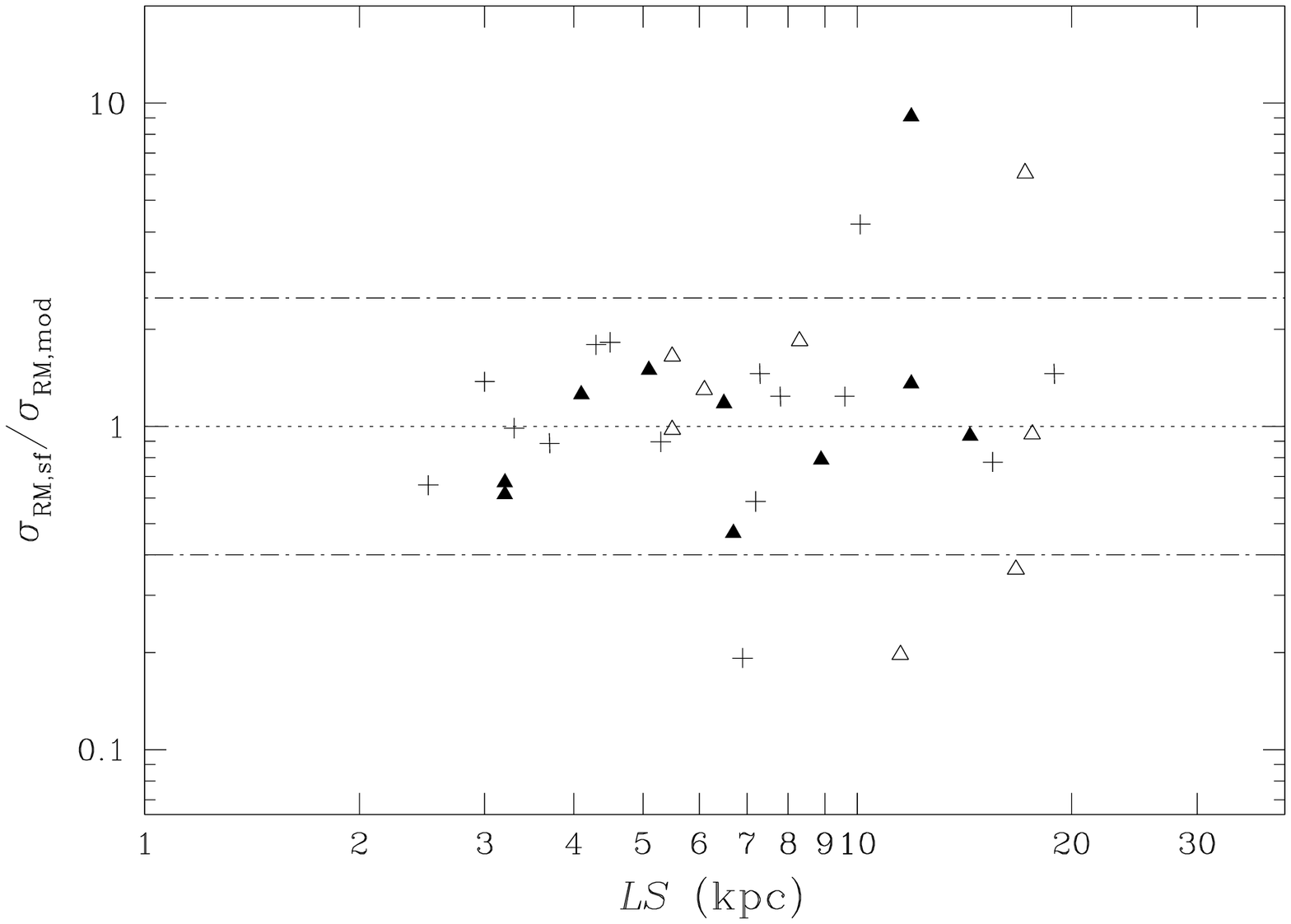}}
      \resizebox{\hsize}{!}
	{\includegraphics{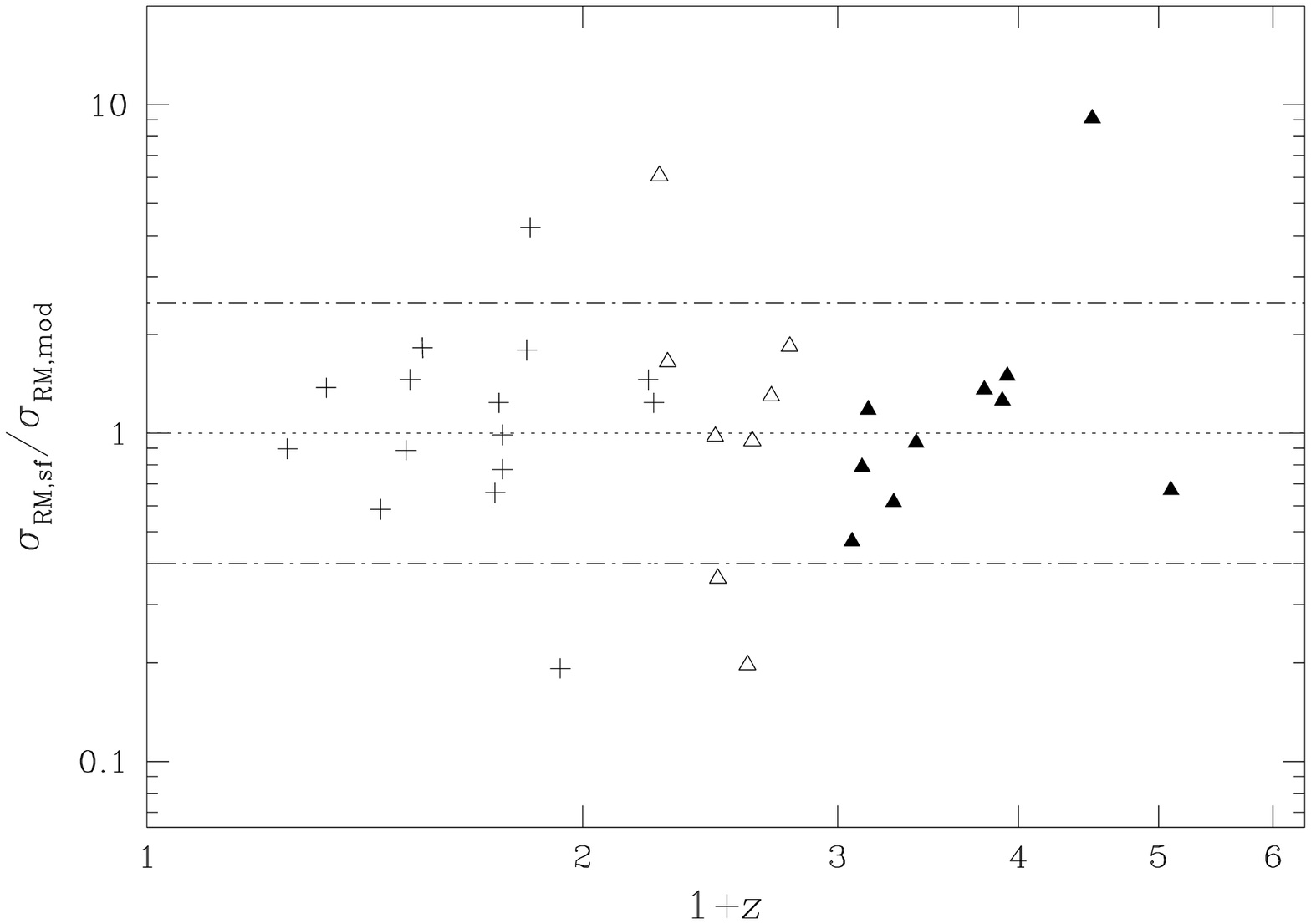}}
      \caption{{\it Top}: Plot of the final model $\sigma_{\rm RM,mod}$
        vs $\sigma_{\rm RM,sf}$. 
        {\it Middle}: Plot of the ratios 
	$\sigma_{\rm RM,sf}/\sigma_{\rm RM,mod}$ vs {\it LS}. {\it
        Bottom}: Plot of 
        the ratios $\sigma_{\rm RM,sf}/\sigma_{\rm RM,mod}$ vs $(1+z)$.
        Symbols are as in Fig.~\ref{fig:sigma_mod_1}. In all three
        panels, the two dash-dotted lines are a factor of 2 above and
        below the model, and include $\ge 80\%$ of the sources. }
      \label{fig:sigma_mod_2}
    \end{figure}

In Figure~\ref{fig:sigma_mod_2},  we compare the data and our adopted
model from Eq.~\ref{eq:sigmafit}: we observe no systematic effects
with respect to $\sigma_{\rm RM,sf}/\sigma_{\rm RM,mod} = 1$, nor any
significant dependence on $LS$ or $(1+z)$. 
The residual dispersion of  $\sigma_{\rm RM,sf}$ about the model is
a factor of approximately 2.5 at $\approx 90\%$ level, and is 
{\bb likely} due to 
intrinsic differences between one object and another. 

\bigskip
The model that we introduced in ~\citeauthor*{Fanti04} was {\bb
 described}  by the following King law:
\begin{eqnarray}\label{eq:oldmod}
  \sigma_{\rm RM,old} &=& \frac{F_{\rm old}}{ [1+(R/
                  r_c)^2]^{\delta_{\rm old}}},
\end{eqnarray}

\noindent
where  $R$ (kpc) is half the source linear size ($R = LS/2$). 
{\bb No  dependence on  $(1+z)$} was introduced, although
the value of $F_{{\rm old}}$ was assumed {\bb for}  the median redshift
$z_{\rm median} = 1$. 
{\bb We derived the following parameter values: $F_{\rm old} =
1.5 \times 10^5$~rad\,m$^{-2}$, $\delta_{\rm old} = 2$,  and assumed the core
radius $r_c = 0.5$~kpc}.  

{\bb For $LS \gg r_c$,  Eq.~\ref{eq:oldmod} can be approximated by a power
law  similar  to Eq.~\ref{eq:sigmafit}}:
\begin{eqnarray}
\label{eq:oldmod2}
  \sigma_{\rm RM,old} \approx F_{\rm old}\,{r_c}^4\, (LS/2)^{-4}  
            = {\cal A}_{\rm old}  LS^{-4}  
\end{eqnarray}
\noindent
and for $r_c = 0.5$~kpc, we find that ${\cal A}_{\rm old} = F_{\rm old}$. 

\smallskip
The two laws differ in their factor ${\cal A}$ and in their dependence on
{\it LS}, the new law being much flatter. They intersect at $LS\approx
3.8$~kpc.

The reason for the large difference between the old and new model
is that, in \citeauthor*{Fanti04}, the $\sigma_{\rm RM,old}(LS)$ model
was derived using the Cotton Effect itself,  namely from the
(visually-estimated) $\lambda$-dependent ``critical $LS$'' below which
the radio sources are almost totally depolarised. The steepness of the
relation was strongly constrained by the ``critical $LS$'' at 21~cm. 
The new data at 13~cm have changed this situation. On the one 
hand, they have shown that, in a significant number of sources, a fraction
of the polarised radiation remains constant at long wavelengths, 
 and, on the other hand,  they have increased the value of
$\sigma_{\rm RM,sf}$ for a number of sources. 
The ``critical sizes''  are now the 
result of a combination of a dependence  of $\sigma_{\rm
  RM,sf}$ on {\it LS}, and of {\it partial coverage} effects.  
{\bb Actually a number of sources are still  polarised at the longer
  wavelengths ($\lambda \ge 13~cm$) as a result of a ``partial coverage''
(which becomes important for $LS \ge 6$ kpc), when the
$\sigma_{\rm RM,sf}$, expected for these sizes at the long
  wavelengths, according to
Eq.~\ref{eq:sigmafit}, would be sufficiently large to totally 
depolarise all radiation for  $f_c = 1$. This is true, in particular,
for 21~cm.}

\subsection{The Cotton Effect Revisited}

    \begin{figure}[t]
      \centering
      \resizebox{0.976\hsize}{!}
      {\includegraphics{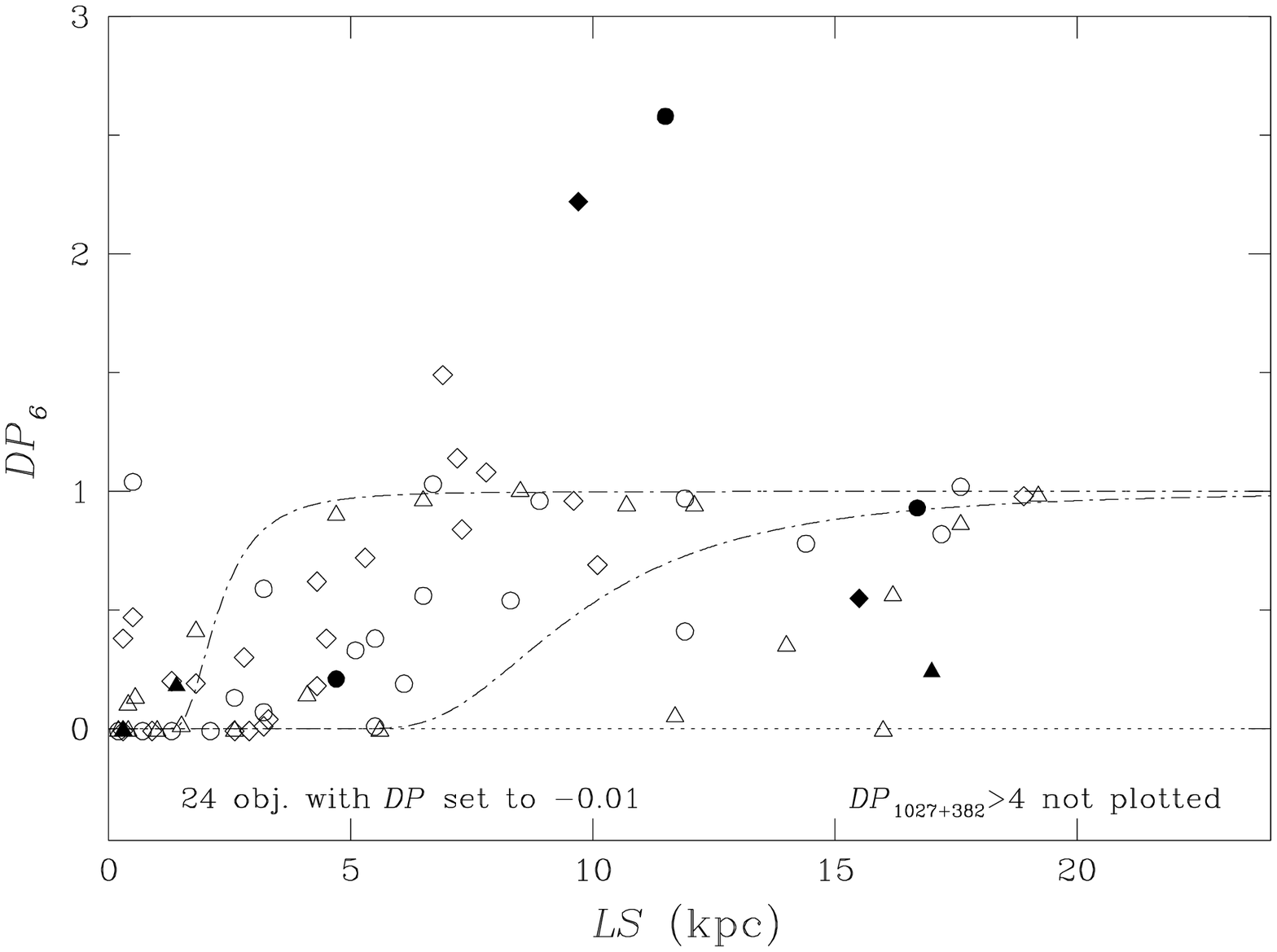}}
      \resizebox{0.976\hsize}{!}
      {\includegraphics{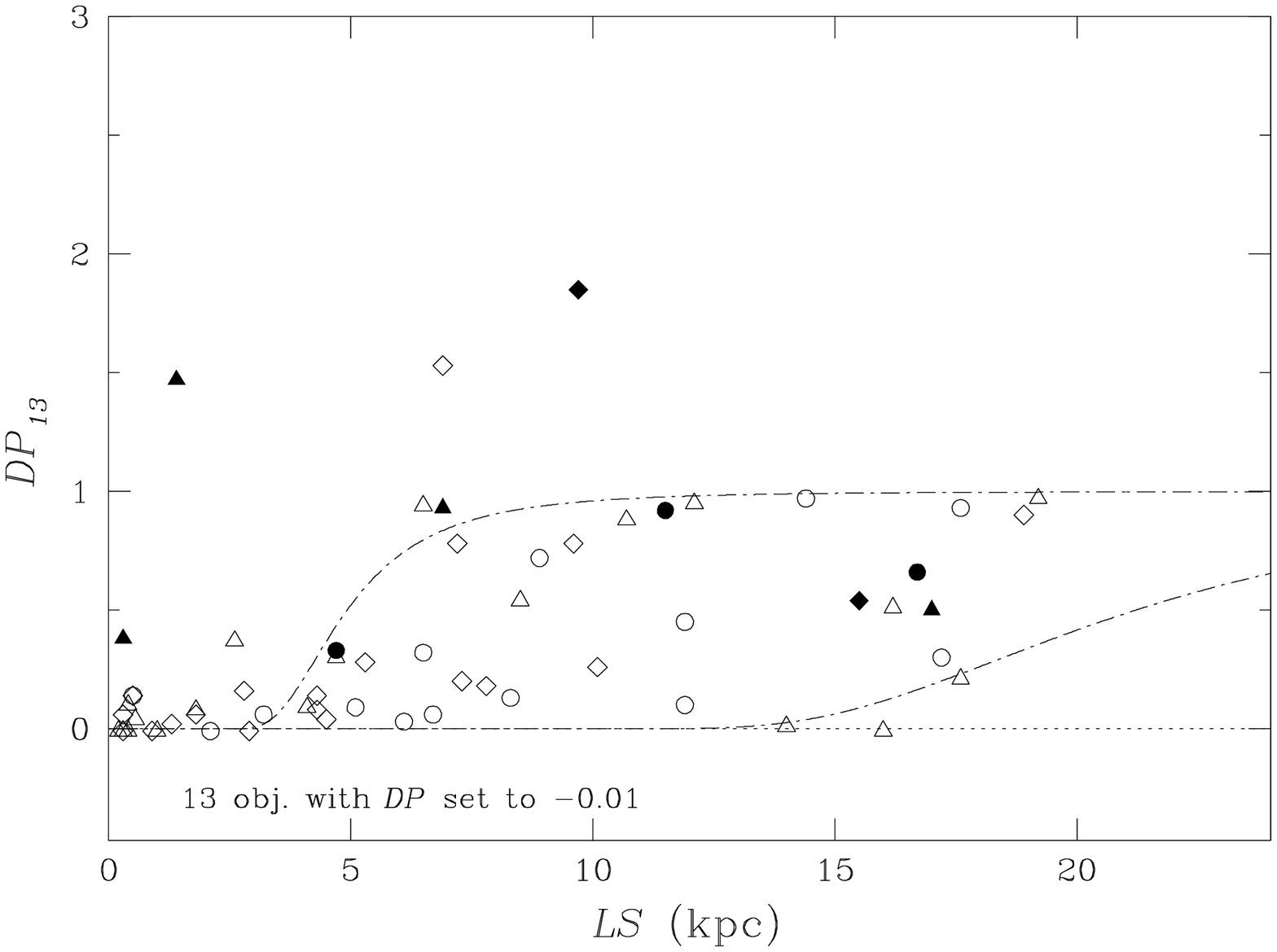}}
      \resizebox{0.976\hsize}{!}
      {\includegraphics{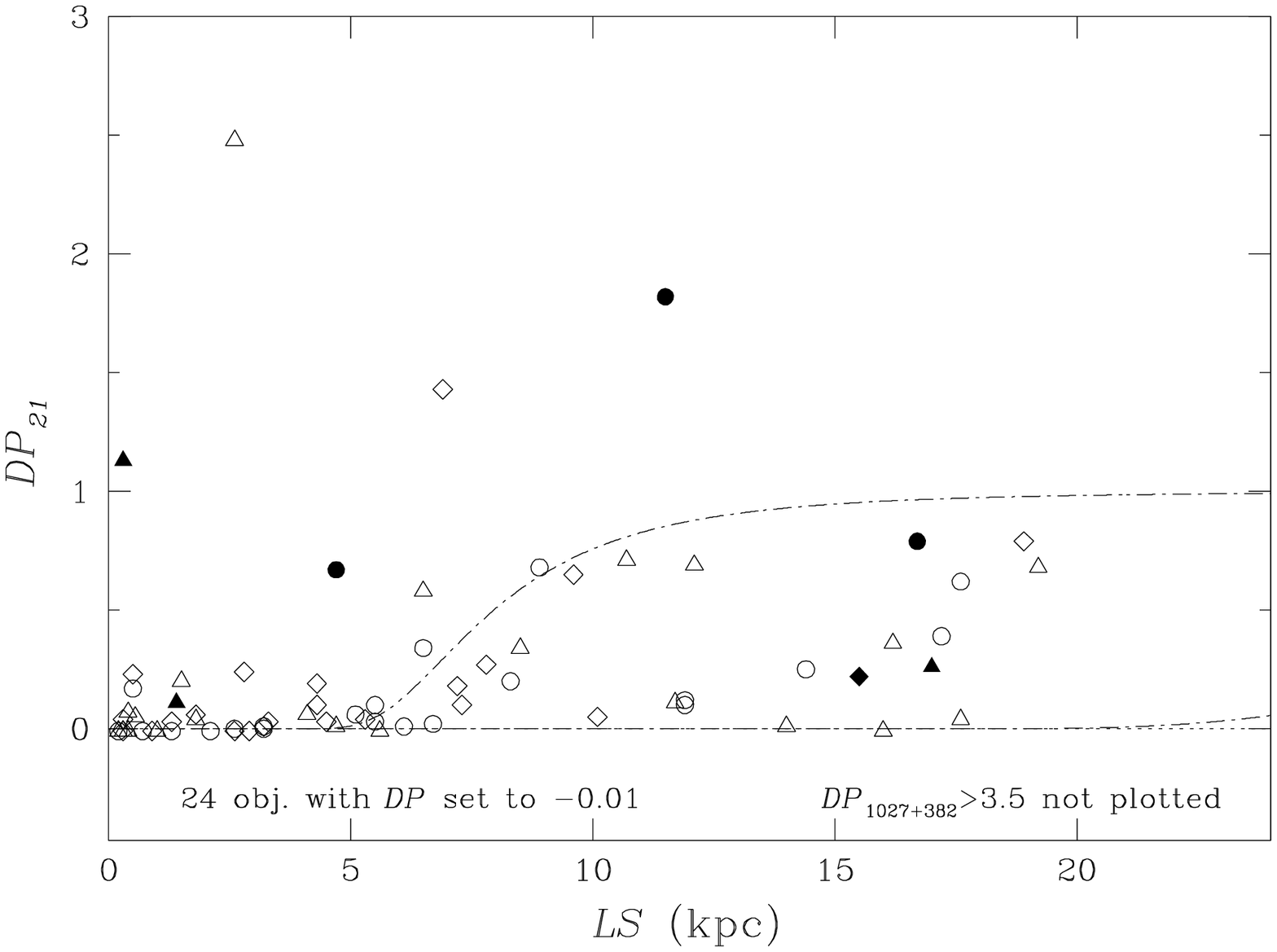}}
\vspace {-0.1 cm}
      \caption{From {\it top} to {\it bottom}: 
      ``$DP_\lambda$'' at 6, 13, 21~cm (see text). In all
	panels ($\Diamond$) are for $z<1.25$, ({\large $\circ$})
       for $z>1.25$, ($\vartriangle$)  for Empty Fields. 
      Sources showing an oscillating behavior are plotted as filled
      symbols. The upper curve is computed for $z=0.3$ and ${\cal
      A}$=500~rad\,m$^{-2}$\,kpc$^{-2}$ in  
      Eq.~\ref{eq:sigmafit}; the lower curve for $z=1.8$ and ${\cal
      A}$=2000~rad\,m$^{-2}$\,kpc$^{-2}$.}
      \label{fig:Cotton_mod}
    \end{figure}

{\bb We re-examined the ``Cotton Effect'' by constructing several model
¡ÈCotton plots¡É. These curves represent the depolarisation, $DP_{\lambda}=
m_{\lambda}/m_0$, that is expected at the observed wavelengths as a
function of {\it LS} and $z$}. 

We used  different values of ${\cal A}$, i.e. $500<{\cal
  A}$~(rad\,m$^{-2}$\,kpc$^{-2})<2000$, to take account of the
dispersion in $\sigma_{\rm RM,sf}/\sigma_{\rm RM,mod}$  (see
Eq.~\ref{eq:sigmafit} and Fig.~\ref{fig:sigma_mod_2}). 
These choices provide values for $\sigma_{\rm RM,mod}$ that are in the
observed range of $\sigma_{\rm RM,sf}$, for the majority  ($\sim 80\%$) 
of sources.

In Fig.~\ref{fig:Cotton_mod}, we plotted only two family 
curves, i.e. those with $z= 0.3$, ${\cal
  A}=500$~rad\,m$^{-2}$\,kpc$^{-2}$,  and $z= 1.8$, ${\cal
  A}=2000$~rad\,m$^{-2}$\,kpc$^{-2}$, which represent a sort of
minimum and maximum range for $DP_{\lambda}$ as a function of {\it LS}. 

{\bb The source depolarisation, $DP_{\lambda}$, lacking a
measure of $m_0$, is  approximated by $m_{\lambda}/m_{3.6}$, with the
assumption that  $m_{3.6}$ is sufficiently close to the intrinsic degree of
polarisation}. These data are plotted in Fig.~\ref{fig:Cotton_mod}
for two redshift bins ($z \le 1.25$, and $z > 1.25$).
Sources without redshift are also plotted.
We label the polarisation as ``$DP_\lambda$'' to highlight that this
is not the true {\bb depolarisation}. This approximation, however,
{\bb rises the following} problems: 
\begin{description}
\item[{\mdseries a)}] When the source is undetected or barely detected
  at 3.6~cm, it  is undetected or barely detected at
  other wavelengths,  and  ``$DP_\lambda$'' would be the ratio of
  two small random numbers {\bb (noise/noise)}. 
  This situation is particularly common for
  sources of   small Linear Sizes. We
  believe that these sources are subject to strong depolarisation at
  all wavelengths, {\bb and their data points are plotted with the 
fictitious value
  ``$DP_\lambda''=-0.01$, to emphasize that they are strongly depolarised}.
\item[{\mdseries b)}] In the few sources of small Linear Size detected
  only at 3.6~cm,  the computed ``$DP_\lambda$''  could
  be overestimated because the measured $m_{3.6}$ {\bb could} already be
  affected by depolarisation.  
\item[{\mdseries c)}] We mentioned in Sect.~\ref{depol} {\bb that some sources
  (perhaps as many as 9)} display an oscillatory behaviour of
  $m_{\lambda}$, which we interpret as internal beating of at least two
  sub-components of different values of {\it RM}. In these sources
  ``$DP_\lambda$'' does not represent the 
  depolarisation. However, for completeness, we plot
  these sources  in Fig.~\ref{fig:Cotton_mod}, using different symbols.
\end{description}
\medskip
The curves plotted in Fig.~\ref{fig:Cotton_mod} include the vast majority
of data points at all wavelengths.
Notable exceptions are:
\begin{itemize}
\item a number of points significantly above the curves
at small {\it LS} (typically $3\div 5$~kpc, depending on $\lambda$). 
These sources, according to the model of Eq.~\ref{eq:sigmafit}, are
expected to have a high $\sigma_{\rm RM,sf}$, and be totally
depolarised, 
but this is not the case. We examined 
these objects one by one, and found that
three with  ``$DP_6$'' significantly different from zero  
have $\sigma_{\rm RM,sf}$  smaller than average. At 21~cm, the four
sources with significantly high ``$DP_{21}$''  have $f_c \ltsim 0.8$,
and therefore part of the polarised emission escapes
unpolarised.

\item Five points, typically at $LS > 10$~kpc, are well below the lower
curve in Fig.~\ref{fig:Cotton_mod}, as seen in the top panel (``$DP_6$''). 
According to Eq.~\ref{eq:sigmafit}, these would be expected to have a
small $\sigma_{\rm RM,sf}$  and hence be depolarised by a small {\bb amount},
which is not the case. They may have an intrinsic value of
$\sigma_{\rm RM, sf}$ that is much larger than average (${\cal A}\gg
2000$~ rad\,m$^{-2}$\,kpc$^{-2}$). Only one source has a redshift and 
{\bb indeed it} has the second highest $\sigma_{\rm RM, sf}$ in the
sample. If this were the case {\bb also} for the remaining four sources,
these should
disappear at other wavelengths.  This effectively happens. The
fact that these are mostly empty fields suggests that they might be at
high redshift. 

\item A few  sources have ``$DP_{\lambda}$''$\gg 1$. These are sources with
an oscillatory behaviour and $m_{3.6}$ depressed by the oscillations. 
\end{itemize}

\subsection {The  Faraday curtain: a physical model}
\label{natureFC}

\begin{figure}[t]
      \centering
      \resizebox{\hsize}{!}
      {\includegraphics{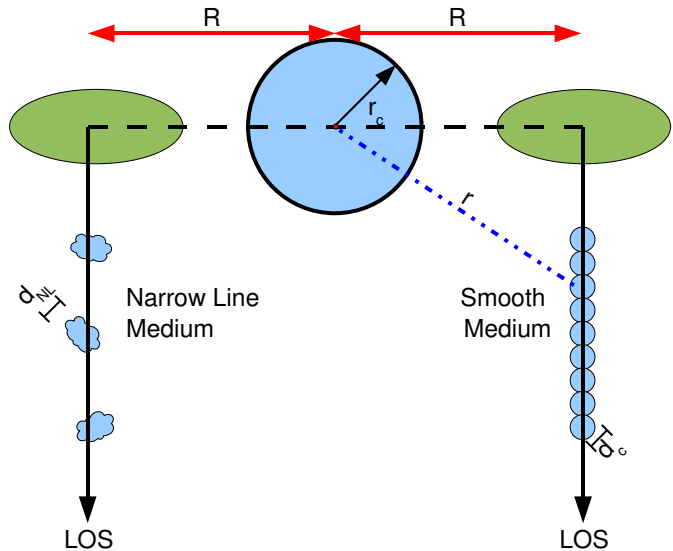}}
      \caption{Sketch of the magneto-ionic ISM model around  the
	sources. {\bb The line of sight (LOS) is the $x$ axis.}}
      \label{fig:sketch}
    \end{figure}

To provide a physical basis to  the empirical model of the 
``Faraday curtain'' (Eq.~\ref{eq:sigmafit}), as in
\citeauthor*{Fanti04}, we adopt the model of 
a magneto-ionic medium (either smooth or clumpy) which is spherically
symmetric about the radio source, described by a King-like profile of the
relevant parameters, and a randomly-oriented magnetic field. We
assume, for simplicity, that the radio-source axis is orthogonal to
the line of sight (non-orthogonality
effects will be considered in Sect.~\ref{Phys-empir}). Any line of
sight (los) to a point of the radio source will pass through several
different elements of the medium, of average size $d$ much smaller
than the source size, in each of which the (randomly-oriented)
magnetic field $B$ and the electron density  $n$ may differ from each 
{\bb others'}
(see a sketch of this model in Fig.~\ref{fig:sketch}). Each
medium element rotates the polarisation, according to the Faraday
law, by an amount $\delta RM = k n B_{||} d$.\footnote{In the
  following, we express the magnetic field $B$ in $\mu$G, 
the electron density  $n$ in cm$^{-3}$, and every linear size ($LS$,
$R$, $d_c$, $d_{\rm NL}$, $r_c$) in kpc. With these units, the value
of the  constant $k$ in the Faraday law is 810~rad\,m$^{-2}\,\mu{\rm
  G}^{-1}$\,cm$^3$\,kpc$^{-1}$.}
As the field orientation changes at random from element to element, 
the overall rotation along the line of sight ($x$ coordinate) 
 is on average zero, with a variance given by

\begin{equation}
\sigma_{\rm RM,los}^2 
\approx 
k^2 \int  d^2 \; n(x)^2 \; B^2_{||}(x)  d{\cal N},
\label{eq:integral}
\end{equation}
\noindent
where ${d \cal N}=dx/d$ is the number of elements crossed by the line of
sight in the interval $dx$. The integration 
is over all elements along that line of sight. For
the smooth medium, this number is roughly the integration length
divided by the element size.

For both $n(r)$ and $B(r)$ we also assume a King-like distribution,  i.e.
\begin{eqnarray}\label{eq:King}
  n(r) =  \frac{n_0}{[1+(r/r_{\rm c})^2]^{3\beta/2}}\nonumber \\
  B(r) =  \frac{B_0}{[1+(r/r_{\rm c})^2]^{3\mu/2}},
\end{eqnarray}

\noindent
where $r_c$ is the {\it core radius} of the medium distribution.
In this model, $\sigma_{\rm RM,los}$ 
is obtained by solving the integral in Eq.~\ref{eq:integral} and, as
shown by \citet{Dolag01}, is given by:

\begin{eqnarray}\label{eq:dispersion}
\sigma_{\rm sf,los}(R) &=& \frac{\displaystyle{F}}
{\displaystyle{[1+(R/ r_{\rm c})^2]^{\delta}}}\nonumber \\
                       &\approx& F\,r_c^{2\delta}\,R^{-2\delta}~~~~
{\rm for}~~ R \gg r_c.
\end{eqnarray}

\noindent
The dependence of $F$ and $\delta$ on the ISM parameters, varies
depending on whether it is smooth or clumpy.
These two components coexist, and we have to understand which
plays the major role. 

\subsubsection{ Smooth medium} 
We assume a ``continuum'' distribution of ISM elements (``cells'') of  low  
density, with scale $d=d_c$, central density $n_0 = n_{\rm sm,0}$ and
central  intensity of the randomly-oriented  magnetic field
$B_0=B_{\rm sm,0}$
in Eq.~\ref{eq:King}.
In this case, the parameters of Eq.~\ref{eq:dispersion} are:
\begin{eqnarray}\label{eq:Fsm}
F &=& K\, B_{\rm sm,0}\, n_{\rm sm,0}\, r_{\rm c}^{1/2}\,d_{\rm
  c}^{1/2} \mbox { rad\,m}^{-2}  \nonumber \\
\delta &=& \frac{6(\beta + \mu)-1}{4},
\end{eqnarray}

\noindent
$K$ is a factor dependent on $\delta$ and, in our units, is in the 
range 300--400.

In the case of equipartition between magnetic and thermal energy
(isothermal case), or of a constant ratio between the two,  $\mu =
\beta/2$. 

\subsubsection {Clumpy medium}
This medium could be, for example, the Narrow Line Region (NLR), which
{\bb consists} of high-density (central value $n_0 = n_{\rm NL,0}$) magnetized
clouds,  $d_{\rm NL}$ in  size,  with a randomly-oriented magnetic 
field of central intensity $B_0=B_{\rm NL,0}$, and volume {\it filling
  factor} $f(r)$.
To ensure pressure equilibrium between the two components, we assume,
in {\bb the first Eq.}~\ref{eq:King}, that $\beta_{\rm NL}$ is {\bb 
equal to $\beta$ of
  the smooth medium} in which the NL clouds are embedded. 
To evaluate the integral of Eq.~\ref{eq:integral}, we have to model
the distribution of the number per unit volume of the clouds, $N_{\rm
  NL}(r)$. We assume that the latter scales with $r$ as a
King-like law: 

\[N_{\rm NL}(r) =  \frac{N_{\rm NL,0}}{[1+(r/r_{\rm c})^2]^{3\epsilon/2}} \]

The cloud filling-factor $f(r)$ is related to $N_{\rm NL}(r)$ by
the  relation:

\[f(r) = \frac{N_{\rm NL,0}\, d_{\rm NL}^3}{[1+(r/r_c)^2]^{3\epsilon/2}} =
{\frac{f_0}{[1+(r/r_c)^2]^{3\epsilon/2}}},\]
\noindent
where $f_0$ is the filling factor at $r = 0$. The  total number of 
clouds along a line
  of sight, at a distance $R$ from the centre, ${\cal N}(R)$, and
the covering factor of these clouds, $f_c^{\rm NL}$, are given by :

\begin{eqnarray*}
{\cal N}(R)\! &=&\! H \frac{\displaystyle{N_{\rm NL}\, d_{\rm NL}^2\,r_c}}
{\displaystyle{[1+(R/r_c)^2]^{(3\epsilon -1)/2}}}   
         \! =\!\frac{\displaystyle{H\cdot (f_0 r_c/d_{\rm NL})}}
{\displaystyle{[1+(R/r_c)^2]^{(3\epsilon -1)/2}}}\nonumber\\
f_c^{\rm NL} &=& 1 - e^{-\cal N},
\end{eqnarray*}

\noindent
where $H$ is a factor dependent on $\epsilon$ and has a typical value
of a few units. 

\smallskip
The parameters of Eq.~\ref{eq:dispersion} {\bb are}:
\begin{eqnarray}\label{eq:FNL}
F &=& K\, B_{\rm NL,0}\, n_{\rm NL,0}\, r_{\rm c}^{1/2}\,d_{\rm
  NL}^{1/2}\, f_0^{1/2} \mbox { rad\,m}^{-2} \nonumber \\
\delta &=& \frac{[6(\beta + \mu) + 3\epsilon] -1}{4}.
\end{eqnarray}

\noindent
As in Eq.~\ref{eq:Fsm},  $K$ is a factor dependent on $\delta$, in the
range 300--400. 

\noindent
 Equations~\ref{eq:FNL} are {\bb similar to  Eqs.~\ref{eq:Fsm}},
 with cloud parameters in place of those 
of the smooth medium, apart for the factor  $f_0$ and
 $\epsilon$. These terms account for the fact that 
along any integration length $\ell$ only a fraction  $f$ is filled
by clouds, and that their number density decreases with increasing $r$.

\subsection{Physical vs empirical model}
\label{Phys-empir}

{\bb If the polarised radiation were produced at the outer edges of
  the lobes only (hot spots), with the approximation $R \gg r_c$,  the second
Eq.~\ref{eq:dispersion}  could be related to Eq.~\ref{eq:sigmafit},
for $R=LS/2$, which is the length of each lobe on the 
assumption that the source is symmetric.} We would derive $\delta
= 1$ and $F={\cal A}~(2 r_c)^{-2}(1+z)^4$.

However, the polarisation does not always originate from hot spots
only. We {\bb made} some simulations that assumed a constant
 polarisation brightness, distributed over a fraction  of the lobe length
from the outer edges towards the centre, ranging in value between
0 (hot spot only) and 1 (entire lobe polarised).  

The simulations illustrate that 
we can describe the effects of polarisation across a fraction of
the source axis by introducing a factor 
$h=R_{\rm eff}/(LS/2)$,
where $R_{\rm eff}$ is the ``effective distance'' between the source
centre and  the ``centroid''  of  the polarised radiation, and 
$LS/2$ is the length of each  lobe for a symmetric source. 
{\bb The range of $h$ is, with good approximation, between $h \approx 1.0$  
(polarised radiation concentrated at the hot spots), and $h \approx
0.5$ (polarisation  uniformly-distributed over the lobe).}  
In other words, one should relate $\sigma_{\rm RM,mod}(LS)$
(Eq.~\ref{eq:sigmafit}) to $\sigma_{\rm RM,los}(R)$ 
(Eq.~\ref{eq:dispersion}) for $R = R_{\rm eff} = h (LS/2)$. 
We should therefore have:

\begin{eqnarray}\nonumber\label{eq:sigmamod2}
{\sigma_{\rm RM,mod}} &=&
    \frac{\displaystyle{F}}{\displaystyle{[1+(h (LS/2)/ r_{\rm
	    c})^2]^{\delta}}}\mbox {      rad\,m}^{-2}\\ 
&\approx& F (2 r_c)^2 (h \cdot LS)^{-2 \delta}~~~~\mbox{ for } LS \gg
    2 r_c. 
\end{eqnarray}

The location and extension of the polarised emission within the lobes
change from source to source; this is likely to be one of the reasons 
for the residual dispersion of the data points with respect to 
Eq.~\ref{eq:sigmafit}. In Fig.~\ref{fig:sigma_mod_2}, we expect that
the  lines $\sigma_{\rm RM,sf}/\sigma_{\rm RM,mod}=1$ 
correspond  to the average value $\langle h\rangle \sim 0.75$, and sources 
with larger (smaller) $h$ will lie above (below) the line. Comparing
Eq.~\ref{eq:sigmamod2} with Eq.~\ref{eq:sigmafit} we find:

\begin{eqnarray}
\delta \approx 1 ~~~{\rm and} ~~~ F = {\cal A}~ (2 r_c)^{-2} \langle h\rangle^{2} (1+z)^4\nonumber\\
   \approx 140~ r_c^{-2} (1+z)^4 \mbox { rad\,m}^{-2}.  
\label{eq:Fmod}
\end{eqnarray}
\noindent

The condition $LS\gg 2r_c$ allows us to take $r_c = 0.5$~kpc as a
fiducial  value for the core radius in what follows.
 
We need to examine some further assumptions that we have made.
In general, the sources are not in the plane of the sky and their
shape may be asymmetric with respect to their {\bb cores}, either
intrinsically or because of an asymmetric ISM distribution. The
orientation and asymmetries of the source structure may {\bb also} cause 
the residual dispersion in the data with respect to Eq.~\ref{eq:sigmafit}.

We simulated different source orientations and
arm-length asymmetries using Eq.~\ref{eq:sigmafit} 
as an approximation to Eq.~\ref{eq:dispersion}. 
For sources that are not perpendicular to the line of sight, we found
that the larger depolarisation of the far component is compensated by
the lower depolarisation of the nearby component; this implies that
the global $\sigma_{\rm RM, sf}$ does not differ significantly
(generally within 15\%) from that expected for a source of similar
{\it LS}, but in the plane of the sky. 
There are, however, some systematic
effects that occur at orientation angles $\ge 40\degr$ in sources of 
large sizes and/or at high redshifts. One consequence is that
the parameter $\delta$ could be underestimated and {\bb be  $\ge$
1.1}. A second consequence concerns the 
interpretation of the ``partial coverage''  and will be discussed in
Sect.~\ref{covering-factor}.

The arm asymmetry does not appear to cause {\bb major} effects.

\subsection{Towards smaller sizes and higher frequencies}

It is interesting to use the depolarisation model of
  Sect.~\ref{natureFC} to predict what
could be found {\bb at  source sizes $\le 2.5$~kpc and/or at
  wavelengths $\le 3.6$~cm.} 

The model shows that at short wavelengths 
the Faraday curtain may stop  being foggy at small {\it LS}. The important
parameter in this regime 
is the core radius, $r_c$, for which we have adopted a tentative value
of 0.5~kpc at large {\it LS} (Sect.~\ref{natureFC}).

Using Eqs.~\ref{eq:sigmamod2} and Eq.~\ref{eq:Fmod} to compute 
$\sigma_{\rm RM,mod}$,  with  $r_c = 1$~kpc,  we find that at 3.6~cm,
for $z\le 1$ and $h = 0.5$, $DP$ never falls below 0.4  for $LS \ge 1$~kpc,
and   $DP \ge 0.35$, even for $LS \le 0.4$~kpc.
This contradicts the low level of polarisation found in
our sources at $LS \le 1$~kpc, which, instead, is compatible with the value
$r_c \le 0.5$ that we have assumed.

At 2~cm, from our model with $r_c = 0.5$~kpc,  we would expect $DP \ge
0.2$ at all sizes for z $\le 1$.   
To our knowledge, integrated fractional polarisation data are rare in
the literature at this frequency. At most, peak polarisation
flux-densities are reported \citep[e.g.][]{Stangh01}. We infer
nevertheless, from these sparse data, that an $r_c \ll 0.5$~kpc may be
required.

\subsection{Physical Properties of the Faraday curtain}
\label{farprop}

From the comparison of the empirical model (Eq.~\ref{eq:sigmafit})
with the physical model (Eq.~\ref{eq:dispersion}), we can estimate the physical
parameters for either the smooth- or the clumpy-medium component,
assuming each to be independently responsible for the {\it RM} 
dispersion, $\sigma_{\rm RM,sf}$. We attempt to determine which
medium is {\bb causing}  the depolarisation.

\medskip\noindent
$\bullet$  {\it The Smooth Component model}.

From the comparison of Eq.~\ref{eq:Fsm} and Eq.~\ref{eq:Fmod} we derive:

\begin{eqnarray}
K\, B_{\rm sm,0}\, n_{\rm sm,0}\, (r_c\, d_c)^{1/2} &\approx&
\frac{140 (1+z)^4}{r_c^2} ~~~{\rm rad\,m}^{-2}\nonumber\\ 
\delta &=& \frac{6(\beta+\mu)-1}{4} \approx 1,  
\label{eq:relsm}
\end{eqnarray}
\noindent
where $K \approx 380$. From the second Eq.~\ref{eq:relsm}, we derive 
$\beta+\mu = 5/6$. If we assume that $\mu = \beta/2$ (see
Sect.~\ref{natureFC}), we derive $\beta \approx 5/9$.
If instead $B$ is constant ($\mu =0$), then  $\beta = 5/6$. 

{\bb Therefore, the density scales with distance as}:
\begin{eqnarray*}
{n_{\rm sm}(r) = \frac{n_{\rm sm,0}}
{[1+(r/r_c)^2]^{0.83 \div 1.25}}}\,.
\end{eqnarray*}
The parameter $\beta$ is in the range of that derived, using
X-ray observations, for the hot component of the interstellar medium 
of nearby early-type galaxies.

\smallskip
\noindent
From the first Eq.~\ref{eq:relsm}, we have:
\begin{eqnarray*}
\frac{n_{\rm sm,0}\, B_{\rm sm,0}}{{\rm cm}^{-3} \mu{\rm G}} 
= 0.37\, (1+z)^4\,\left(\frac{d_c}{\rm kpc}\right)^{-1/2}\,
\left(\frac{r_c}{\rm kpc}\right)^{-5/2}.
\end{eqnarray*}
\noindent
Introducing a constant ratio, $g_{\rm sm}$, between thermal ($2 n_{\rm
  sm,0}\, k_{\rm B}\,T$) and magnetic ($B_{\rm sm,0}^2/8\pi$) energy
  densities, we derive:  
\begin{eqnarray*}
\frac{n_{\rm sm,0}}{{\rm cm}^{-3}} = 1.4 \cdot 10^{-5}\, g_{\rm
    sm}\,\left(\frac{B_{\rm sm,0}}{\mu {\rm G}}\right)^2\, T_7^{-1},
\end{eqnarray*}
\noindent
where $T_7$ is the temperature of the smooth medium in units of
$10^7$\,K. Combining these two expressions, we obtain:
\begin{eqnarray*}
\frac{n_{\rm sm,0}}{{\rm cm}^{-3}} &\approx&  1.3\cdot 10^{-2}\, g_{\rm
  sm}^{1/3}\, (1+z)^{8/3}
\left(\frac{d_c \,r_c^5}{{\rm kpc}^6}\right)^{-1/3}  T_7^{-1/3} \\
\frac{B_{\rm sm,0}}{\mu {\rm G}} &\approx&  30\, g_{\rm sm}^{-1/3}\, 
\left(\frac{d_c\, r_c^5}{{\rm kpc}^6}\right)^{-1/6} T_7^{1/3}
\, (1+z)^{4/3}.
\end{eqnarray*}
We have no information about $d_{\rm c}$, apart from that it should be
sufficiently small, compared to the sizes 
of the smallest sources, to produce the observed strong depolarisation;
therefore presumably  it is on parsec scales.  
Hence, for $r_c \le 0.5$~kpc, $d_c \approx~ 1$~pc, and $T_7 \le 1$,  
we have $n_{\rm sm,0}(z=0)\approx~ 0.4\, g_{\rm sm}^{1/3}$~cm$^{-3}$.  
For $g_{\rm sm} \approx 1$ (magnetic and thermal energy close to 
equipartition), $n_{\rm sm,0}$ is in the range of the values {\bb derived
from}  X-ray observations for the hot component of the interstellar medium
in the central regions of nearby early-type galaxies.

In addition, the dependence of $\sigma_{\rm sf,mod}$ on ($1+z$) 
(Sect.~\ref{farCurt}) indicates that the density of the medium increases 
strongly with redshift and then saturates. This occurs close to  the
``magic'' redshift at which radio sources reach their maximum space 
density and perhaps ``star formation'' is  close to its maximum.   

\medskip\noindent
$\bullet$ {\it The NL region model}.

In ~\citeauthor*{Fanti04}, we assumed that the steep decline of
$\sigma_{\rm RM,sf}$ was due only to the  decrease in the number of
clouds per unit volume, as a function of $r$; {\bb we  found} that the
implied covering factor of the clouds becomes negligible at sizes
greater than a few kpc (the {\it NL covering problem}), making the
model unrealistic.  
Using the new parameters from the revised empirical model, and
 assuming  a King-like distribution for cloud density and
magnetic-field strength (Eq.~\ref{eq:King}), not considered in 
~\citeauthor*{Fanti04}, we re-examine the case.

From the comparison of Eq.~\ref{eq:Fmod} and Eq.~\ref{eq:FNL}, 
 we derive the following relations:
\begin{eqnarray}
K B_{\rm NL,0}\, n_{\rm NL,0}\, (r_c d_{\rm NL}\, f_0)^{1/2}
\approx \frac{140 (1+z)^4}{r_c^2}~~~ {\rm rad\, m}^{-2} \nonumber\\
\delta = \frac{[6(\beta + \mu) + 3\epsilon] -1}{4} \approx 1
 \label{eq:relNL}
\end{eqnarray}
\noindent
($K\approx 380$).

To keep the number of parameters as low as possible, we assume that
$\epsilon = \beta$ and $\mu = \beta/2$ (as assumed for the smooth medium);
hence, from the second  Eq.~\ref{eq:relNL}, we obtain $\beta  = 5/12$.
If we had assumed $B$ to be constant, we would have derived $\delta =
[(6 \beta + 3\epsilon) - 1)]/4$ and   $\beta = 5/9$. 

We observe that the density of the ambient diffuse medium, in which clouds
are embedded, decreases in a way that is similar to
the  ``smooth medium only'' model, for a slightly lower value of $\beta$
($0.43 \le \beta \le 0.56$).  
For these values of $\beta$, the number of clouds along the line of
sight, ${\cal N}(R)$, decreases slowly with $R$ ($\propto R^{-1/4}$),
outside the core radius, and  $f_c^{\rm NL}$ is almost
constant. 
If we assume, {\bb instead}, that $n_{\rm NL}$ and $B_{\rm NL}$ are
both constant,  {\bb as in ~\citeauthor*{Fanti04}},
then ${\cal N}(R)$ drops quickly as a function of $R$, and $f_c^{\rm
  NL}$ becomes negligible, {\bb thus causing}  the {\it NL
covering problem}.

\smallskip
We now  evaluate the strength of the magnetic field.  
From the first of Eq.~\ref{eq:relNL} we derive: 
 
\begin{eqnarray*}
\frac{B_{\rm NL,0}}{\mu {\rm G}}\approx 0.4 \cdot
\left(\frac{n_{\rm NL,0}\, r_c^2\, d_c}{{\rm cm}^{-3} {\rm kpc}^3}\right)^{-1}
\, (f_0\, r_c/d_{\rm NL})^{-1/2}\,
(1+z)^{4}.
\label{Bnl}
\end{eqnarray*}

We first consider the situation for $z = 0$.
We take  $n_{\rm NL,0}(z=0)  \approx 5 \times 10^3$~cm$^{-3}$, which
is {\bb in the range of} values quoted in the literature  
~\citep[e.g][]{peterson,koski}.
The quantity $(f_0\, r_c/d_{\rm NL})$ is the average number of NL
clouds along the line of sight, within the core radius. To have a covering
factor $\ge 0.9$ for $LS \le 4$~kpc, as required by the data
  (Fig.~\ref{fig:cf_LS}),   
$(f_0\, r_c/d_{\rm NL})$ has to be $\le 1$. Taking $r_c \le 0.5$~kpc and 
$f_0 \approx 10^{-4}$ (a value generally quoted in the literature,
e.g.~\citealt{peterson}), we
derive  $d_{\rm NL} \gtsim 0.05$~pc, of the order of what quoted from
spectroscopic observations ~\citep{peterson}. We obtain:
\[B_{\rm NL,0}(z = 0) \approx 6~\mu{\rm G}. \]
The ratio between thermal energy and magnetic energy is:
 \[g_{\rm NL} \approx 10^4.\]

This model is not affected by the {\it NL covering problem}
we experienced in ~\citeauthor*{Fanti04}. It does, however, have {\bb some
 implications}: 
\begin{description}
\item[{\mdseries i)}]The density inside the clouds, $n_{\rm NL}$,
  decreases with $r$, and at 5~kpc is already a factor 10 below the
central value.
\item[{\mdseries ii)}] The magnetic field is quite  low in these dense 
clouds, because the magnetic energy is a negligible fraction
($\approx10^{-4}$) of the  thermal energy.
 \item[{\mdseries iii)}] The quantity $n_{\rm NL} B_{\rm NL}$ is a 
strong function of $z$.
If $g_{\rm NL}$ were independent of $z$,  $n_{\rm NL}$ and
$B_{\rm NL}$ would increase  with redshift in proportion to
$(1+z)^{8/3}$, and $(1+z)^{4/3}$ 
respectively, out to $z = 1.8$ (where  $\sigma_{\rm RM,sf}$
saturates). At this redshift,  the cloud
density $n_{\rm NL}$ would be larger by more than a factor 10. 
If instead the magnetic-field strength were independent of $z$,
the increase of $n_{\rm NL}$ up to $z = 1.8$ would be a factor
$\approx 60$. 
\end{description}
\noindent
\smallskip
{\bb Whether or not these implications of the NL model are acceptable 
is not clear to us. }

\subsection{Implications for young radio source evolution}

CSS sources are {\bb considered to be mostly} {\it young radio sources}
with typical ages far smaller than $ 10^6$~years. They nevertheless
form a large fraction of sources in radio catalogues. {\bb This old  
problem} can be overcome by assuming that in early life-stages, the
GPS/CSS phase, radio sources are luminous {\bb and then dim, 
due to adiabatic expansion moderately balanced 
by a continuous energy-injection from the ``nuclear
engine''  ~\citep[see, e.g., ][]{Scheuer74, Baldwin82, Fanti95,
  Readhead96, Begelman96}}.

In these simple physical models 
the evolution of the source luminosity depends on
 the density distribution of the ambient medium, which is assumed  to
 be  a power law [$n(r) \propto r^{-\eta}$, with $\eta = 3
   \beta$]. The range of values required by the models is $1.5 \le
 \eta \le 2.0$ ~\citep[see, e.g.,][and references therein]{crfan}. 
 
In our analysis of depolarisation parameters for a ``smooth medium'',
the density distributions derived are consistent with our above
estimates. The density distribution for the $NL~ model$ is flatter
($1.2 \le \eta \le 1.6$), but perhaps in agreement, within the
uncertainties.

\subsection{Origin of ``Partial Coverage'' } 

\label{covering-factor}

\citet{Burn66} discussed a variant of his own model 
(Eq.~\ref{eq:delta}) that considered ``partial coverage''.
He assumed that the Faraday depolarisation is due to discrete
clouds\footnote{Burn supposed that the clouds are in our Galaxy, but the
  model is easily usable for a location around the radio source} 
 whose average number along the line of sight is ${\cal N}$.
If ${\cal N}~\gg 1$, the depolarisation is similar to that of the
Gaussian model (Eq.~\ref{eq:delta}). If ${\cal N}\ll 1$, a number of
lines of sight, however, will not intersect any cloud. Therefore 
a fraction of the source's polarised radiation will emerge undepolarised
through  ``holes'' in the ``curtain''
keeping a constant level of fractional polarisation 
at long wavelengths. 
The relevant equation is:
\begin{equation}
m = m_0 e^{-{\cal N}(1-e^{2F_c^2 \lambda^4})},
\label{eq:Burn_mod}
\end{equation}
\noindent
where $F_c$ is the $RM$ of a single cloud. 

\begin{figure}[t]
      \centering
      \resizebox{0.976\hsize}{!}
      {\includegraphics{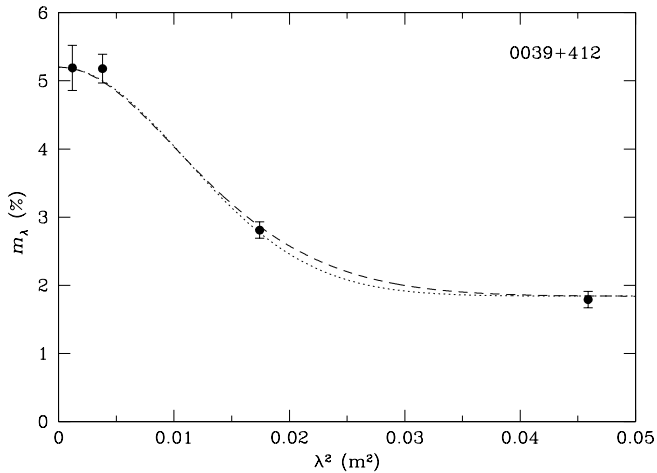}}
      \caption{B3\,0039+412: example  of data fit with Burn's
	Eq.~\ref{eq:Burn_mod} (dashed line) and with our
	Eq.~\ref{eq:cover_f} (dotted line).} \label {fig:2_mods}
\end{figure}

This equation, although {\bb formally different} from the
empirical one that we have used (Eq.~\ref{eq:cover_f}), {\bb is similar in
shape}. The differences cannot be discerned using the available data,
because of the limited  wavelength coverage and the accuracy with
which $m_{\lambda}$ are measured (Fig.~\ref{fig:2_mods}). 

The revised Burn model would appear appropriate if 
depolarisation is due to clouds of the NLR. However, as discussed in
Sect.~\ref{farprop}, if the NL model that we have described is tenable, we
{\bb expect}  that  ${\cal N}(R)$ slowly decreases with $R$, and we expect 
covering factors $\ge 0.9$, {\bb while we also find } values as low as 
0.6--0.5 or less.  

Therefore we have considered other possible interpretations of the
``partial coverage'', namely
orientation effects, intrinsic asymmetries in the radio-source
structure and/or asymmetries in the distribution of the ambient medium.

\begin{figure}[t]
      \centering
      \resizebox{\hsize}{!}
      {\includegraphics{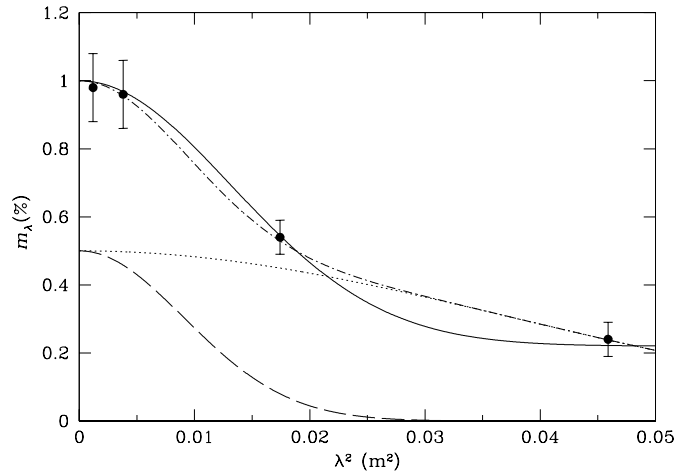}} 
      \caption{Simulation of a symmetric radio source at $45\degr$
	from the plane of the sky  with $LS = 8$~kpc and $z = 0.6$.  
	The two Gaussians with  $m_0=0.5$ (dashed and dotted
	lines) represent the depolarisation of the far and near lobes
	(respectively strongly and weakly depolarised) as a function
	of $\lambda^2$. The dash-dotted line is the sum of the two
	contributions.  The ``experimental'' data points are values
	read from the  model curve at the observed wavelengths of the
	present experiment. The continuous line is the fit to the four
	data points using Eq.~\ref{eq:cover_f} with $f_c =0.78$,
	i.e. assuming a partial source coverage.} 
      \label{fig:es_part_cov}
\end{figure}

A  promising alternative is  related to effects of source
orientation with respect to the line of sight.   
When the source axis is not perpendicular
to the line of sight  the two lobes suffer 
different depolarisations. 
In small-size sources of typical size $\leq 4$~kpc, the
two lobes are rapidly depolarised at $\lambda \ll 13$~cm, for any
orientation and redshift. 
The overall depolarisation corresponds to the
average $\sigma_{\rm RM,mod}$ of the two lobes, which is  close 
(within $\sim 10$\%) to that for the
case of orthogonality to the line of sight.
For large source-sizes ($\ge 8$~kpc),deviations {\bb  $\ge 30\degr$ }
from the plane of the sky, and moderately high
redshifts, cause the far component to be depolarised at $\lambda \le
13$~cm, and the {\bb near} component {\bb little depolarised} even 
at 21~cm. The effect 
increases with increasing {\it LS}, inclination angle to the plane of the
sky, and $z$. The overall
behaviour of depolarisation is a  drop at short wavelengths, followed 
by a much slower decrease (see e.g. dash-dotted line in
Fig.~\ref{fig:es_part_cov}) that, when allowance is made for the errors,
is indistinguishable from a flattening.
 {\bb We made}  simulations for $h=0.75$ in Eq.~\ref{eq:sigmamod2}, and
 found that, in a number of situations, the depolarisation, with 
only the four wavelengths available to us, is well fitted by
Eq.~\ref{eq:cover_f}, with $0.5 \le f_c \le 0.8$.

We analyzed {\bb also} the possible effects of intrinsic asymmetries in the
source arm length. The two lobes suffer different 
depolarisations because the shorter arm,  located closer to the
centre, is inside the inner denser 
region of the medium (see Eq.~\ref{eq:sigmamod2}). 
We {\bb made simulations, based on the arm-ratio distribution
of ~\citet{alex}, for the } 21 B3-VLA CSS sources for which a core is
detected at 15~GHz.  
We found that, under the assumption of a  spherically-symmetric
Faraday medium, no significant effects are expected.
The results might be different if the source asymmetries were caused by 
density asymmetries in the ISM ~\citep{Jeyak}, the denser ISM
preventing one of the two lobes to grow like the other. In this case, the
$\sigma_{\rm RM, sf}$ of the two lobes would differ {\bb not only} because
the shorter lobe is closer to the source centre (as before)   
but because, in addition, the ISM is  denser on the short
lobe side.
We did not model this situation because it requires too many
parameters that cannot be constrained easily. We {\bb used, instead},
an experimental approach. For 13 of 21 sources observed at 15~GHz
~\citep{alex} that belong to the WSRT sub-sample, we plotted
$f_c$ versus arm ratio. 
No significant relation has been
found, suggesting that, if radio source asymmetries are due to ISM
asymmetries, the latter are not {\bb very important} in simulating
a partial  coverage effect.

In conclusion, we suggest that  covering factors smaller than 0.8, are
possibly justified by orientation effects.

\subsection {Rotation Measures: A large-scale ordered magnetic field?}
\label{rot-depol}

\begin{figure}[t]
      \centering
      \resizebox{\hsize}{!}
      {\includegraphics{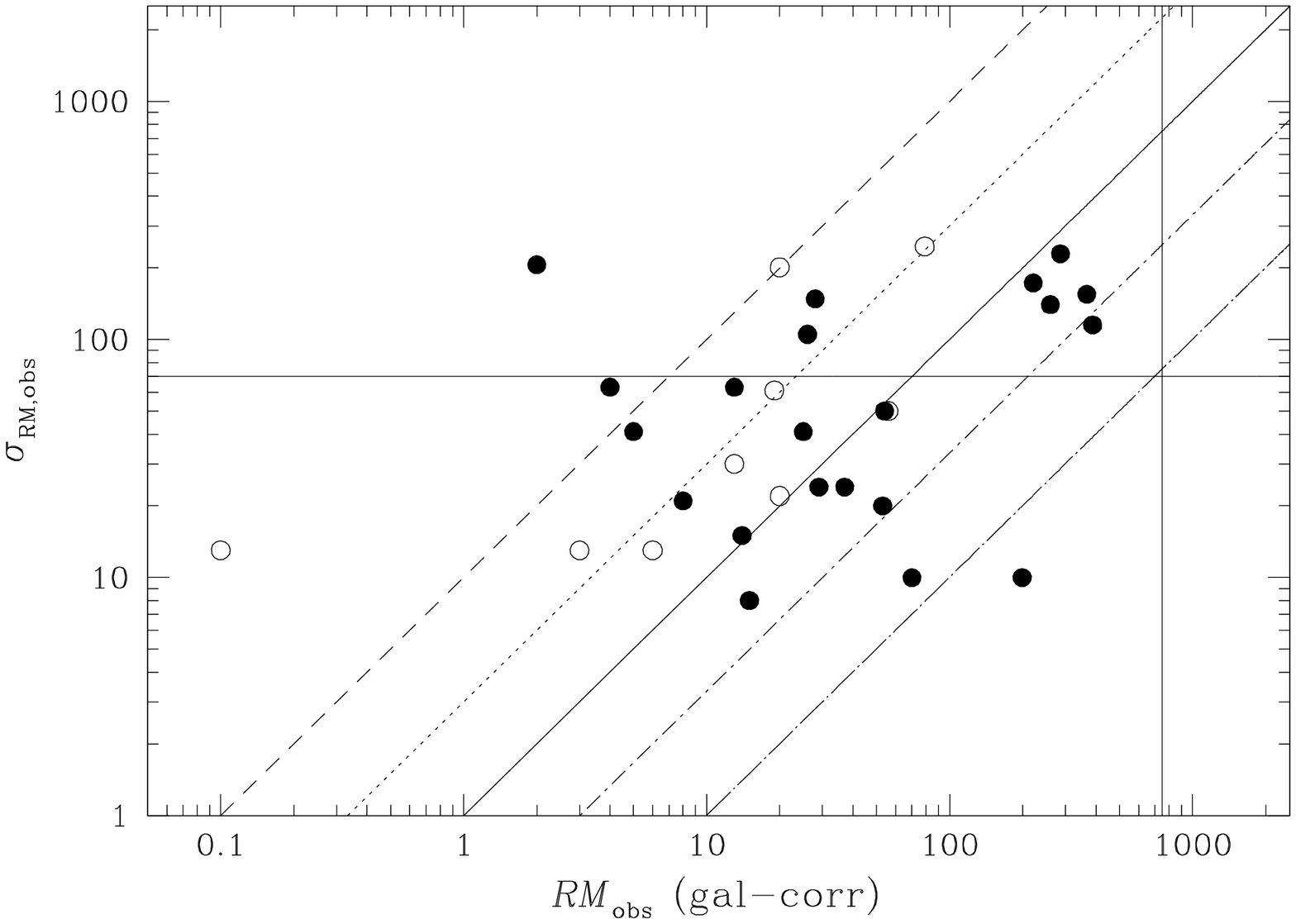}}
      \resizebox{\hsize}{!}
      {\includegraphics{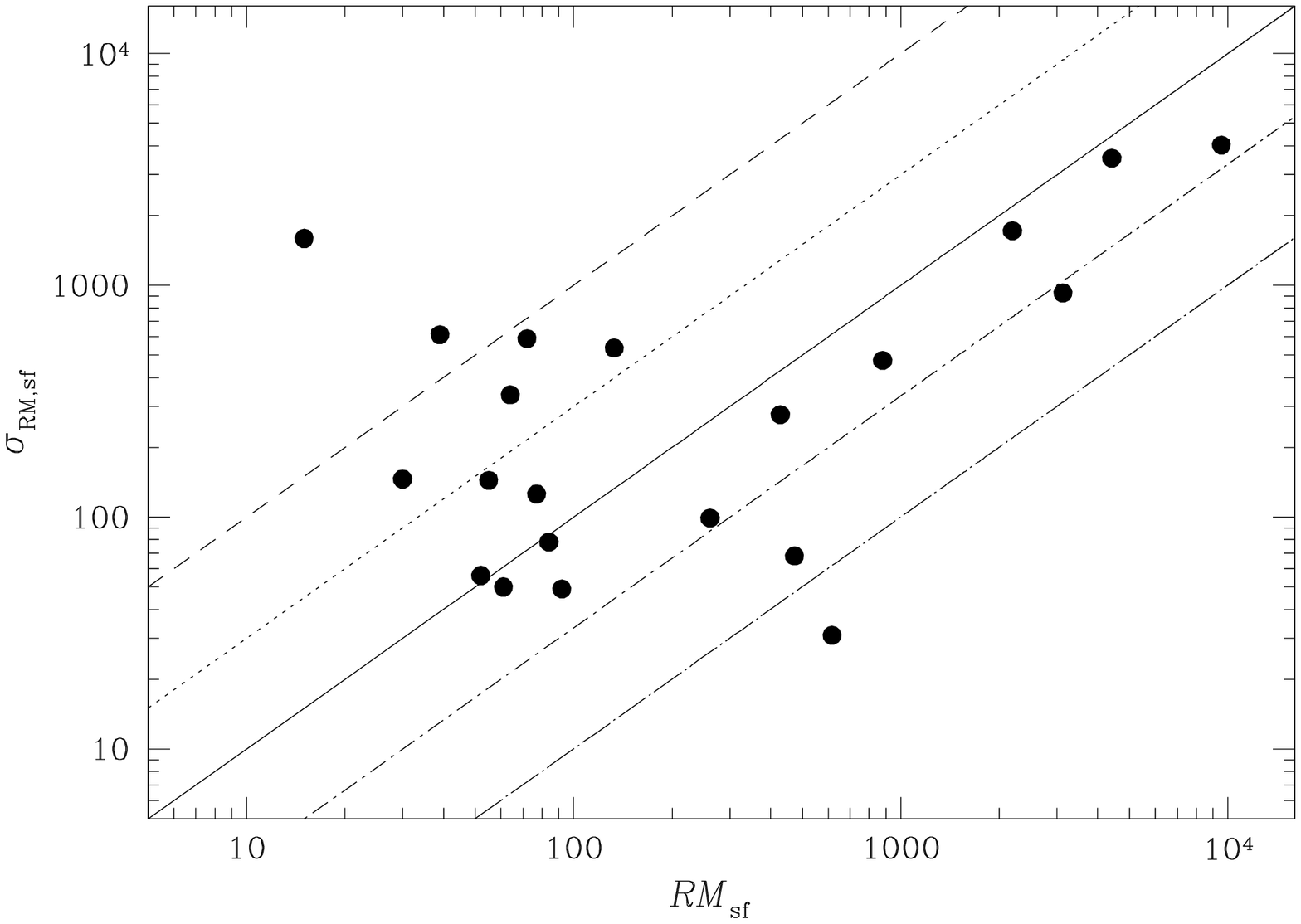}}

\caption{Plot of $\sigma_{\rm RM}$ vs  $RM$ in the observed ({\it top}
  panel) and 
  in the source  ({\it bottom} panel) frame. Open symbols in the {\it top}
 panel indicate Empty Fields. The tilted lines represent 
  ($\sigma_{\rm RM,obs}/RM_{\rm obs}$)  and ($\sigma_{\rm RM,sf}/RM_{\rm sf}$) = 10,
  3, 1, 0.3, 0.1.} 
\label{fig:RM_sigma}
\end{figure}

In the {\it top} panel of Fig.~\ref{fig:RM_sigma}, we plot
$\sigma_{\rm RM,obs}$ versus the observed, Galaxy-corrected, Rotation
Measures, $RM_{\rm obs}$ for the 30 sources (21 with $z$) for which
both data are available\footnote{We recall that 15 other  sources (10
  with $z$) with derived $\sigma_{\rm RM}$ are unpolarised at 13~cm so
  that their {\it RM} is unknown (see Sect.~\ref{rot-angl}). In
  addition, 2 sources (1 with $z$) have {\it RM} but not $\sigma_{\rm
    RM}$}. 
We discuss possible  selection effects in the plot. Sources for which
{\bb $\sigma_{\rm RM,obs} \ge 70$~rad\,m$^{-2}$  and $f_c \ge0.9$ should be
strongly depolarised at 13~cm (Eq. \ref{eq:delta}), therefore their
{\it RM}s are not} measurable because the
detectability at this wavelength is the condition {\bb we} adopted to compute
{\it RM} (Sect.~\ref{rot-angl}). The 10 sources in the
plot that have $\sigma_{\rm RM,obs} \ge 70$~rad\,m$^{-2}$, have $f_c
\le 0.9$.

Because of band-width
depolarisation, most sources with $RM_{\rm obs} \gtsim 750$~rad\, 
m$^{-2}$ would be depolarised at 13~cm  and their {\it RM} would not
be measurable. The vertical and horizontal lines show these limits. 
In the  bottom panel of Fig.~\ref{fig:RM_sigma}, we plot similar data,
with the source frame parameters.

Both figures show that about half of the sources have $RM \ge
\sigma_{\rm RM}$. 
This is unexpected for a model in which the magnetic field of
the Faraday screen is random on small scales. Our 
Monte Carlo simulation (Sect.~\ref{depol}) shows that in these models
the  $E$ 
vector is expected to  have, statistically, a small global rotation, with
$|RM| \le \sigma_{\rm RM} / \sqrt{N_c}$  (where $N_c$ is the ratio between
source area and cell  area) as long as $\sigma_{\rm RM} \lambda^2
<2$~rad ($DP\gtsim 5$\%). If $N_c \gg 100$, the  
data points in Fig.~\ref{fig:RM_sigma} would be expected to cluster
around or above the uppermost  
line, and, given the values we find for $\sigma_{\rm RM}$ (see
Table~\ref{bt}),  
no large rotations of the polarisation angle with $\lambda^2$, the
residual of  the {\it RM} dispersion, are expected. In other words,
with the possible exception of a  minority of sources ($\le15 \%$),
the {\it RM}s are not  the residuals of the random rotations that
depolarise the radiation. They are {\bb likely} generated on a larger
scale by an ordered {\bb magnetic field component}.

 \begin{figure}[t]
      \centering
      \resizebox{\hsize}{!}
     {\includegraphics{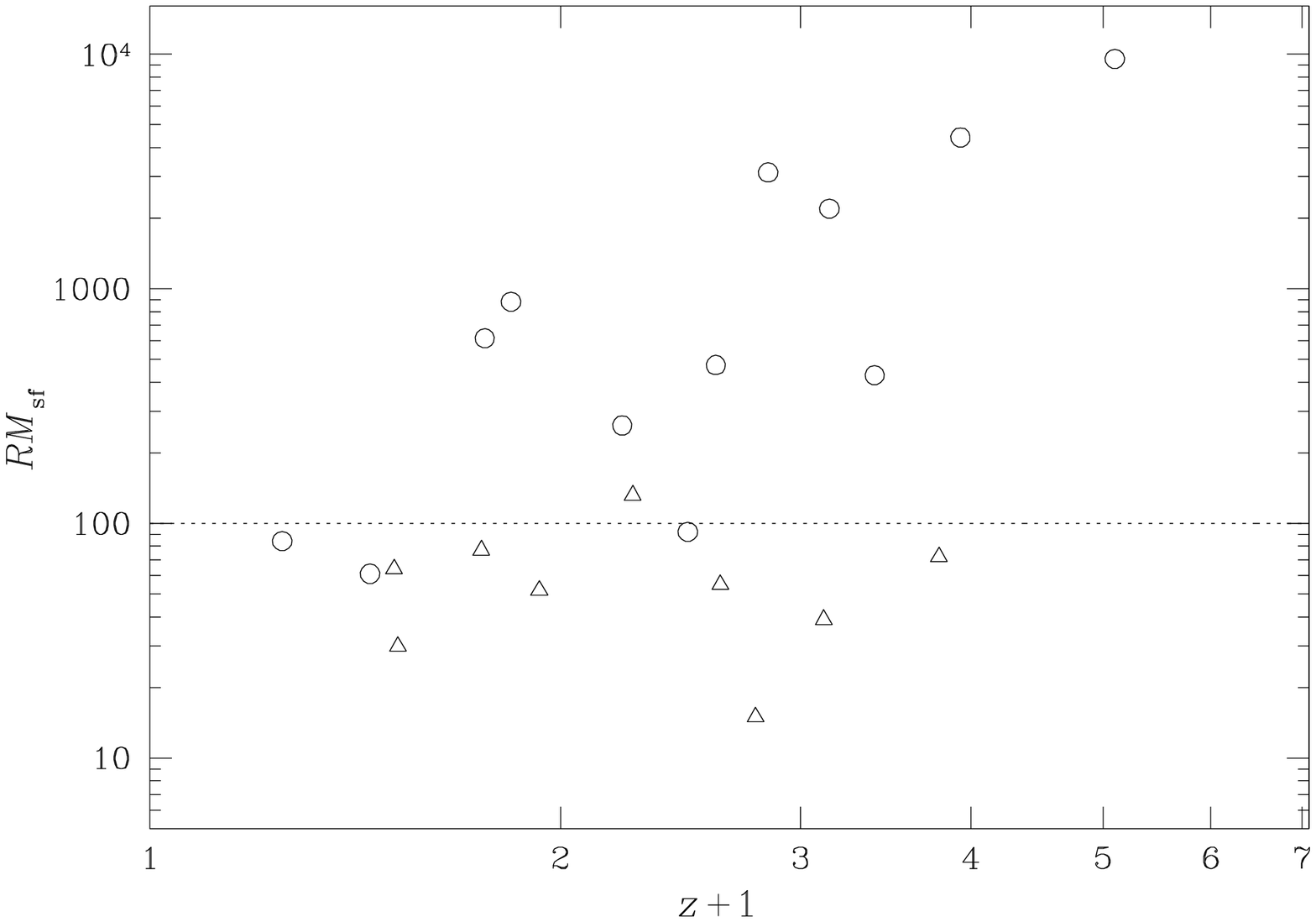}}
      \resizebox{\hsize}{!}
     {\includegraphics{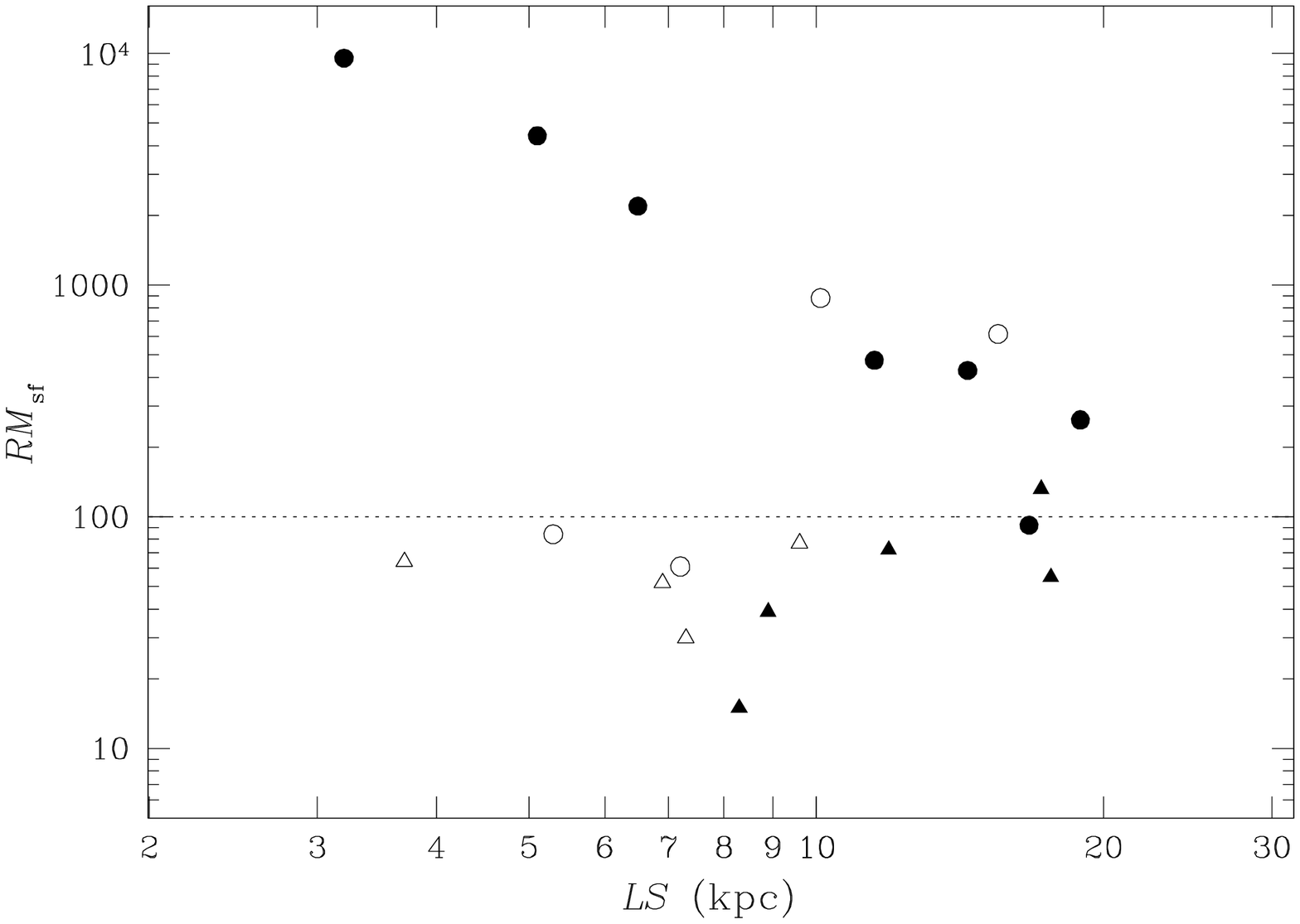}}

\caption{Plot of $RM_{\rm sf}$ vs (1+$z$)~ ({\it top} panel) and of
  $RM_{\rm sf}$ vs {\it LS} ({\it bottom} panel). 
In both panels ($\vartriangle$)
  and ({\large $\circ$}) represent  objects with 
$RM_{\rm sf} < $ and respectively $ > \sigma_{\rm sf}$; filled  
symbols in the bottom panel represent objects with $z \ge 1$.}
\label{fig:RM_z_LS}
\end{figure}

\smallskip
In~\citeauthor*{Fanti04}, we found a marginal correlation between
$RM_{\rm sf}$ and $z$.  
In the {\it top} panel of Fig.~\ref{fig:RM_z_LS}, we plot $RM_{\rm
  sf}$ as a function of 
$(1+z)$. For $RM_{\rm sf}< 100$~rad\,m$^{-2}$, the distribution in $z$
is uniform, while for $RM_{\rm sf}> 100$~rad\,m$^{-2}$, 8 sources out
of 10 have $z>1$. There is a hint of an upper bound that increases
with $(1+z)$. The probability that the observed distribution is a
random selection from a uniform redshift distribution is $\approx 7 \%$.
We {\bb further} note that sources for which $RM_{\rm sf} \ge \sigma_{\rm RM,
  sf}$ (circles in Fig.~\ref{fig:RM_z_LS}, {\it top} panel) are
distributed in a broad band such that $RM_{\rm sf}\propto
(1+z)^{3}$. The probability that this is caused by a random selection
from an uniform distribution is $\le 1.5 \%$. We have to examine if
there are any biases which can affect the source 
distribution in the plot {\it RM} vs $(1+z)$. There are 10 sources
for which we do not have a measured {\it RM} because they are
depolarised at 13~cm. If these sources had an {\it RM} above the 
apparent upper bound in Fig.~\ref{fig:RM_z_LS}, ({\it top} panel),  their
{\it RM} would be  high ($\ge 300$~rad\,m$^{-2}$) in the 
observer's frame; we would expect therefore to see a  rotation of the
polarisation angle between 3.6 and 6~cm. 
At first sight there appear to be no candidates for these ``missing
{\it RM}'' among the unpolarised sources  at 13~cm. Although   this
statement has to be taken with some caution, we conclude 
that there is a real correlation between  $RM_{\rm sf}$  and $(1+z)$,
at least for sources for which $RM_{sf} \ge \sigma_{\rm RM,sf}$.  

\smallskip
In the {\it bottom} panel of Fig.~\ref{fig:RM_z_LS}, we plot $RM_{\rm
  sf}$ versus {\it LS}. 
The distribution of data points is peculiar. There seems to be a
  ``sequence'' of sources, for $RM_{\rm sf} > 100$~rad\, m$^{-2}$, for
  which $RM_{\rm sf}\propto LS^{-2.2}$, which is a dependence on {\it
  LS} that is similar to that of  $\sigma_{\rm RM,sf}$.
This ``sequence'' (9 sources out of 21) is composed mainly
of objects for which $RM_{\rm sf} > \sigma_{\rm RM, sf}$ (8 out of 9) and 
$z > 1$ (7 out of 9). Of the remaining objects, $RM_{\rm sf}$ does not
appear to show correlation with {\it LS}. 

We  conclude  that for 50\% of the CSS (mainly at high
redshifts, $\langle z\rangle \approx 1.5$), 
a large-scale ordered magnetic field is present and produces the
higher values of {\it RM}s. Its scale must be approximately 10--20~kpc
to justify the correlation observed in the {\it bottom} panel of
Fig.~\ref{fig:RM_z_LS}.

Applying a  model similar to that for $\sigma_{\rm RM,sf}$
(Sect.~\ref{natureFC}) to
the data of the ``sequence'' $RM_{\rm sf}$ versus {\it LS}, we find that 
the required ambient density and magnetic-field dependence on $r$, are
similar to that for $\sigma_{\rm RM,sf}$ ($\beta+\mu \approx 1$), and
$B_{||,0} \approx 40\,  r_c^{-3}\, n_{0}^{-1} \approx 10~ \mu$G,
at the median $z \approx 1.9$ of the sources in the ``sequence''.
We observe that {\bb not all} high-redshift objects lie on the
$RM_{\rm sf} \propto LS^{-2.2}$ relation. A possible explanation is
that in a large-scale magnetic-field, the orientation effects are
important for {\it RM}. High-redshift sources 
that do not belong to the sequence, should have a line of sight that
is at significant angles to the magnetic field. 

If  $B_{||,0}\, n_0 \propto (1+z)^4$, as for  $\sigma_{\rm RM,sf}$,
at $z \le 1$ the {\it RM} would drop to $\le 150$~rad\,m$^{-2}$.
Most of the low-redshift sources would fail to show the correlation of
the high-redshift ones.

\section{Summary and Conclusions}
\label {concl}

We have observed 65 radio sources from the
B3-VLA CSS sample (\citeauthor*{Fanti01}) using the WSRT at 13~cm,
to study the source polarisation properties.
At the WSRT resolution, the radio sources are all unresolved, and we
can therefore only discuss their global properties.
The new 13-cm data, combined with
earlier low-resolution VLA data at 3.6, 6, and 21~cm, have 
improved the determination of Rotation Measures ($RM$) and the
Rotation Measure Dispersions ($\sigma_{\rm RM}$).
This has allowed their properties to be defined more carefully as
a function of both redshift and Linear Size, and the characteristics
of the surrounding ISM to be modelled. 

The main results of the paper are the following:

\begin{description}
\item[{\mdseries 1)}] Radio sources $\le 5$~kpc are mostly unpolarised 
at 13~cm. This result is similar to results found at 20~cm
\citep{Cotton03} and  at 3.6 and 6~cm (\citeauthor*{Fanti04}), and the
``critical size'' that we find is intermediate between those found earlier.

\item[{\mdseries 2)}] The 13~cm polarisation angles have led to a
  revision of the earlier $RM$s (\citeauthor*{Fanti04}) for $\approx 30\%$ 
of the sources. 

\item[{\mdseries 3)}] The integrated fractional polarisation
  $m_{\lambda}$, as a function of wavelength shows in general a
  decrease between 3.6 and 13~cm, followed by a flattening between 
13 and 20~cm. This is the major observational result of the present
  paper.
  At variance with the conclusions of
  ~\citeauthor*{Fanti04}, the ~\citet{Burn66} and ~\citet{Tribble91}
  models are 
  not adequate to describe this behaviour. However, an empirical
  variant of the ~\citet{Burn66} model, which introduces  a 
``partial coverage'' by the depolarising curtain, appears to reproduce
  the data well. 
The adopted formula allows the Rotation Measure Dispersion,
  $\sigma_{\rm RM}$, and the fraction, $f_{\rm c}$, of source covered
  by the depolarising curtain, to be determined.

\item[{\mdseries 4)}] For a {\bb minority of sources}, $m_{\lambda}$ shows an
irregular, possibly oscillatory, behaviour with $\lambda$. We propose
that these sources contain sub-components with different Rotation Measures
that produce beats {\bb with} $\lambda$ in the integrated fractional 
polarisation.

\item[{\mdseries 5)}]  $\sigma_{\rm RM}$ shows a clear dependence on
  redshift (up to $z \approx 1.8$),  and on projected Linear Size. 

\item[{\mdseries 6)}] We have analysed these dependences using a model 
  similar to that presented in ~\citeauthor*{Fanti04}, in which a
  depolarising Faraday curtain, with a King-like distribution of the 
  magneto-ionic medium, is produced either by a smooth medium
  with an irregular magnetic field on small scales, or by  a
  clumpy magnetised medium (NL region). The parameters
  derived in either case  differ from those we 
  obtained in ~\citeauthor*{Fanti04}, because of the use of the new
  13~cm data and of the new model adopted {\bb for} the
  depolarisation behaviour with  $\lambda$.
  In both models the core radius of the curtain has {\bb to be
 $\le 0.5$~kpc.}

  In the smooth medium model, the required central density, at $z = 0$,
  is in the range of that found using X-ray observations in the
  centre of early-type galaxies. Beyond the core radius, it decreases
  according to $r^{-\eta}$, where $1.5 \le \eta \le 2.0$. The central
  magnetic field, at $z = 0$, is $\approx 150$~$\mu$G, and the magnetic
  field energy  density is close to the thermal energy. 
 
In  the clumpy medium model, {\bb identified with} the NLR, we
assumed pressure equilibrum between the clouds and the diffuse medium in
which they are embedded. The required 
magnetic field, inside the clouds at $z =0$, is $\approx 6~\mu$G, and the 
magnetic energy is a negligible ($\approx 10^{-4}$) fraction of the
thermal energy. Beyond the core radius, the density of the
smooth medium, which confines the clouds, should decline as
$r^{-\eta}$, with  $1.2 \le \eta \le 1.6$.
The product $B_{\rm NL,0} \times n_{\rm NL,0}$  is a 
strong function ($\propto (1+z)^4$) of redshift up to $z \approx 1.8$.

The $NL~ model$ requires more parameters than the {\it smooth medium
  model}, and these parameters need tuning to avoid the
  ``NL covering  problem'' discussed in Sect.~\ref{farprop}.
For these reasons, we favour the {\it smooth medium model}.

\item[{\mdseries 7)}] The ISM density profiles required by the
  depolarisation models  are consistent with the expectations of 
the evolutionary models of young radio sources.

\item[{\mdseries 8)}]  The partial coverage, that we introduced to
describe the long wavelength behaviour of $m_{\lambda}$ vs $\lambda^2$,
is probably due to {\bb orientations effects} of the source with
respect to the line of sight. For sources that are almost orthogonal to the
line of sight, the two lobes are depolarised similarly  by a
spherically-symmetric medium, and $m_{\lambda}$ is fitted by the Burn
model. However, when the source
is at large angles with respect to the line of sight, the far lobe is more
depolarised causing the rapid decrease of $m_{\lambda}$ at
short wavelengths, while the {\bb nearby one} is less depolarised giving rise
to a  lower decrement of $m_{\lambda}$ with wavelength. The total
fractional polarisation $m_{\lambda}$ then appears to be almost
constant  at long wavelengths. 

\item[{\mdseries 9)}]  The total source-frame Rotation Measures 
are generally too large to be the residual of random rotations across
the source due to a totally irregular magnetic field, and require a
large-scale ordered magnetic-field component.
The Rotation Measures show hints of  a correlation with both
redshift and 
Linear Size in a way similar to the Rotation Measure dispersions.
If these correlations are real, the  ordered
magnetic-field component would have  parameters similar to those of the
random field component.
\end{description}

\begin{acknowledgements}
We thank the referee, Prof. U. Klein, for carefully reading the
paper and for several comments which improved its presentation.
 
The WSRT is operated by ASTRON (The Netherlands Foundation for
Research in Astronomy) with support from the Netherlands Foundation
for Scientific research (NWO).

\end{acknowledgements}



\appendix
\section{Polarisation Data}
\label{app_1}
\begin{table*}[p]
\begin{center}
\caption{Polarisation Data }
\scriptsize
\medskip
\begin{tabular}{lccccrrrrrrrc}
\hline
Name&$S_{13}$
&$m_{13}$&$\chi_{13}$&$RM_{\rm obs}$&$RM_{\rm sf}$&$\sigma_{\rm RM,obs}$&$
\sigma_{\rm RM,sf}$&$m_0$&$f_c$&z&LS& Notes \\
\hline
    (1) & (2)
& (3)& (4) & (5) & (6) & (7) & (8) & (9) & (10) & (11) & (12) & (13) \\
\hline
0034+444&0.42&1.40$\pm$0.11&--33& --85&--72&41&589&3.1&0.9&2.79&11.9& {\it P1, N} \\
0039+373&0.60&0.10$\pm$0.10&    &             &    &243&982&1.0&0.85&1.01&0.5& {\it P5} \\
0039+398&0.48&0.34$\pm$0.11  &42&           &      &  &     &   &&    &16 &{\it P6}\\
0039+412&0.25&2.81$\pm$0.12 &--2&--100 & &22& &5.2&0.64&    &
8.5& {\it P1, N, m}\\
0041+425&0.30&0.28$\pm$0.13 &(--24) &           &      &$>$368& &3.0&0.93&   &4.1 & {\it P4} \\
0049+379&0.46&0.27$\pm$0.11 & (17)  &           &      &254&1852&9.0&0.98&1.7&6.1 & {\it P3} \\
0110+401&0.42&3.54$\pm$0.11& 54&--65 & --92  &$\sim 8$
&$\sim49$&5.4&[1]&1.48&16.7&{\it P1, N, m} \\
0120+405&0.38&1.41$\pm$0.11&--102&--340 &--880  &140&474&5.5&0.75&0.84&10.1&{\it P1, N}  \\
0123+402&0.16&1.32$\pm$0.15&--48&--136 a& &$\le$50& &4.5&$\ge$0.95& &5.2 &{\it P2}  \\
0128+394&0.20&0.55$\pm$0.14&--87&--150&--473&$\sim
10$&$\sim68$&0.8&[1]&1.6&11.5 &{\it P5}, {\it N, m} \\
0137+401&0.19&1.53$\pm$0.14&--30&--88
a&--55&$\le$21&144&1.7&$\ge0.4$&1.62&17.6& {\it P1, N}   \\
0144+432&0.23&1.05$\pm$0.12&--40&--54&
$+$132&105&536&3.5&0.66&1.26&17.2&{\it P1, N}  \\
0147+400&0.51&0.01$\pm$0.10&  &           &      &  &     &   &   &&0.4 &{\it P6}\\
0213+412&0.39&0.63$\pm$0.11&  10&--67 a &  $+$30 &63&146&3.1&0.89&0.52&7.3 & {\it P1}        \\
0222+422&0.16&0.39$\pm$0.14& (60) &               &&195&3949&4.0&0.9&3.5&11.9& {\it P3} \\
0228+409A&0.24&1.60$\pm$0.12&  38&--159& &246& &4.0&0.65&&16.2&{\it P1}  \\
0254+406&0.33&1.53$\pm$0.12&   3&--27 a &$+$262&$\le$20&$\le$99&1.7&$\ge$0.25&1.22&18.9& {\it P1}    \\
0255+460&0.45&0.10$\pm$0.11&    &           &      &  &  & &  &1.21&3.2& {\it NP}\\
0701+392&0.35&0.15$\pm$0.12&    &           &      &102&512 &1.0 &0.8   &1.24&7.8& {\it P3}\\
0722+393A&0.64&0.10$\pm$0.10&    &           &      &  &     &   &   &    &1.0 & {\it NP}\\
0744+464&0.32&0.50$\pm$0.11&--3&--277
a&--4417&229&3537&5.5&0.9&2.93&5.1& {\it P1, N}   \\
0748+413B&0.13&1.40$\pm$0.13&  13&$+$29 &  &201& &18&0.94&&1.8&{\it P1}    \\
0754+396&0.33&8.32$\pm$0.13&--97&$+$5 a& --39 & 63&613&11.6&0.32&2.12& 8.9& {\it P1}    \\
0800+472&0.60&0.33$\pm$0.10&--54&--19 &--64&148&337&2.4&0.83&0.51&3.7&  {\it P1}   \\
0805+406&0.37&5.75$\pm$0.12&   2&$+$22 a&  & 30& &6.5&0.3&&10.7& {\it P1}    \\
0809+404&0.75&0.15$\pm$0.10&    &           &         &216&519&4.5&0.97& 0.55&4.5 & {\it P3}\\
0810+460B&0.66&0.11$\pm$0.10&    &           &&$>$270&$>$478&2.0&0.94&0.33&3.0 & {\it P4}\\
0814+441&0.17&1.93$\pm$0.12 &--64&$+$352&   & &  &   &   &&17.0&{\it
  P5,  N} \\
0822+394&0.77&0.02$\pm$0.10 &    &           &      &  &     &   &&1.18&0.2 & {\it NP}\\
0840+424A&1.03&0.06$\pm$0.10&    &           &      &  &     &   &   & &0.3 &  {\it P6}\\
0902+416&0.35&1.90$\pm$0.11 &--65&         &      &  &     &   &
&&1.4 &{\it P5, N}\\
0930+389 &0.17&2.79$\pm$0.12&--45&$+$46&$+$428&$\le$24&$\le$277&2.8&$\ge0.8$ &2.4 &14.4& {\it P1, N}   \\
0951+422 &0.29&0.44$\pm$0.11&--3&$+7$   & --15&206&1592&4.5&0.88&1.78&8.3 & {\it P1}            \\
0955+390 &0.31&6.27$\pm$0.13&  15&$+$15 a& &$\le$13& &6.6&$\ge0.6$&  &19.2& {\it P1}  \\
1007+422 &0.29&0.13$\pm$0.13&    &           & &400& &6.5&$\ge$0.95& &0.55 & {\it P5} \\
1025+390B&0.49&2.98$\pm$0.11&--101&--20  & --54 & &   & & &0.36&
9.7&{\it P2,  N, m}     \\
1027+392 &0.28&0.26$\pm$0.11& (91)  &              &      & && &    &   &6.9& {\it P5,  N}\\
1039+424 &0.17&6.90$\pm$0.15&--57&$+$12 a& &$\le$13&&7.3&$\ge0.8$&
& 6.5& {\it P1}     \\
1044+404A&0.25&0.35$\pm$0.11&14&$+$376 &(9546)&155&($+$4032)&8.5&0.95&4.1&3.2& {\it P2, N}\\
1055+404A&0.25&5.69$\pm$0.12&   0&$+$8.9 a& &$\le$13& &5.9&$\ge0.6$&    &12.1& {\it P1}  \\
1128+455 &1.35&0.00$\pm$0.10&    &                &    &    &     & &   &0.40&2.9 &{\it NP} \\
1136+420 &0.32&0.20$\pm$0.11&    &             &    &$>$324&$>$1085&2.4&0.9&0.83&4.3 & {\it P4}\\
1201+394 &0.32&2.29$\pm$0.11&  26& $+$38 &$+$61&$\le$24&$\le$51&3.0&$\ge0.90$ &0.45& 7.2& {\it P1, N, m}  \\
1204+401 &0.15&0.31$\pm$0.15& (--1)   &                &  & 68&641   &5.2&$\ge$0.95&2.07&6.7 & {\it P3}\\
1216+402 &0.23&2.03$\pm$0.12&  64& --190
&--616&$\le$10&$\le30$&4.0&[1]&0.76&15.5& {\it P1, N, m}    \\
1220+408B&0.28&1.47$\pm$0.11&--100& $+$28 a&  &61&&6.9&0.95& &17.6&
{\it P1}  \\
1225+442 &0.24&0.09$\pm$0.12&    &           &      &  &     &   & &0.22&0.9&{\it P6}  \\
1233+418 &0.49&1.15$\pm$0.11&--52& --45 &-84&50&78&4.3&$\ge$0.9
&0.25& 5.3 & {\it P2, N, m}   \\
1242+410 &1.04&0.04$\pm$0.10&    &           &      &  &     &   &&0.81&0.3&{\it NP} \\
1340+439 &0.32&0.23$\pm$0.11& (--21) &           &      &  &     &   &
& &0.3 &  {\it P6, N} \\
1343+386 &0.66&0.38$\pm$0.11&--15& --379 &--3129 &
115&928&3.8&0.86&1.84&0.5& {\it P1, N, m} \\
1350+432 &0.09&3.18$\pm$0.16&--86& $+$230   &$+$2193
&150&1490&10.0&0.55&2.15& 6.5& {\it P1, N, m}   \\
1432+428B&0.66&0.15$\pm$0.10&    &           &      &  &     &   &   &  &0.2&{\it P6} \\
1441+409 &0.67&0.10$\pm$0.10&    &           &      &$>$280& &1.4&$\ge$0.9   &    &0.4 & {\it P4}\\
1449+421 &0.43&0.12$\pm$0.10&    &           &      &  &     &   &   &    &0.3 & {\it NP}\\
1458+433  &0.29&2.70$\pm$0.11&  12& --5 a &
--52&$\le$15&$\le$55&2.7&$\ge0.2$ &0.93& 6.9&{\it P1, N}  \\
2301+443 &0.69&0.03$\pm$0.10&    &           &      &  &     &   & &1.7&2.1 & {\it P6}\\
2302+402 &0.83&0.10$\pm$0.10&    &           &      &  &     &   &   &
&2.6 & {\it P6, N}\\
2304+377 &1.07&0.19$\pm$0.10&    &           &      &182&357 &3.2&0.96 &0.4&0.3 & {\it P5}\\
2311+469  &1.38&4.22$\pm$0.11&--70& --8  & $+$77 &  41& 126&5.5&0.36&0.75 & 9.6& {\it P1, N}\\
2322+403 &0.22&0.03$\pm$0.12&    &           &      &207&&6.5&0.98& &14.0 & {\it P3}\\
2330+402 &0.60&0.06$\pm$0.10&    &           &      &  &     &   &   & &0.3& {\it P6}\\
2348+450 &0.50&0.04$\pm$0.11&    &           & &$>$290&1137&2.5&0.95&0.98&1.3 & {\it P4}\\
2349+410 &0.28&0.24$\pm$0.11& (76)   &           &      &   &    &
&&2.05&4.7 & {\it P5, N}\\
2358+406 &0.99&0.02$\pm$0.11&    &           &      &  &     &   &   &    &0.3 & {\it NP}\\
\hline
\end{tabular}
\label{bt}
\end{center}
\small\baselineskip 0.5 cm

- Columns 1, 2 and 3: Source Name, 13 cm flux density (Jy), fractional polarisation and error (\%);

- Column 4: 13 cm Electric vector p.a. (degree) for sources with 
            $m_{13} \ge 3 \sigma_{m}$; values in parenthesis 
	    are for sources with $2 \sigma_{m} \le m_{13} \le 3 \sigma_{m}$.
            Errors, computed as in Sect.~\ref{data-red}, are $\le 9\degr$ 
	    for $m_{13} \ge 3\sigma_{m}$
	    
- Columns 5 and 6: Observed and source frame Rotation Measure (rad\,m$^{-2}$).
           Formal errors of $RM_{\rm obs}$ are $\le
           10$~rad\,m$^{-2}$. An ``a'' near $RM_{\rm obs}$ means a good
           $\lambda ^2$-linear fit (see Sect.~\ref{rot-angl}). 
	   
- Columns 7 and 8:  Observed and source frame $\sigma_{\rm RM}$
                   (rad\,m$^{-2}$); formal errors are typically $\le 20$\%
		   
- Columns 9 and 10: Intrinsic Fractional polarisation (\%) and Covering factor 

- Columns 11, 12 and 13: Redshift (photometric one  decimal digit
  only), Largest (projected) Linear Size (kpc), Notes

{\it P1}: the source is detected at $\ge 3 \sigma_{\rm P}$ at all four wavelengths;
{\it P2}: the source is detected at $\ge 3 \sigma_{\rm P}$ at 3.6, 6, and 13~cm;
{\it P3}: the source is detected at $\ge 3 \sigma_{\rm P}$ at 3.6 and  6~ cm;
{\it P4}: the source is detected at $\ge 3 \sigma_{\rm P}$ at 3.6~ cm only;
{\it P5}: the source is detected at $\ge 3 \sigma_{\rm P}$ at two or three non
          contiguous wavelengths;
{\it P6}: the source is detected at $\ge 3 \sigma_{\rm P}$ at 21~ cm only;
{\it NP}: the source is undetected ($\le 3 \sigma_{\rm P}$) at all frequencies.
{\it N}: see notes in Sect.~\ref{notes}.
{\it m}: data from the ``two polarised component model'' (see Notes and 
Appendix \ref{pol_mod}).

\end{table*}
\normalsize

\subsection {Notes to individual sources}
\label{notes}

\noindent
{\bf 0034+444:} See Fig.~\ref{fig:es_0034} and Sect.~\ref{rot-angl}. 
The $\chi_{3.6}$ is discrepant from the $\chi(\lambda^2)$ behaviour at
the other frequencies, causing a poor overall fit. 
 We have not been able to improve the fit 
using the ``two polarised component model'' of ~\ref{pol_mod}.

\noindent
{\bf 0039+412:} The $RM_{\rm obs}$ obtained by a  $\lambda^2$-linear fit, using also the 
individual IFs of the 13~cm band, does not properly account  
for $\chi_{21}$. 
We have used the ``two polarised component model'', constrained by the 
polarisation structure seen by the VLA at 3.6 and 6~ cm, which introduces
a small modulation in the polarisation angles over the $\lambda^2$ behaviour
which better fits  all the data. The $RM_{\rm obs}$ in Table~\ref{bt} is from this model.

\noindent
{\bf 0110+401:} The source shows a deep minimum in fractional polarisation 
at $\approx  11$~cm (data from the Effelsberg telescope, see
\citet{Klein03}). A  $\lambda^2$-linear fit, using also the  
individual IFs of the 13~cm band, does not account for $\chi_{21}$.
We have used the ``two polarised component model'' which  fits very well
both behaviours of $m_{\lambda}$ and $\chi(\lambda)$ vs $\lambda^2$
(see Fig.~\ref{fig:es_0110}). The $RM_{\rm obs}$ and $\sigma_{\rm
  RM,obs}$ in Table \ref{bt} are  from this model.

\noindent
{\bf 0120+405:} An alternative value of $RM_{\rm obs}$ is $\approx - 154$~rad\,
m$^{-2}$. The adopted $RM_{\rm obs}$ justifies  the lack of polarisation at 21~
cm as a bandwidth depolarisation effect. Both the reported $RM_{\rm
  obs}$ and the  alternative one are in bad agreement with $RM_{13}$. 
 
\noindent
{\bf 0128+394:} This source shows a possible oscillatory behaviour
 of $m$ vs $\lambda^2$.
We have used  the ``two polarised component model'', constrained by 
the polarisation structure seen by the VLA at 3.6 and 6~cm, which fits
both $m$ and $\chi$ vs $\lambda^2$.
The $RM_{\rm obs}$ and $\sigma_{\rm RM,obs}$ in Table~\ref{bt} are from this model.

\noindent
{\bf 0137+401:} The angle $\chi_{3.6}$ is in poor agreement with a 
$\lambda^2$-linear fit.

\noindent
{\bf 0144+432:} The fitted $RM_{\rm obs}$ has a poor chi-square, mainly due to 
$\chi_{21}$. No acceptable two component model has been obtained to 
improve the fit.

\noindent
{\bf 0744+464:} According to our $RM_{\rm obs}$ the fractional polarisation
at 21~cm is depressed by a large factor because of bandwidth depolarisation.

\noindent
{\bf 0814+441:} The source is undetected at 6~cm and well detected at the
other three wavelengths, suggesting a possible oscillatory behaviour of $m$ 
vs $\lambda^2$. The two polarised component model quite fits
the data except $\chi_{21}$ well.  The fitted $RM_{\rm obs}$,
which is very large,  
would imply virtually no polarisation at 21~ cm because of bandwidth 
depolarisation, contrary to what is observed. However, the model shows
strong modulations of $RM_{\rm obs}$, and 
at 21~cm the ``local $RM_{\rm obs}$'' is lower than the average, implying only
a reduction of $m_{21}$ of a factor $\approx 2$, in agreement
with the data.

\noindent
{\bf 0902+416:} This is another source with a possible $m_{\lambda}$
oscillatory behaviour, being well detected at 3.6 and 13~ cm but not at the
other two wavelengths. We have modelled it with the two polarised
component model; however, because of the small amount of data and lack of
information on polarisation sub-structure, we consider the results too 
uncertain. No polarisation parameters are reported in Table~\ref{bt}.

\noindent
{\bf 0930+389:} The $RM_{\rm obs}$ quoted in Table ~\ref{bt} is strongly
constrained by $RM_{13}$. The angle $\chi_{3.6}$ is in bad agreement
with the overall fit and causes a poor chi-square. 

\noindent
{\bf 1025+390B:} No data is available at 21~cm. The fractional polarisation
$m_{3.6}$ is a factor $\approx 2$ lower than at 6 and 13~cm.
It may be another case of oscillatory behaviour of $m_{\lambda}$.  
We have modelled the source with a two polarised component model, but,
due to the limited amount of data, only  $RM_{\rm obs}$ is given in
Table~\ref{bt}. 

\noindent
{\bf 1027+393:} The source is strongly polarised at 6 and 21~cm
but not at the other wavelengths. Therefore it is another case
of oscillatory  
behaviour of $m_{\lambda}$. Although the two polarised component
model fits  both $m_{\lambda}$ and $\chi(\lambda)$, because of
the limited amount of data the results are not properly constrained.

\noindent
{\bf 1044+404A:} The very large $RM_{\rm obs}$ is likely the cause of
the lack of polarisation at 21~cm because of bandwidth depolarisation.

\noindent
{\bf 1201+394:} An improved fit of $\chi(\lambda)$ vs $\lambda^2$ is 
obtained by the  ``two polarised component model'' which introduces a
small modulation of the $\chi(\lambda)$ behaviour. Also the fit of the
amplitudes is improved. The data in Table~\ref{bt} are from this model.

\noindent
{\bf 1216+402:} A possible oscillatory behaviour of $m_{\lambda}$ 
is well described by the ``two polarised component model'' which well fits 
$m_{\lambda}$ and $\chi(\lambda)$, except for $\chi_{21}$. $RM_{13}$ is
somewhat lower than the adopted $RM_{\rm obs}$,
but the model justifies it 
as a result of modulation of $\chi(\lambda)$ because of the
beat between the two components. Data in Table~\ref{bt}
are from this model.

\noindent
{\bf 1233+418:}  The ``two polarised component model'', based also on
the polarisation substructure seen by the VLA at 3.6 and 21~cm,
gives an improved fit to both  $m_{\lambda}$ and $\chi(\lambda)$.
The parameters  in Table~\ref{bt} are from this model.

\noindent
{\bf 1340+439:} This source is well detected at 21~cm and possibly
detected at $\ge 2 \sigma_{\rm P}$ at 3.6~cm. It may be another example of
the oscillatory behaviour of $m_{\lambda}$, which is easily
fitted by the ``two polarised component model''. However, because of
to the poorness of the data we do not report any polarisation parameter. 

\noindent
{\bf 1343+386:} The angle $\chi_{3.6}$ is very discrepant with respect to
the overall fit. \citet{orie} showed that the polarisation is
from the bright southern hot spot, and has substructures, with $RM_{\rm
  obs}$ 
differences $\approx 100$~rad\, m$^{-2}$. We have applied the ``two
polarised component  model'' and obtained a much better fit for
$m_{\lambda}$ and $\chi(\lambda)$, except for $\chi_{3.6}$ which
 we consider an outlier. The data in Table~\ref{bt} are
from this model. 

\noindent
{\bf 1350+432:} At a first sight there is a large discrepancy between
$RM_{13}$ and the global $RM_{\rm obs}$. However, the ``two polarised component
model'' properly fits the data, and justifies the low value of $RM_{13}$
as a result of modulation because of the beating of the two components. 
Data in Table~\ref{bt} are from this model.

\noindent
{\bf 1458+433:} $m_{3.6} = 1.8 \%$ is low compared to
$m_{6}  = 2.6$\%, $m_{13} = 2.7$\% and $m_{21}= 2.4$\%.  
We have no obvious explanation for this behaviour
and have fitted $\sigma_{\rm RM,obs}$ with the exclusion of $m_{3.6}$.

\noindent
{\bf 2302+402:} This source is detected at 21~cm only with 
a S/N = 17. The fractional polarisation is much larger than the upper
limits at the other wavelengths.

\noindent
{\bf 2311+469:} the angle $\chi_{3.6}$ is quite discrepant from the global
$\chi(\lambda)$ fit.

\noindent
{\bf 2349+410:} Undetected in polarisation at 6 and 13~cm. Another
possible case of oscillatory behaviour of $m_{\lambda}$, which
can be fitted  
by the ``two polarised component model''. However, because of the
poorness of the data we do not report any polarisation parameter.

\section{The two polarised component model}
\label{pol_mod}

The polarisation  status of a source can be represented by a vector of
amplitude $S_P$ and position angle $2 \chi$, defined in Sect.~\ref{data_red}.
When two sub-structures are present, different  {\it RM}s in the two components
may cause a differential rotation of the two polarisation vectors as a 
function of $\lambda^2$.
The result is that the total source polarisation  and  $E$ vector
position angle, under vectorial addition
of the  two component polarisation vectors, may present signs of
interference. This produces, for instance, minima in
$S_P(\lambda^2)$ when the two polarisation vectors happen to be opposite.

We have applied this {\it two polarised component model} to a number of
sources having one of the following characteristics:
\begin{description}
\item[{\mdseries i)}] discrepancies between the $m_{\lambda}$ and the
  fitted model (Eq.~\ref{eq:cover_f}), mostly possible oscillations
  of $m_{\lambda}$ vs $\lambda^2$, which could be indications of 
  a beat between  polarised subcomponents with different {\it RM};
\item[{\mdseries ii)}] poor fit of the polarisation angles with the
  $\lambda^2$-linear law, again a possible indication of modulation
  which are due to polarised subcomponents with different {\it
  RM}. These sources should  also exhibit beats, but there is only a
  marginal evidence for them. However, the oscillations might have
  been lost because of the limited $\lambda^2$ sampling or might have
  strongly been damped by depolarisation. 
\end{description}
The parameters involved in the model are: the intrinsic fractional 
polarisation $m_0$, intrinsic position angle $\chi_0$ and {\it RM} of each 
component, and a
common $\sigma_{\rm RM}$ and $f_c$ which accounts for the depolarisation.
The available data  are  the $m_{\lambda}$ and $\chi(\lambda)$,
(typically  $2 \times 4$ data), and $RM_{13}$ (Sect.~\ref{in_band})
when available, so that 
we mostly have a comparable number of parameters and data 
points. It makes no sense to use more than two components in the model 
because the number of parameters would exceed the number of data.

In order to constrain the model as much as possible we used as
starting parameters, when available, the data of the two more
polarised components from the high resolution VLA data (\citeauthor*{Fanti04}):
{\it i)} $m_{3.6}$, {\it ii)} $RM^{3.6}_{6}$, {\it iii)}  $\chi_{3.6}$
corrected to zero wavelength for  $RM^{3.6}_{6}$. It has to be realized
that $RM^{3.6}_{6}$ has an uncertainty $\ge 50$~rad\, m$^{-2}$. 
Furthermore it is also possible that the beat is due to subcomponents
unresolved even at our best VLA resolution, so that the input data
would not be the relevant ones.

Sources of  group i) are mostly fitted by the
model, ranging from  good and well  constrained cases (see
e.g. B3\,0110+401) to less secure cases (e.g. B3\,1340+439 and B3\,2349+410
which are detected at two wavelengths  only). 
The success is due to the fact that the fluctuations in $m_{\lambda}$ help 
in constraining some of the
parameters, such as the difference between the two  $RM$s 
and the intrinsic fractional polarisation $m_0$.
For instance, undetected points are taken as deep minima in $m$.
The level of success is definitely lower ($\le 30$\%) in the second group.

The results of the model have to be taken with caution because of the 
limited number of data points. When reasonably secure, we give the
average {\it RM}, the common $\sigma_{\rm RM,obs}$ and $f_c$ and the intrinsic
fractional polarisation $m_0$ (vectorial sum of $m(\lambda=0)$ of
the two components).
 
Comments on individual fits are contained in the notes to
individual sources (\ref{notes}).

\end{document}